\documentclass[aps,rmp,showpacs,twocolumn,nofootinbib,superscriptaddress,
reprint,amsmath,amssymb]{revtex4-1}

\usepackage{epsfig}
\usepackage{psfrag}
\usepackage{times}
\usepackage{amsfonts}
\usepackage{array}
\usepackage{stmaryrd,latexsym,bm,bbm}
\usepackage{color}
\usepackage{epstopdf}
\usepackage[
colorlinks=true,
linkcolor=black,
breaklinks=true,
urlcolor=blue,
citecolor=green]{hyperref}
\usepackage[utf8]{inputenc}

\definecolor{darkblue}{rgb}{0.,0.,0.7}
\definecolor{light-blue}{rgb}{0.8,0.85,1}
\definecolor{green}{rgb}{0,0.6,0}

\newcolumntype{L}{>{$}l<{$}} 
\newcolumntype{C}{>{$}c<{$}} 
\newcolumntype{R}{>{$}r<{$}}

\newcommand{\br}[1]{\mathcal{B}\left(#1\right)}
\newcommand{\order}[1]{\mathcal{O}\!\left(#1\right)\!\,}

\newcommand{\be}{\begin{equation}}
\newcommand{\ee}{\end{equation}}
\newcommand{\beq}{\begin{equation}}
\newcommand{\eeq}{\end{equation}}
\newcommand{\ba}{\begin{eqnarray}}
\newcommand{\ea}{\end{eqnarray}}
\newcommand{\beqa}{\begin{eqnarray}}
\newcommand{\eeqa}{\end{eqnarray}}

\newcommand{\Ham}{\mathcal{H}}

\newcommand{\Amp}{\mathcal{A}}

\newcommand{\Y}{Y(4260)}
\newcommand{\Z}{Z_c(3900)}
\newcommand{\nreft}{NREFT$_\text{I}$}
\newcommand{\nreftii}{NREFT$_\text{II}$}
\newcommand{\bd}{ \bm{  D  } }
\newcommand{\bdbar}{ \bm{  {\bar D } } }

\newcommand{\X}{X(3872)}
\newcommand{\Zc}{Z_c(3900)}
\newcommand{\Zcp}{Z_c(4020)}
\newcommand{\Zb}{Z_b(10610)}
\newcommand{\Zbp}{Z_b(10650)}
\newcommand{\mev}{\mathrm{MeV}}
\newcommand{\gev}{\mathrm{GeV}}

\newcommand{\ev}{\mathrm{eV}}

\newcommand{\itp}{\affiliation{CAS Key Laboratory of Theoretical Physics,
            Institute of Theoretical Physics, Chinese Academy of Sciences,
            Beijing 100190, China}}

\newcommand{\bonn}{\affiliation{Helmholtz-Institut f\"ur Strahlen- und
             Kernphysik and Bethe Center for Theoretical Physics,
             Universit\"at Bonn,  D-53115 Bonn, Germany}}

\newcommand{\fzj}{\affiliation{Institute for
           Advanced Simulation, Institut f\"ur Kernphysik and
           J\"ulich Center for Hadron Physics, Forschungszentrum J\"ulich,
           D-52425 J\"ulich, Germany}}

\newcommand{\ihep}{\affiliation{Institute of High Energy Physics,
           Chinese Academy of Sciences, Beijing 100049, China}}

\newcommand{\ucas}{\affiliation{School of Physical Sciences,
            University of Chinese Academy of Sciences,
            Beijing 100049, China}}

\newcommand{\TPCSF}{\affiliation{Theoretical Physics Center for Science Facilities,
         Chinese Academy of Sciences, Beijing 100049, China}}

\date{\today}

\begin{document}

\title{{Hadronic molecules} }

\itp
\fzj
\bonn
\ihep
\ucas

\author{Feng-Kun~Guo} \email{fkguo@itp.ac.cn}\itp\ucas

\author{Christoph~Hanhart} \email{c.hanhart@fz-juelich.de}\fzj

\author{Ulf-G.~Mei\ss{}ner} \email{meissner@hiskp.uni-bonn.de}\bonn\fzj

\author{Qian~Wang} \email{wangqian@hiskp.uni-bonn.de}\bonn

\author{Qiang~Zhao} \email{zhaoq@ihep.ac.cn}\ihep\ucas\TPCSF

\author{Bing-Song~Zou} \email{zoubs@itp.ac.cn}\itp\ucas

\begin{abstract}

A large number of experimental discoveries especially in the heavy quarkonium
sector that did not at all fit to the expectations of the until then very
successful quark model led to a renaissance of hadron spectroscopy.
Among various explanations of the internal structure of these excitations,
hadronic molecules, being analogues of light nuclei, play a unique role since
for those predictions can be made with controlled uncertainty. We review
experimental evidences of various candidates of hadronic molecules, and methods
of identifying such structures. Nonrelativistic effective field theories are the
suitable framework for studying hadronic molecules, and are discussed in both
the continuum and finite volumes. Also pertinent lattice QCD results are
presented. Further, we discuss the production mechanisms and decays of hadronic
molecules, and comment on the reliability of certain assertions often made in
the literature.

\end{abstract}

\maketitle
\newpage

\tableofcontents

\newpage

\section{Introduction}
\label{sec:1}

With the discovery of the deuterium in 1931 and the neutron in 1932, the first
bound state of two hadrons, {\sl i.e.}, the deuteron composed of one proton and
one neutron, became known.  The deuteron is very shallowly bound, by a mere MeV
per nucleon, i.e. it is located just below the neutron-proton continuum
threshold. Furthermore, it has a sizeable spatial extension. These two features
can be used for defining a hadronic molecule. A more precise definition will be
given in the course of this review.

Then the first meson, the pion, as the carrier particle of the nuclear force
proposed in 1935 by  Yukawa was discovered in 1947, followed by the discovery of
a second meson, the kaon, in the same year.
Since then, many different hadrons have been observed. Naturally hadronic
molecules other than the deuteron have been expected.  The first identified
meson-baryon molecule, {\sl i.e.}, the $\Lambda(1405)$ resonance composed of one
kaon and one nucleon, was predicted by Dalitz and Tuan in
1959~\cite{Dalitz:1959dn} and observed in the hydrogen bubble chamber at
Berkeley in 1961~\cite{Alston:1961zzd} several years before the  quark model was
proposed.
With the quark model developed in the early 1960s, it became clear that  hadrons
are not elementary particles, but composed of quarks and antiquarks. In the
classical quark model, a baryon is composed of three quarks and a meson is
composed of one quark and one antiquark.
In this picture, the $\Lambda(1405)$ resonance would be an excited state of a
three-quark ($uds$) system with one quark in an orbital $P$-wave excitation. Ten
years later, the theory of the strong interactions, Quantum Chromodynamics
(QCD), was proposed to describe the interactions between quarks as well as
gluons. The gluons  are the force carriers of the theory that also exhibit
self-interactions due to the non-abelian nature of the underlying gauge group,
SU(3)$_C$, where $C$ denotes the color degree of freedom.
In QCD the basic constituents of the hadrons are both quarks and gluons.
Therefore, the structure of hadrons is more complicated than the classical quark
model allows. There may be glueballs (which contain only valence gluons),
hybrids (which contain valence quarks as well as gluons) and multiquark states
(such as tetraquarks or pentaquarks). Note, however, that in principle the quark
model also allows for certain types of multiquark states~\cite{GellMann:1964nj}.

While the classical quark model is very successful in explaining properties of
the spatial ground states of the flavor SU(3) vector meson nonet, baryon octet
and decuplet, it fails badly even for the lowest spatial excited states in both
meson and baryon sectors.

In the meson sector, the lowest spatial excited SU(3) nonet is supposed to be
the lowest scalar nonet which includes the $f_0(500)$, the $\kappa(800)$, the
$a_0(980)$ and the $f_0(980)$. In the classical constituent quark model, these
scalars should be $q\bar q~(L=1)$ states, where $L$ denotes the orbital angular
momentum, with the $f_0(500)$ as an  $(u\bar u+d\bar d)/\sqrt{2}$ state, the
$a_0^0(980)$ as an $(u\bar u-d\bar d)/\sqrt{2}$ state and the $f_0(980)$ as
mainly an $s\bar s$ state. This picture, however, fails to explain why the mass
of the $a_0(980)$ is degenerate with the $f_0(980)$ instead of being close to
the$f_0(500)$,
 as it is the case of the $\rho$ and the $\omega$ in the vector nonet. Instead,
this kind of mass pattern can be easily understood in the tetraquark
picture~\cite{Jaffe:1976ig} or in a scenario where these states are dynamically
generated from the meson-meson
interaction~\cite{Weinstein:1982gc,Janssen:1994wn,Oller:2000ma}, with the
$f_0(980)$ and the $a_0(980)$ coupling strongly to the $\bar KK$ channel with
isospin 0 and 1, respectively.

In the baryon sector, a similar phenomenon seems also to be
happening~\cite{Zou:2007mk}.
In the classical quark model, the lowest spatial excited baryon is expected to
be a ($uud$) $N^*$ state with one quark in an orbital angular momentum $L=1$
state to have spin-parity $1/2^-$.
However, experimentally, the lowest negative parity $N^*$ resonance is found to
be the $N^*(1535)$, which is heavier than two other spatial excited baryons:
the $\Lambda^*(1405)$ and the $N^*(1440)$. This is the long-standing mass
reversal problem for the lowest spatial excited baryons. Furthermore, it is also
difficult to understand the strange decay properties of the $N^*(1535)$, which
seems to couple strongly to the final states with strangeness~\cite{Liu:2005pm},
as well as the strange decay pattern of another member of the $1/2^-$-nonet, the
$\Lambda^*(1670)$, which has a coupling to $\Lambda\eta$ much larger than to
$NK$ and $\Sigma\pi$ according to its branching ratios listed in the tables in
the Review of Particle Physics by the Particle Data Group
(PDG)~\cite{Olive:2016xmw}.
All these difficulties can be easily understood by assuming large five-quark
components in them~\cite{Zou:2007mk,Liu:2005pm,Helminen:2000jb} or considering
them to be dynamically generated  meson-baryon
states~\cite{Oller:2000ma,Kaiser:1995eg,Oset:1997it,Oller:2000fj,Inoue:2001ip,
GarciaRecio:2003ks,Hyodo:2002pk,Magas:2005vu,Huang:2007zza,Bruns:2010sv}.

No matter which configurations are realized in multiquark states, such as
colored diquark correlations or colorless hadronic clusters, the mass and decay
patterns for the lowest meson and baryon nonets strongly suggest that one must
go beyond the classical, so-called quenched, quark model. The unquenched picture
has been further supported by more examples of higher excited states in the
light quark sector, such as the $f_1(1420)$ as a $K^*\bar K$
molecule~\cite{Tornqvist:1993ng}, and by many newly observed states with heavy
quarks in the first decade of the new century, such as the $D^*_{s0}(2317)$ as a
$DK$ molecule or tetraquark state, $X(3872)$ as $D^*\bar D$ molecule or
tetraquark state~\cite{Chen:2016qju}. In fact, the possible existence of
hadronic molecules composed of two charmed mesons was already proposed 40 years
ago by Voloshin and Okun~\cite{Voloshin:1976ap} and supported by T{\"o}rnqvist
later within a one-pion exchange model~\cite{Tornqvist:1993ng}.

However, although many hadron resonances were proposed to be dynamically
generated states from various hadron-hadron interactions or multiquark states,
most of them cannot be clearly distinguished from classical quark model states
due to tunable ingredients and possible large mixing of various configurations
in these models.
A nice example is the already mentioned $\Lambda(1405)$.
Until 2010, {\sl i.e.}, 40 years after it  was predicted and observed as the
$\bar KN$ molecule, the PDG~\cite{Nakamura:2010zzi} still claimed that ``the
clean $\Lambda_c$ spectrum has in fact been taken to settle the decades-long
discussion about the nature of the $\Lambda(1405)$ --- true 3-quark state or
mere $\bar KN$ threshold effect? --- unambiguously in favor of the first
interpretation."  Only after many delicate analyses of various relevant
processes, the PDG~\cite{Olive:2016xmw} now acknowledges the two-pole structure
of the $\Lambda(1405)$ \cite{Oller:2000fj} and thus a dynamical generation is
most probable.

One way to unambiguously identify a multiquark state (including hadronic
molecular configurations) is the observation of resonances decaying into a heavy
quarkonium plus a meson with nonzero isospin made of light quarks or plus a
baryon made of light quarks. Since 2008, several such states have been claimed,
six $Z_c$ states, two $Z_b$ states and two $P_c$ states --- details on the
experimental situation are given in the next section. Among these newly claimed
states, the two $P_c$ states are quite close to the predicted hadronic molecular
states~\cite{Wu:2010jy,Wang:2011rga,Yang:2011wz,Xiao:2013yca}. However, many of
those states are challenged by some proposed kinematic explanations, such as threshold cusp
effects~\cite{Bugg:2011jr,Swanson:2014tra}, triangle singularity
effects~\cite{Chen:2013coa,Wang:2013cya,Guo:2015umn}, etc.
Some of these claims were challenged in Ref.~\cite{Guo:2014iya} where strong
support is presented that at least some of the signals indeed refer to
$S$-matrix poles.

Further experimental as well as theoretical studies are necessary to settle the
question which of the claimed states indeed exist.
Nevertheless the observation of at least some of these new states opens a new
window for the study of multiquark dynamics. Together with many other newly
observed states in the heavy quarkonium sector, they led to a renaissance of
hadron spectroscopy.
Among various explanations of the internal structure of these excitations,
hadronic molecules, being analogues of the deuteron, play a unique role since
for those states predictions can be made with controlled uncertainty, especially
for the states with one of or both hadrons containing heavy quark(s).
In fact most of these observed exotic candidates are indeed closely related to
open flavor $S$-wave thresholds. To study these hadronic molecules, both
nonrelativistic effective field theories and pertinent lattice QCD calculation
are the suitable frameworks. Especially,  Weinberg's famous compositeness
criterion~\cite{Weinberg:1962hj,Weinberg:1963zza} (and extensions thereof),
which pinned down the nature of the deuteron as a proton-neutron bound state, is
applicable here. The pole location in the corresponding hadron-hadron scattering
$S$-matrix could also shed light on the nature of the resonances as extended
hadronic molecules or compact states.

The revival of hadron spectroscopy is also reflected in a number of review
articles.
A few years ago, Klempt and his collaborators have given two broad reviews on
exotic mesons~\cite{Klempt:2007cp} and baryons~\cite{Klempt:2009pi}.
Other more recent pertinent reviews
include~\cite{Brambilla:2010cs,Olsen:2014qna,Oset:2016lyh,Chen:2016qju,
Chen:2016spr,Esposito:2016noz, Lebed:2016hpi,Hosaka:2016pey,Dong:2017gaw,Olsen:2017bmm}. Among
various theoretical models for these new hadrons, we
mainly cite those focusing on hadronic molecules and refer the interesting
readers to the above mentioned comprehensive reviews for more references on
other models.

This paper is organized as follows: In Sec.~\ref{sec:2}, we discuss the
experimental evidences for states that could possibly be hadronic molecules. In
Sec.~\ref{sec:3}, after a short review of the basic $S$-matrix properties, we
give a general definition of hadronic molecules and discuss related aspects.
Then, in Sec.~\ref{sec:4}, nonrelativistic effective field theories tailored to
investigate hadronic molecules are formulated, followed by a brief discussion of
hadronic molecules in lattice QCD in Sec.~\ref{sec:lattice}. Sec.~\ref{sec:6} is
devoted to the discussion of phenomenological manifestations of hadronic
molecules, with a particular emphasis on clarifying certain statements from the
literature that have been used to dismiss certain states as possible hadronic
molecules. We end with a short summary and outlook in Sec.~\ref{sec:sum}.  We
mention that this field is very active, and thus only references that appeared
before April 2017 are included.

\section{Candidates of hadronic molecules --- experimental evidences}
\label{sec:2}

\begin{table*}[tbh]
 \caption{Mesons that contain at most one heavy quark that cannot be easily
 accommodated in the $q\bar q$ quark model.
  Their quantum numbers $I^G(J^{PC})$, masses, widths,
 the nearby $S$-wave thresholds, $m_{\text{threshold}}$, where we add in 
 brackets $M-m_{\text{threshold}}$, and the observed decay modes are listed in 
 order. The data without references are taken from the 2016 edition of the 
 Review of Particle Physics~\cite{Olive:2016xmw}.  
 }
 \begin{ruledtabular}
  \begin{tabular}{l c c c c c}
State & $I^G(J^{PC})$ & $M \ [\mathrm{MeV}]$ & $\Gamma[\mathrm{MeV}]$ & $S$-wave
threshold(s) [MeV] & Decay mode(s) [branching ratio(s)]\tabularnewline
\hline
$f_{0}(500)$~\cite{Pelaez:2015qba}~\footnote{The mass and width
are derived from the pole position quoted in Ref.~\cite{Pelaez:2015qba} via
$\sqrt{s_p}=M-i\Gamma/2$. } & $0^{+}(0^{++})$ & $449^{+22}_{-16}$ & $550\pm 24$
& $\pi\pi(173^{+22}_{-16})$ & $\pi\pi$ [dominant]\tabularnewline &  &  &  &  &
$\gamma\gamma$\tabularnewline
\hline
$\kappa(800)$ & $\frac{1}{2}(0^{+})$ & $682\pm29$ & $547\pm24$ & $K\pi(48\pm 29)$ & $\pi K$\tabularnewline
\hline
$f_{0}(980)$ & $0^{+}(0^{++})$ & $990\pm20$ & $10\sim100$ & $K^+K^-(3\pm20)$ &
$\pi\pi$ [dominant]\tabularnewline &  &  &  & $K^0\bar{K}^0(-5\pm20)$ &
$K\bar{K}$\tabularnewline &  &  &  &  & $\gamma\gamma$\tabularnewline
\hline
$a_{0}(980)$ & $1^{-}(0^{++})$ & $980\pm20$ & $50\sim100$ & $K\bar{K}(-11\pm20)$
& $\eta\pi$ [dominant]\tabularnewline &  &  &  &  & $K\bar{K}$\tabularnewline
 &  &  &  &  & $\gamma\gamma$\tabularnewline
\hline
$f_{1}(1420)$ & $0^{+}(1^{++})$ & $1426.4\pm0.9$ & $54.9\pm2.6$ & $K\bar{K}^{*}(39.1\pm0.9)$ & $K\bar{K}^{*}$(dominant)\tabularnewline
 &  &  &  &  & $\eta\pi\pi$ [possibly seen]\tabularnewline
 &  &  &  &  & $\phi\gamma$\tabularnewline
\hline
$a_{1}(1420)$ & $1^{-}(1^{++})$ & $1414_{-13}^{+15}$ & $153_{-23}^{+8}$ & $K\bar{K}^{*}(27^{+15}_{-13}$) &
$f_{0}(980)\pi$ [seen]\tabularnewline
\hline
$X(1835)$ & $?^{?}(0^{-+})$ & $1835.8_{-3.2}^{+4.0}$ & $112\pm40$ & $p\bar{p}(-40.7^{+4.0}_{-3.2})$ & $p\bar{p}$\tabularnewline
 &  &  &  &  & $\eta^{\prime}\pi\pi$\tabularnewline
 &  &  &  &  & $K_{S}^{0}K_{S}^{0}\eta$\tabularnewline
\hline
$D_{s0}^{*}(2317)^{+}$ & $0(0^{+})$ & $2317.7\pm0.6$ & $<3.8$ & $DK(-45.1\pm 
0.6)$ & $ D_{s}^{+}\pi^{0}$\tabularnewline
\hline
$D_{s1}(2460)^{+}$ & $0(1^{+})$ & $2459.5\pm0.6$ & $<3.5$ & $D^{*}K(-44.7\pm 
0.6)$ & $D_{s}^{*+}\pi^{0}\,[(48\pm 11)\%]$\tabularnewline
 &  &  &  &  & $D_{s}^{+}\gamma\, [(18\pm 4)\%]$\tabularnewline
 &  &  &  &  & $D_{s}^{+}\pi^{+}\pi^{-}\, [(4\pm 1) \%]$\tabularnewline
 &  &  &  &  & $D_{s0}^*(2317)^{+}\gamma\, [(4^{+5}_{-2})\%]$\tabularnewline
 \hline
$D_{s1}^{*}(2860)^{+}$ & $0(1^{-})$ & $2859 \pm 27$ & $159\pm 80 $ & 
$D_1(2420)K(-59 \pm 27)$ & $ DK$\tabularnewline
 &  &  &  &  & $D^*K$ \tabularnewline
\end{tabular}
\end{ruledtabular}
\label{tab:1}
\end{table*}

 \begin{table*}
 \caption{Same as Table~\ref{tab:1} but in the 
 charmonium and bottomonium sectors. A blank in the fifth column means that 
 there is no relevant nearby $S$-wave threshold. 
}
\begin{ruledtabular}
 \begin{tabular}{l  c c c c c }
State & $I^G(J^{PC})$ & $M\,[\mathrm{MeV}]$ & $\Gamma\,[\mathrm{MeV}]$ &
$S$-wave threshold(s) [$\mathrm{MeV}$] & Observed mode(s) (branching
ratios)\tabularnewline
\hline
$X(3872)$ & $0^{+}(1^{++})$ & $3871.69\pm0.17$ & $<1.2$ & $D^{*+}D^{-}+c.c.(-8.15\pm0.20)$ & $B\to K[\bar{D}^{*0}D^{0}](>24\%)$\tabularnewline
 &  &  &  & $D^{*0}\bar{D}^{0}+c.c.(0.00\pm0.18)$ & $B\to
 K[D^{0}\bar{D}^{0}\pi^{0}](>32\%)$\tabularnewline &  &  &  &  & $B\to K[J/\psi\pi^{+}\pi^{-}](>2.6\%)$\tabularnewline
 &  &  &  &  & $B\to K[J/\psi\pi^{+}\pi^{-}\pi^{0}]$\tabularnewline
 &  &  &  &  & $p\bar{p}\to[J/\psi\pi^{+}\pi^{-}]...$\tabularnewline
 &  &  &  &  & $pp\to[J/\psi\pi^{+}\pi^{-}]...$\tabularnewline
 &  &  &  &  & $B\to K[J/\psi\omega](>1.9\%)$\tabularnewline
 &  &  &  &  & $B\to[J/\psi\gamma](>6\times10^{-3})$\tabularnewline
 &  &  &  &  & $B\to[\psi(2S)\gamma](>3.0\%)$\tabularnewline
\hline
$X(3940)$ & $?^{?}(?^{??})$ & $3942.0\pm9$ & $37_{-17}^{+27}$ & $D^{*}\bar{D}^{*}(-75.1\pm 9)$ & $e^{+}e^{-}\to
J/\psi[D\bar{D}^{*}]$\tabularnewline
\hline
$X(4160)$ & $?^{?}(?^{??})$ & $4156_{-25}^{+29}$ & $139_{-60}^{+110}$ & $D^{*}\bar{D}^{*}(139_{-25}^{+29})$ & $e^{+}e^{-}\to
J/\psi[D^{*}\bar{D}^{*}]$\tabularnewline
\hline
$Z_{c}(3900)$ & $1^{+}(1^{+-})$ & $3886.6\pm2.4$ & $28.1\pm2.6$ & $D^{*}\bar{D}(10.8\pm2.4)$ &
$e^{+}e^{-}\to\pi[D\bar{D}^{*}+c.c.]$\tabularnewline
 &  &  &  &  & $e^{+}e^{-}\to\pi[J/\psi\pi]$\tabularnewline
\hline
$Z_{c}(4020)$ & $1(?^{?})$ & $4024.1\pm1.9$ & $13\pm5$ & $D^{*}\bar{D}^{*}(7.0\pm 2.4)$ & $e^{+}e^{-}\to\pi[D^{*}\bar{D}^{*}]$\tabularnewline
 &  &  &  &  & $e^{+}e^{-}\to\pi[h_{c}\pi]$\tabularnewline
 &  &  &  &  & $e^{+}e^{-}\to\pi[\psi'\pi]$\tabularnewline
\hline
$Y(4260)$ & $?^{?}(1^{--})$ & $4251\pm9$ & $120\pm12$ & 
$D_{1}\bar{D}+c.c.(-38.2\pm 9.1)$ & $e^{+}e^{-}\to 
J/\psi\pi\pi$\tabularnewline
 &  &  &  & $\chi_{c0}\omega(53.6\pm 9.0)$ & $e^{+}e^{-}\to
 \pi D\bar D^* +c.c.$\tabularnewline 
  &  &  &  &  & $e^{+}e^{-}\to \chi_{c0}\omega$\tabularnewline
 &  &  &  &  & $e^{+}e^{-}\to
 X(3872)\gamma$\tabularnewline
\hline
$Y(4360)$ & $?^{?}(1^{--})$ & $4346\pm6$ & $102\pm10$ & $D_{1}\bar{D}^{*}+c.c.(-85\pm6)$ &
$e^{+}e^{-}\to\psi(2S)\pi^{+}\pi^{-}$\tabularnewline
\hline
$Y(4660)$ & $?^{?}(1^{--})$ & $4643\pm9$ & $72\pm11$ & $\psi(2S)f_{0}(980)(-33\pm21)$ &
$e^{+}e^{-}\to\psi(2S)\pi^{+}\pi^{-}$\tabularnewline
 &  &  &  & $\Lambda_{c}^{+}\Lambda_{c}^{-}(70\pm 6)$ & \tabularnewline
\hline
$Z_c(4430)^{+}$ & $?(1^{+})$ & $4478_{-18}^{+15}$ & $181\pm31$ & 
$\psi(2S)\rho(17^{+15}_{-18})$ & $B\to
K[\psi(2S)\pi^{+}]$\tabularnewline
 &  &  &  &  & $B\to K[J/\psi\pi^{+}]$\tabularnewline
\hline
$Z_{c}(4200)^{+}$ & $?(1^{+})$ & $4196_{-32}^{+35}$ & $370_{-32}^{+100}$ &  & $\bar{B}^{0}\to K^{-}[J/\psi\pi^{+}]$\tabularnewline
\hline
$Z_{c}(4050)^{+}$ & $?(?^{?})$ & $4051_{-40}^{+24}$ & $82_{-28}^{+50}$ & $D^*\bar{D}^*(34^{+24}_{-40})$ & $\bar{B}^{0}\to K^{-}[\chi_{c1}\pi^{+}]$\tabularnewline
\hline
$Z_{c}(4250)^{+}$ & $?(?^{?})$ & $4248_{-50}^{+190}$ & $177_{-70}^{+320}$ & $\chi_{c1}\rho(-37^{+24}_{-50})$ & $\bar{B}^{0}\to K^{-}[\chi_{c1}\pi^{+}]$\tabularnewline
\hline
$X(4140)$\cite{Aaij:2016nsc,Aaij:2016iza} & $0^{+}(1^{++})$ & $4146.5\pm4.5_{-2.8}^{+4.6}$ & $83\pm21_{-14}^{+21}$ &
$D_{s}\bar D_{s}^{*}(-66.1_{-3.2}^{+4.9})$ & $B^{+}\to K^{+}[J/\psi\phi]$\tabularnewline
\hline
$X(4274)$\cite{Aaij:2016nsc,Aaij:2016iza} & $0^{+}(1^{++})$ & $4273.3\pm8.3_{-3.6}^{+17.2}$ & $56\pm11_{-11}^{+8}$ &
$D_{s}^{*}\bar D_{s}^{*}(-49.1_{-9.1}^{+19.1})$ & $B^{+}\to K^{+}[J/\psi\phi]$\tabularnewline
\hline
$X(4500)$\cite{Aaij:2016nsc,Aaij:2016iza} & $0^{+}(0^{++})$ & $4506\pm11_{-15}^{+12}$ & $92\pm21_{-20}^{+21}$ & $D_{s0}^*(2317)\bar D_{s0}^*(2317)(-129^{+16}_{-19})$ & $B^{+}\to
K^{+}[J/\psi\phi]$\tabularnewline
\hline
$X(4700)$\cite{Aaij:2016nsc,Aaij:2016iza} & $0^{+}(0^{++})$ & $4704\pm10_{-24}^{+14}$ & $120\pm31_{-33}^{+42}$ & $D_{s0}^*(2317)\bar D_{s0}^*(2317)(69^{+17}_{-26})$  & $B^{+}\to
K^{+}[J/\psi\phi]$\tabularnewline
\hline
$Z_{b}(10610)$ & $1^{+}(1^{+})$ & $10607.2\pm2.0$ & $18.4\pm2.4$ & $B\bar{B}^{*}+c.c.(4.0\pm3.2)$ &
$\Upsilon(10860)\to\pi[B\bar{B}^{*}+c.c.]$\tabularnewline
 &  &  &  &  & $\Upsilon(10860)\to\pi[\Upsilon(1S)\pi]$\tabularnewline
 &  &  &  &  & $\Upsilon(10860)\to\pi[\Upsilon(2S)\pi]$\tabularnewline
 &  &  &  &  & $\Upsilon(10860)\to\pi[\Upsilon(3S)\pi]$\tabularnewline
 &  &  &  &  & $\Upsilon(10860)\to\pi[h_{b}(1P)\pi]$\tabularnewline
 &  &  &  &  & $\Upsilon(10860)\to\pi[h_{b}(2P)\pi]$\tabularnewline
\hline
$Z_{b}(10650)$ & $1^{+}(1^{+})$ & $10652.2\pm1.5$ & $11.5\pm2.2$ & $B^{*}\bar{B}^{*}(2.9 \pm1.5)$ &
$\Upsilon(10860)\to\pi[B^{*}\bar{B}^{*}]$\tabularnewline
 &  &  &  &  & $\Upsilon(10860)\to\pi[\Upsilon(1S)\pi]$\tabularnewline
 &  &  &  &  & $\Upsilon(10860)\to\pi[\Upsilon(2S)\pi]$\tabularnewline
 &  &  &  &  & $\Upsilon(10860)\to\pi[\Upsilon(3S)\pi]$\tabularnewline
 &  &  &  &  & $\Upsilon(10860)\to\pi[h_{b}(1P)\pi]$\tabularnewline
 &  &  &  &  & $\Upsilon(10860)\to\pi[h_{b}(2P)\pi]$\tabularnewline
\end{tabular}
\end{ruledtabular}
\label{tab:2}
\end{table*}

In this section we briefly review what is known experimentally about some of the
most promising candidates for exotic states.
Already the fact that those are all located close to some two-hadron continuum
channels indicates that the two-hadron continuum is of relevance for their
existence. {We will show  that many of those states are located near
$S$-wave thresholds in both light and heavy hadron spectroscopy, which is
not only a natural property of hadronic molecules, which are QCD bound states
of two hadrons (a more
proper definition will be given in Sec.~\ref{sec:wein}), but also a
prerequisite for their identification as will be discussed in Sec.~\ref{sec:3}.} In the course of this
review, we will present other arguments why many of these states should be
considered as hadronic molecules and what additional experimental inputs are
needed to further confirm this assignment.

In Tables~\ref{tab:1} and ~\ref{tab:2} we present the current status for exotic
candidates in the meson sector. Exotic candidates in the baryon sector are
listed later in Tab.~\ref{tab:baryon}. Besides the standard properties we also
quote for each state the nearest relevant $S$-wave threshold  as well as its
distance to that threshold.
Note that only thresholds of narrow states are quoted since these are the only
ones of relevance here~\cite{Guo:2011dd}. Otherwise, the bound system would also
be broad~\cite{Filin:2010se}. In addition, as a result of the centrifugal
barrier one expects that if hadronic molecules exist, they should first of all
appear in the $S$-wave which is why in this review we do not consider $P$- or
higher partial waves although there is no principle reason for the non-existence
of molecular states in the $P$-wave.

\subsection{Light mesons}
\label{sec:2-lightmesons}

\subsubsection{Scalars below 1 GeV }

The lowest $S$-wave two-particle thresholds in the hadron sector are those for
 two pseudoscalar mesons, $\pi\pi$, $\pi K$, $\eta\pi$, and $K\bar{K}$.
Those channels carry scalar quantum numbers. The pion pair is either in an
isoscalar or an isotensor state, and the isovector state is necessarily in a
$P$-wave. It turns out that there is neither a resonant structure in the
isotensor $\pi\pi$ nor in the isospin $3/2$ $\pi K$ $S$-wave, however, there are
resonances observed experimentally in all other channels. According to the
conventional quark model, a scalar meson of $q\bar{q}$ with $J^P=0^+$ carries
one unit of orbital angular momentum. Thus, the mass range of the  lowest
scalars is expected to be higher than the lowest pseudoscalars or vectors of
which the orbital angular momentum is zero. However, the lightest scalars have
masses below those of the lightest vectors. Moreover, the mass ordering of the
lightest scalars apparently violates the pattern of other $q\bar{q}$ nonets:
Instead of having the isovectors to be the lowest states, the isovector
$a_0(980)$ states are almost degenerate with one of the isoscalar states,
$f_0(980)$, and those are the heaviest states in the nonet. The other isoscalar,
$f_0(500)$, also known as $\sigma$, has the lightest mass of the multiplet and
an extremely large width. The strange scalar $K_0^*(800)$, also known as
$\kappa$, has a  large width as well. All these indicate some nontrivial
substructure beyond a simple $q\bar{q}$ description.

The mass ordering of these lightest scalars is seen as a strong evidence for the
tetraquark scenario proposed by Jaffe in the
1970s~\cite{Jaffe:1976ig,Jaffe:1976ih}. Meanwhile, they can also be described
as dynamically generated states through meson-meson
scatterings~\cite{Pennington:1973xv,Au:1986vs,Morgan:1993td,Pelaez:2015qba}.
For a theoretical understanding of the $f_0(500)$ pole it is crucial to
recognize that as a consequence of the chiral symmetry of QCD the scalar isoscalar $\pi
\pi$ interaction is proportional to $(2s-M_\pi^2)/F_\pi^2$ at the leading order
(LO) in the chiral expansion. Here, $M_\pi (F_\pi)$ denotes the pion mass (decay
constant).
As a result, the LO scattering amplitude has already hit the
unitarity bound for moderate energies necessitating some type of unitarization,
which at the same time generates  a resonance-like
structure~\cite{Meissner:1990kz}.
This observation, deeply nested in the symmetries of QCD, has indicated the
significance of the $\pi\pi$ interaction for the light scalar mesons. The
history and the modern developments regarding the $f_0(500)$ was recently very
nicely reviewed in~\cite{Pelaez:2015qba}.
Similar to the isoscalar scalar $f_0(500)$ generated from the $\pi\pi$
scattering, the whole light scalar nonet appears naturally from properly unitarized chiral
amplitudes for pseudoscalar-pseudoscalar
scatterings~\cite{Oller:1997ng,Oller:1998hw,GomezNicola:2001as}. Similar
conclusions also follow from more phenomenological
studies~\cite{Weinstein:1990gu,Janssen:1994wn}.
One of the most interesting observations about $a_0(980)$ and $f_0(980)$ is that
their masses are almost exactly located at the $K\bar{K}$ threshold. The
closeness of the $K\bar{K}$ threshold to $a_0(980)$ and $f_0(980)$ and their
strong $S$-wave couplings makes both states good candidates for $\bar KK$
molecular states~\cite{Weinstein:1990gu,Baru:2003qq}.

\subsubsection{Axial vectors $f_1(1420)$, $a_1(1420)$ and implications of the 
triangle singularity}
\label{sec:TS}

The $S$-wave pseudoscalar meson pair scatterings can be extended to $S$-wave
pseudoscalar-vector scatterings and vector-vector scatterings where again
dynamically generated states can be investigated. The $S$-wave
pseudoscalar-vector scatterings can access the quantum numbers $J^P=1^+$, while
the vector-vector scatterings give $J^P=0^+$, $1^+$ and $2^+$.
This suggests that some of the states with those quantum numbers can be affected
by the $S$-wave open thresholds if their masses are close enough to the
thresholds. Or, it might be possible that such scatterings can dynamically
generate states as discussed in the
literature~\cite{Lutz:2003fm,Roca:2005nm,Geng:2008gx}.
Note that not all states found in these studies survive once a more
sophisticated and realistic treatment as outlined in Ref.~\cite{Gulmez:2016scm}
is utilized.

In addition, the quark model also predicts regular $q\bar q$ states in the same
mass range such that it appears difficult to identify the most prominent
structure of the states.

Let us focus on the lowest $1^{++}$ mesons. Despite that these states could be
dynamically generated from the resummed chiral
interactions~\cite{Lutz:2003fm,Roca:2005nm}, there are various experimental
findings consistent with a usual $q\bar q$ nature of the members of the lightest
axial nonet, $f_1(1420)$, $f_1(1285)$, $a_1(1260)$, and
$K_{1A}(1^3P_1)$~\cite{Olive:2016xmw}.
 However, two recent experimental observations expose novel features
in their decay mechanisms which illustrate the relevance of their couplings to
the  two-meson continua. The BESIII Collaboration observed an anomalously large
isospin symmetry breaking in $\eta(1405)/\eta(1475)\to
3\pi$~\cite{BESIII:2012aa}, which could be accounted for by the so-called
triangle singularity (TS) mechanism as studied in Ref.~\cite{Wu:2011yx,Aceti:2012dj}.
This special threshold phenomenon {arises in triangle (three-point loop)
diagrams with special kinematics which} will be detailed in Sect.~\ref{sec:4-pc}.
Physically, it emerges when all the involved vertices in the triangle diagram
can be interpreted as classical processes. For it to happen, one necessary
condition is that all intermediate states in the triangle diagram,
$\bar{K}K^*(K)+c.c.$ for the example at hand, should be able to reach their
on-shell condition simultaneously. As a consequence, the $f_1(1420)$,
which is close to the $\bar K K^*$ threshold and couples to $\bar K K^*$ in an
$S$-wave as well, should also have large isospin violations in $f_1(1420)\to
3\pi$.
This contribution has not been included in the BESIII analysis~\cite{BESIII:2012aa}.  However, a detailed partial wave analysis
suggests the presence of the $f_1(1420)$ contribution via the TS
mechanism~\cite{Wu:2012pg}. Moreover, the TS mechanism predicts structures in
different $C$-parity and isospin (or $G$ parity) channels via the
$\bar{K}K^*(K)+c.c.$ triangle diagrams. The $f_1(1420)$ was speculated long time
ago to be a $\bar K^* K$ molecule from a dynamical study of the $K\bar K\pi$
three-body system~\cite{Longacre:1990uc}.

Apart from the $I=0$, $J^{PC}=1^{++}$ state $f_1(1420)$, one would expect that
the TS will cause enhancements in $I=1$ channels with $C=\pm$. It provides a
natural explanation for the newly observed
 $a_1(1420)$ by the COMPASS Collaboration~\cite{Adolph:2015pws} in $\pi^-p\to
\pi^-\pi^-\pi^+ p$ and $\pi^-\pi^0\pi^0 p$~\cite{Liu:2015taa,Ketzer:2015tqa}. It
should be noted that in
Refs.~\cite{Aceti:2016yeb,Cheng:2016hxi,Debastiani:2016xgg}  the $a_1(1420)$
enhancement is proposed to be caused by the $a_1(1260)$ together with the TS
mechanism and similarly $f_1(1420)$ is produced by $f_1(1285)$.
However, as shown by the convincing experimental data from MARK-III, BESII,
BESIII, and the detailed partial wave analysis of Ref.~\cite{Wu:2012pg}, the
$f_1(1420)$ matches the behavior of a genuine state in the $K\bar{K}\pi$ channel
that is distorted  in other channels by an interference with
 the TS.
This appears to be a more consistent picture to explain the existing data and
underlying mechanisms~\cite{Zhao:2017wey}. These issues are discussed further in
Sec.~\ref{sec:6}.

\subsection{{Open heavy-flavor mesons}}

Since 2003, quite a few {open heavy-flavor} hadrons have been observed
experimentally. Some of them are consistent with the excited states predicted in
the potential quark model, while the others are not (for a recent review,
see~\cite{Chen:2016spr}). Particular interest has been paid to the
positive-parity charm-strange mesons $D_{s0}^*(2317)$ and $D_{s1}(2460)$ observed
in 2003 by the BaBar~\cite{Aubert:2003fg} and CLEO~\cite{Besson:2003cp}
Collaborations.
The masses of $D_{s0}^*(2317)$ and $D_{s1}(2460)$ are below the $DK$ and $D^*K$
thresholds, respectively, by about the same amount, only 45~MeV (see
Table~\ref{tab:1} and references therein), which makes them natural candidates
for hadronic molecules~\cite{Barnes:2003dj,vanBeveren:2003kd,Szczepaniak:2003vy,
Kolomeitsev:2003ac,Hofmann:2003je,
Guo:2006fu,Guo:2006rp,Gamermann:2006nm,Faessler:2007gv,Flynn:2007ki,
Cleven:2010aw,Wu:2011yb,Cleven:2014oka, Albaladejo:2016hae}, while also other
explanations such as $P$-wave $c\bar s$ states and tetraquarks exist in the
literature.
We will come back to the properties of these states occasionally in this review.
Here, we collect the features supporting the $DK/D^*K$ molecular hypothesis:
\begin{itemize}
  
\item Their masses are about $160~\mev$ and  $70~\mev$, respectively, below the
predicted $0^+$ and $1^+$ charm-strange mesons by the Godfrey--Isgur quark
model~\cite{Godfrey:1985xj,DiPierro:2001dwf}, making them not easy to be
accommodated by the conventional $c\bar{s}$ states.

\item {The mass difference between these two states is equal to the energy
difference between the corresponding $D^{(*)}K$  thresholds. 
This appears to be
 a natural
consequence in the hadronic molecular scenario, since the involved interactions
is approximately heavy quark spin symmetric~\cite{Guo:2009id}.}

\item {The small width of both $D_{s0}^*(2317)$ and $D_{s1}(2460)$
can only be understood if the are isoscalar states~\footnote{Negative result was reported in a search for
the isospin partner of the $D_{s0}^*(2317)$~\cite{Choi:2015lpc}}, for then,
since both of them are below the $DK/D^*K$  thresholds, the only possible
hadronic decay modes are the isovector channels $D_s^+\pi^0$ and
$D_s^{*+}\pi^0$, respectively.} The molecular nature together with the proximity
to the $DK/D^*K$ thresholds leads to a prediction for the width of the states
above 100~keV while other approaches give a width about
10~keV~\cite{Colangelo:2003vg,Godfrey:2003kg}.
These issues are discussed in detail in Sections~\ref{sec:latres}
and~\ref{sec:isospinviol}.

\item  Their radiative decays, {\sl i.e.}, $D_{s0}^*(2317)\to D_s \gamma$ and 
$D_{s1}(2460)\to D^{(*)}_s \gamma$, and production in $B$ decays proceed via
short-range interactions~\cite{Lutz:2007sk,Chen:2013upa,Cleven:2014oka}.
They are therefore insensitive to the molecular component of the states.

\item As will be discussed in Secs.~\ref{sec:wein} and \ref{sec:latres}, the
$DK$ scattering length extracted from LQCD {calculations}~\cite{Liu:2012zya} is
compatible with the result extracted in the molecular scenario for
$D_{s0}^*(2317)$ based on Weinberg's compositeness theorem.

\end{itemize}

The $D_{sJ}(2860)$ observed by the BaBar Collaboration~\cite{Aubert:2006mh} 
presents another example of an interesting charm-strange meson. It decays into 
both $DK$ and $D^*K$ with similar branching fractions~\cite{Olive:2016xmw}.  
One notices that the difference between the $D_{sJ}(2860)$ mass and the 
$D_1(2420)K$ threshold is similar to that between the $D_{s0}^*(2317)$ and 
$DK$. Assuming the $D_{s0}^*(2317)$ to be a $DK$ hadronic molecule, an $S$-wave 
$D_1(2420)K$ bound state with quantum numbers $J^P=1^-$ was predicted to have a 
mass $(2870\pm9)$~MeV, consistent with that of the $D_{sJ}(2860)$, 
in~\cite{Guo:2011dd}, where the ratio of its partial widths into the $DK$ and 
$D^*K$ also gets naturally explained. As a result of heavy quark spin symmetry, 
a $D_2(2460)K$ hadronic molecule with $J^P=2^-$ and a mass of around 2.91~GeV 
was predicted in~\cite{Guo:2011dd}.  A later analysis by the LHCb Collaboration 
suggests that this structure corresponds to two states: $D_{s1}^*(2860)$ with 
$J^P=1^-$ and $D_{s3}^*(2860)$ with $J^P=3^-$~\cite{Aaij:2014xza}. Regular 
$c\bar s$ interpretations for these two states have been nicely summarized 
in~\cite{Chen:2016spr}.

The most recently reported observation of an exotic singly-heavy meson candidate
is a narrow structure in the $B_s^0\pi^\pm$ invariant mass distribution, dubbed
as $X(5568)$, by the D0 Collaboration~\cite{D0:2016mwd}.
Were it a hadronic state, it would be an {isovector} meson containing four
different flavors of valence quarks $(\bar b s\bar u d)$. However, the peak is
located at only about 50~MeV above the $B_s\pi$ threshold. The existence of a
tetraquark, whether or not being a hadronic molecule, at such a low mass is
questioned from the quark model point of view in~\cite{Burns:2016gvy}, and, more
generally, from chiral symmetry and heavy quark flavor symmetry
in~\cite{Guo:2016nhb}. Both the LHCb~\cite{Aaij:2016iev} and
CMS~\cite{CMS:2016fvl} Collaborations quickly reported negative results on the
existence of $X(5568)$ in their data sets.
An alternative explanation for the $X(5568)$ observation is necessary. One
possibility is provided in Ref.~\cite{Yang:2016sws}. Because of these
controversial issues with the $X(5568)$, we will not discuss this structure any
further.

\subsection{Heavy quarkonium-like states: {$XYZ$}}

The possibility of hadronic molecules in the charmonium mass region was 
suggested in~\cite{Voloshin:1976ap,DeRujula:1976zlg} only a couple of years  
after the ``November Revolution'' due to the discovery of the $J/\psi$. Such 
an idea became popular after the discovery of the famous $\X$ by Belle in
2003~\cite{Choi:2003ue}.

Since then, numerous other exotic candidates have been found in the heavy
quarkonium sector as listed in Table~\ref{tab:2}. In fact, it is mainly due to 
the observation of these structures that the study of hadron spectroscopy 
experienced a renaissance.
The naming scheme currently used in the literature for these $XYZ$ states 
assigns isoscalar $J^{PC}=1^{--}$ states as $Y$ and
the isoscalar (isovector) states with other quantum numbers are named as $X 
(Z)$. Note that the charged heavy quarkonium-like states $\Zc^\pm$, 
$\Zcp^\pm$, $\Zb^\pm$, $\Zbp^\pm$ and $Z_c
(4430)^\pm$ are already established as being exotic, since they should contain 
at least two quarks and two anti-quarks with the hidden pair of $c\bar c$ or 
$b\bar b$ providing the dominant parts of their masses.

\begin{figure}[tbh]
\begin{center}
 \includegraphics[width=\linewidth]{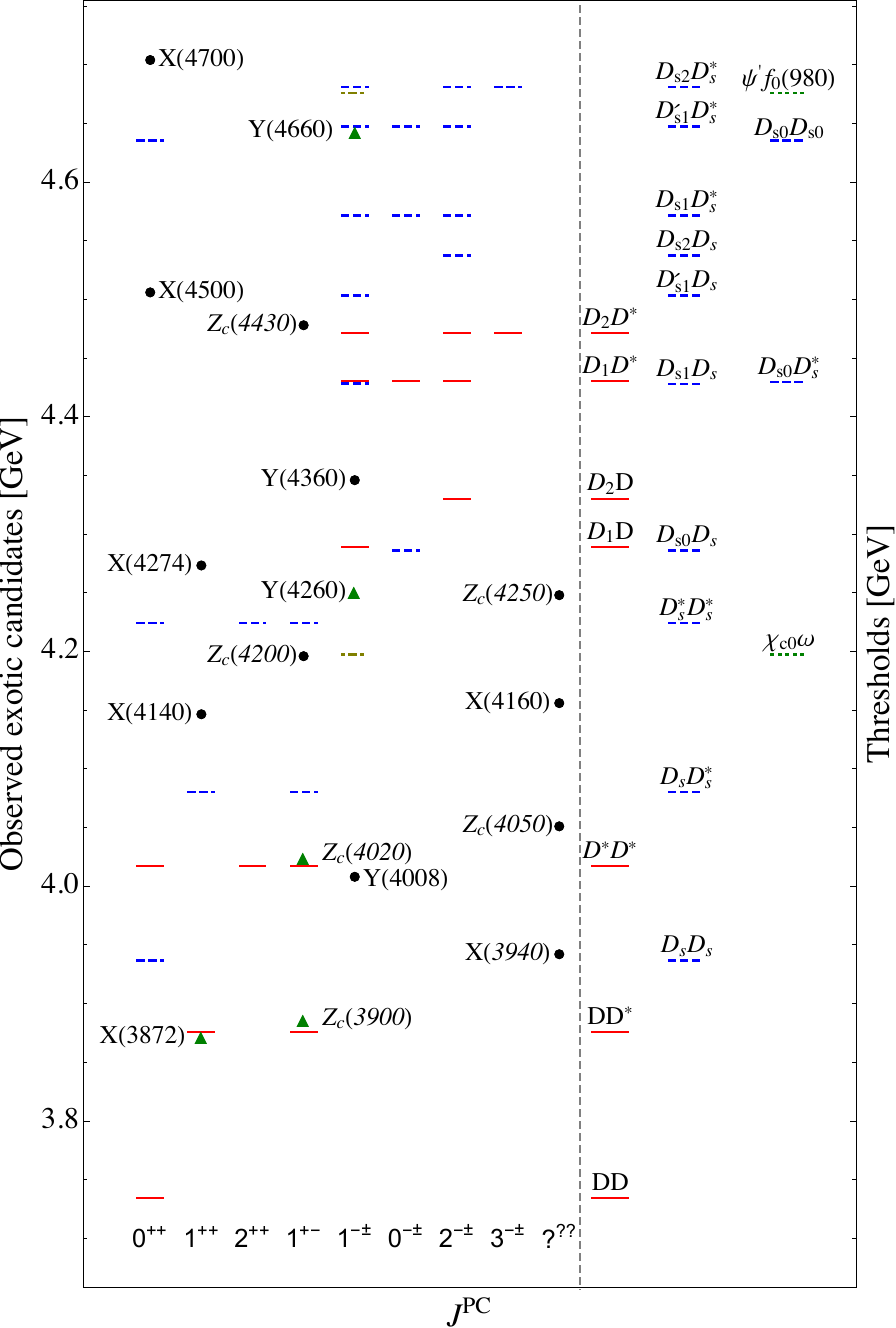}
\caption{$S$-wave open charm thresholds and candidates for exotic states in 
charmonium sector. Red solid (blue dashed)
 horizontal lines indicate the thresholds for nonstrange (strange) meson pairs.
 Two additional thresholds involving a charmonium $\chi_{c0}\omega$ and
 $\psi^\prime f_0(980)$ are also shown in the figure as green dotted lines.
The data are taken from Ref.~\cite{Olive:2016xmw}. The exotic candidates
are listed as black dots and green triangles with the latter marking the states 
to be discussed here. Here $D_{s0}^{}$,
$D_{s1}^{}$, $D_{s1}'$ and $D_{s2}^{}$ mean $D_{s0}^*(2317)$, $D_{s1}^{}(2460)$, 
$D_{s1}^{}(2536)$ and $D_{s2}^{}(2573)$, respectively.}
\label{fig:charmoniumlike}
\end{center}
\end{figure}
In the heavy quarkonium mass region, there are quite a few $S$-wave thresholds 
opened by narrow heavy-meson pairs. In the charmonium mass region, the
lowest-lying thresholds are  $D\bar{D}$, $D\bar{D}^*$ and 
$D^*\bar{D}^*$. They are particularly interesting for understanding the 
$X$ and $Z$ states which can couple to them in an $S$-wave. The relevant 
quantum numbers are thus $J^{PC}=1^{+-}$ and $(0,1,2)^{++}$ (for more details, 
see Section~\ref{sec:4-interactions}). 
The $S$-wave thresholds for the $XYZ$ exotic candidates are also shown in 
Table~\ref{tab:2}. In addition,
the exotic candidates in the charmonium sector and the $S$-wave open-charm 
thresholds are shown in Fig.~\ref{fig:charmoniumlike}. Here, the thresholds 
involving particles with a large width, $\gtrsim100$~MeV, have been neglected.

Since only $S$-wave hadronic molecules with small binding energies are 
well-defined (Sec.~\ref{sec:wein}), in the following, we will focus on 
those candidates, {\sl i.e.}, $\X$, $\Zc$, $\Zcp$, $Y(4260)$ in the charmonium 
sector and $\Zb$, $\Zbp$ in the bottomonium sector. All of them have extremely
close-by $S$-wave thresholds except for the $Y(4260)$, as will be discussed 
below.  For the experimental status and phenomenological models of other exotic
candidates, we refer to several recent  
reviews~\cite{Swanson:2006st,Eichten:2007qx,Brambilla:2014jmp,
Esposito:2014rxa,Lebed:2016hpi, Richard:2016eis,Chen:2016qju,Esposito:2016noz}
and references therein.

\subsubsection{$\X$}

In 2003, the Belle Collaboration reported a narrow structure $\X$ in the 
$J/\psi\pi^+\pi^-$ invariant mass distribution in $B^\pm\to K^\pm  
J/\psi\pi^+\pi^-$~\cite{Choi:2003ue} process. It was confirmed shortly after by 
BaBar~\cite{Aubert:2004ns,Aubert:2008gu} in $e^+e^-$ collisions, and by 
CDF~\cite{Acosta:2003zx,Abulencia:2005zc,Abulencia:2006ma,Aaltonen:2009vj} and 
D0~\cite{Abazov:2004kp} in $p\bar{p}$  collisions. Very recently 
LHCb also confirmed its production in $pp$  
collisions~\cite{Aaij:2011sn,Aaij:2013zoa,Aaij:2014ala,Aaij:2015eva} and pinned 
down its quantum numbers to $J^{PC}=1^{++}$, 
which are consistent with the observations of its radiative 
decays~\cite{Abe:2005ix,Aubert:2006aj,Bhardwaj:2011dj} and 
multipion transitions~\cite{Abulencia:2005zc,Abe:2005ix,delAmoSanchez:2010jr}.
The negative result of searching for its
charged partner in $B$ decays~\cite{Aubert:2004zr} indicates that the $\X$ is an 
isosinglet state.

The most salient feature of the $X(3872)$ is that its mass coincides exactly 
with the $D^0\bar{D}^{*0}$ threshold~\cite{Olive:2016xmw}\footnote{Here we use
the updated ``OUR AVERAGE'' values in PDG2016 for the masses:
$M_{D^0}=(1864.84\pm0.05)$~MeV, $M_{D^{*0}}=(2006.85\pm0.05)$~MeV, and 
$M_{X}=(3871.69\pm0.17)$~MeV from the $J/\psi \pi^+\pi^-$ and $J/\psi\omega$
modes~\cite{Olive:2016xmw}. } 
\begin{equation}
M_{D^0}+M_{D^{*0}}-M_{\X}=(0.00\pm0.18)~\mev\,,
\end{equation}
which indicates the important role of the $D^0\bar{D}^{*0}$ in 
the $X(3872)$ dynamics. That this should be the case can be seen most clearly 
from the large branching fraction~\cite{Gokhroo:2006bt,Adachi:2008sua} (see 
Table~\ref{tab:2})
\begin{eqnarray}
  \mathcal{B}(\X\to \bar D^{0} D^0\pi^0) > 32\% \, ,
\end{eqnarray}
although the $\X$ mass is so close to the $D^0\bar D^{*0}$ and $\bar D^0
D^0\pi^0$ thresholds.
These experimental facts lead naturally to the interpretation 
of the $\X$ as a $D\bar D^{*}$ hadronic
molecule~\cite{Tornqvist:2003na},\footnote{
See also,
e.g.,~\cite{Tornqvist:2004qy,Swanson:2003tb,Close:2003sg,Pakvasa:2003ea,
Wong:2003xk,Voloshin:2003nt,Swanson:2004pp,AlFiky:2005jd,Braaten:2007dw, 
Fleming:2007rp,Liu:2008tn,Dong:2009yp,Ding:2009vj,Zhang:2009vs,Wang:2009aw, 
Lee:2009hy,Gamermann:2009uq,Mehen:2011ds,Nieves:2011vw, Lee:2011rka,
Nieves:2012tt,Li:2012ss, Li:2012cs,Sun:2012zzd, 
Sun:2012sy,Guo:2013sya,Hidalgo-Duque:2013pva,Wang:2013kva,Yamaguchi:2013ty, 
Guo:2014hqa,He:2014nya, Zhao:2014gqa,Karliner:2015ina,
Baru:2015tfa,Jansen:2015lha,Baru:2015nea,Molnar:2016dbo,
Yang:2017prf}.}
which had been predicted by T\"ornqvist with the correct mass a 
decade earlier~\cite{Tornqvist:1993ng}.
As will be discussed in Section~\ref{sec:6}, precise measurements of the 
partial widths of the processes $X(3872)\to D^0\bar D^0\pi^0$ and $X(3872)\to 
D^0\bar D^0\gamma$ are particularly important in understanding the 
long-distance structure of the $X(3872)$.
In the $D^0\bar{D}^{*0}$ hadronic molecular scenario, one gets a tremendously 
large 
$D^0\bar{D}^{*0}$ scattering length of $\geq10$~fm, {\sl c.f.} 
Eq.~\eqref{eq:arwein}. 
However, a precision measurement of its mass is necessary
to really distinguish a molecular $\X$ from, e.g., a tetraquark 
scenario~\cite{Maiani:2004vq,Esposito:2014rxa}. This will be discussed further 
in Sec.~\ref{sec:poletrajectories} and in Sec.~\ref{sec:lineshapes}.

Other  observables are also measured which could be sensitive to the internal
structure of the $X(3872)$. The ratio of branching fractions 
\[ R^I\equiv
\frac{\br{\X\to J/\psi \pi^+\pi^-\pi^0}}{\br{\X \to J/\psi\pi^+\pi^-}}
\] 
was
measured to be $1.0\pm 0.4\pm 0.3$ by Belle~\cite{Abe:2005ix} and $0.8\pm 0.3$ by
BaBar~\cite{delAmoSanchez:2010jr}. The value about unity means a significant
isospin breaking because the three and two pions are  from the isoscalar
$\omega$~\cite{Abe:2005ix,delAmoSanchez:2010jr} and from the isovector
$\rho$~\cite{Abulencia:2005zc}, respectively.
Notice that there is a strong  phase space suppression on the isospin conserved
three-pion transition through the $J/\psi\omega$ channel. The fact that the
molecular scenario of $\X$ provides a natural explanation for the value of $R^I$
will be discussed in Sec.~\ref{sec:isospinviol}.

The experimental information available about the radiative decays of the 
$X(3872)$ is~\cite{Aaij:2014ala} 
\begin{equation}
\frac{\br{\X\to \psi^\prime\gamma}}{\br{\X \to J/\psi\gamma}} = 2.46\pm 
0.64\pm 0.29 \ . 
\label{eq:sec2X3872R}
\end{equation}
A value larger than 1 for this ratio was argued to favor the 
$\chi_{c1}(2^3P_1)$ 
interpretation~\cite{Swanson:2004pp} 
over the $D^0\bar{D}^{*0}$ hadronic molecular picture. This, however, is not 
the case~\cite{Mehen:2011ds,Guo:2014taa} as will be demonstrated in 
Sec.~\ref{sec:6}.

The production  rates of $\X$ in $B^0$ and $B^-$ decays was measured by 
BaBar~\cite{Aubert:2005zh}, {\sl i.e.}, 
\begin{eqnarray}
&& \frac{\mathcal{B}(B^0\to \X K^0\to J/\psi \pi^+\pi^-
K^0)}{\mathcal{B} (B^-\to\X K^-\to J/\psi \pi^+\pi^- K^-)}  \nonumber\\
&=& 0.50\pm 0.30\pm
0.05 \ .
\label{eq:sec2X3872RB}
\end{eqnarray}
We show in Sec.~\ref{sec:6} that this value is also consistent with a   
molecular nature of the $X(3872)$.

One expects mirror images of charmonium-like states to be present in the
bottomonium sector. The $Z_c$ and $Z_b$ states to be discussed in the next
subsection suggest that such phenomena do exist. The analogue of the $\X$ in the
bottom sector, $X_b$, has not yet been identified.
A search for the $X_b$ was carried out by the CMS Collaboration, but no signal
was observed in  the $\Upsilon \pi^+\pi^-$ channel~\cite{Chatrchyan:2013mea}.
However, as pointed out in Ref.~\cite{Guo:2013sya} before the experimental
results and stressed again in Refs.~\cite{Guo:2014sca,Karliner:2014lta}
afterwards, the $X_b\to \Upsilon \pi^+\pi^-$ decay requires an isospin breaking
which should be strongly suppressed due to the extremely small mass differences
between the charged and neutral bottomed mesons and the large difference between
the $B\bar B^*$ threshold and the $\Upsilon(1S)\omega$, $\Upsilon(1S)\rho$
thresholds.
In contrast, other channels such as $X_b\to \Upsilon\pi^+\pi^-\pi^0$, $X_b\to
\chi_{bJ}\pi^+\pi^-$~\cite{Guo:2013sya,Guo:2014sca,Karliner:2014lta} and $X_b\to
\gamma\Upsilon(nS)$~\cite{Li:2014uia} should be a lot more promising for an
$X_b$ search.

\subsubsection{$\Zb$, $\Zbp$ and $\Zc$, $\Zcp$}
\label{sec:zc}

\begin{figure*}[tb]
\begin{center}
 \includegraphics[width=0.24\textwidth]{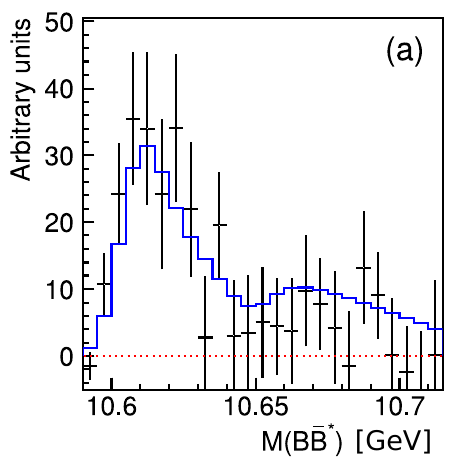}\hfill
 \includegraphics[width=0.24\textwidth]{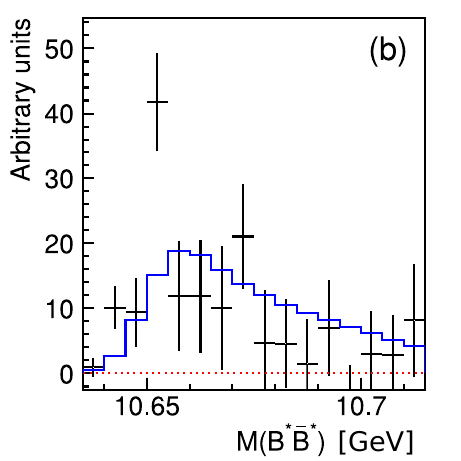}\hfill
 \includegraphics[width=0.24\textwidth]{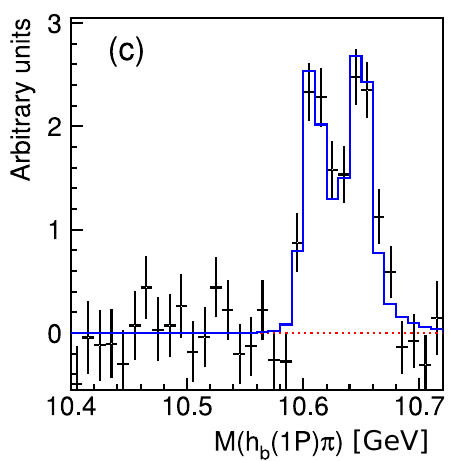}\hfill
 \includegraphics[width=0.24\textwidth]{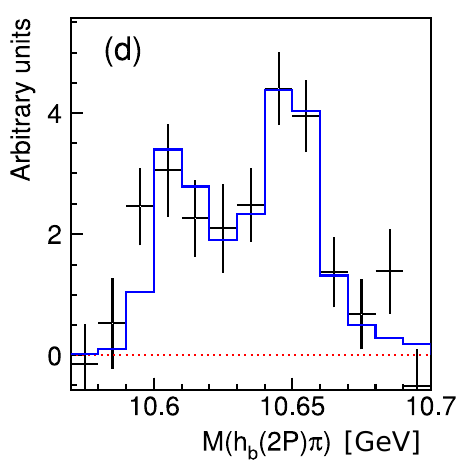}\hfill
\end{center}
\caption{Measured line shapes of the two $Z_b$ states in the
$B\bar{B}^*$, $B^*\bar{B}^*$ and $h_b(1P, 2P)\pi$
channels~\cite{Garmash:2015rfd} and a fit using the parameterization of 
Refs.~\cite{Hanhart:2015cua,Guo:2016bjq}. } \label{fig:Zb5Sfull}
\end{figure*}

\begin{figure}[tb]
\begin{center}
 \includegraphics[height=5cm]{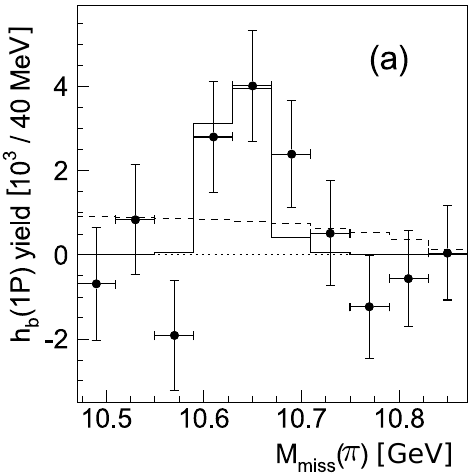}\hfill
 \includegraphics[height=5cm]{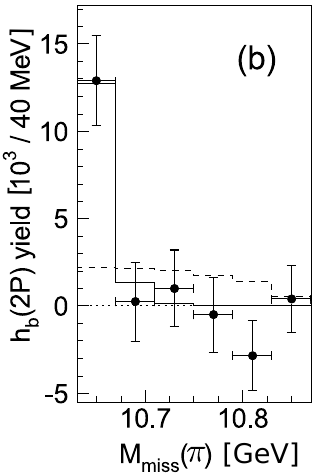}
\caption{The missing mass spectra for $h_b(1P)\pi^+\pi^-$ and 
$h_b(2P)\pi^+\pi^-$ channels  in the 
$\Upsilon(11020)$ region. The solid and dashed histograms are the fits with the
$Z_b$ signal fixed from the $\Upsilon(10860)$ analysis and with only a phase 
space distribution, respectively.
Taken from Ref.~\cite{Abdesselam:2015zza}.
} \label{fig:Zb6S}
\end{center}
\end{figure}

From an analysis of the $\Upsilon(10860)\to \pi^+\pi^-(b\bar b)$ processes in
2011 the Belle Collaboration reported the discovery of two charged states
decaying into $\Upsilon(nS)\pi$  with $n=1,2,3$ and $h_b(mP)\pi$ with
$m=1,2$~\cite{Belle:2011aa}. Their line shapes in a few channels are shown in
Fig.~\ref{fig:Zb5Sfull}. A later analysis at the same experiment allowed
for  an amplitude analysis where the quantum numbers $I^G(J^{P})=1^+(1^{+})$
were strongly favored~\cite{Garmash:2014dhx}.\footnote{The existence of an
isovector $b\bar b q\bar q$ state with exactly these quantum numbers was
speculated long time ago for explaining the puzzling
$\Upsilon(3S)\to\Upsilon(1S)\pi\pi$
transition~\cite{Voloshin:1982ij,Anisovich:1995zu}. The $Z_b$ effects in dipion
transitions among $\Upsilon$ states were reanalyzed using the dispersion
technique recently in~\cite{Chen:2015jgl,Chen:2016mjn}. } This, together with
the fact that the $\Zb$ and $\Zbp$ have masses very close  to the $B\bar B^*$
and $B^*\bar B^*$ thresholds, respectively, makes both excellent candidates for
hadronic molecules~\cite{Bondar:2011ev}\footnote{See also,
e.g.,~\cite{Zhang:2011jja,
Yang:2011rp,Danilkin:2011sh,Sun:2011uh,Cleven:2011gp,Ohkoda:2011vj, Li:2012as, 
Dong:2012hc,Wang:2013daa,Wang:2014gwa, Dias:2014pva,Karliner:2015ina}.}  This
statement finds further support in the observation that both states also decay by far most probably into $B\bar B^*$ and $B^*\bar B^*$,
respectively~\cite{Garmash:2015rfd} (see Tab.~\ref{tab:ZbBr}).\footnote{The branching
fractions were measured by assuming that these channels saturate the decay modes and using the Breit--Wigner (BW) parameterization for the $Z_b$ structures~\cite{Garmash:2015rfd}.
However, there could be non-negligible modes such as the $\eta_b\rho$, and the
branching fractions measured in this way for near-threshold states should not be
used to calculate partial widths by simply multiplying with the BW width.
This point is discussed in detail in~\cite{Chen:2015jgl} for the $Z_b$ case.}
The neutral partner is so far observed only for the lighter
state~\cite{Krokovny:2013mgx}. Very recently, the Belle Collaboration reported
the invariant mass distributions  of $h_b(1P)\pi$ and $h_b(2P)\pi$ channels at
$\Upsilon(11020)$ energy region~\cite{Abdesselam:2015zza}, see
Fig.~\ref{fig:Zb6S}, clearly showing a resonant enhancement in the $Z_b$ mass
region. However, due to the limited statistics it is impossible to judge whether
there are two peaks or just one.

\begin{table}
\caption{The reported branching fractions of the known decay modes of $\Zb^+$
and $\Zbp^+$~\cite{Garmash:2015rfd} with the
statistical and systematical uncertainties in order.}
\begin{ruledtabular}
\begin{tabular}{l c c}
channel & $\mathcal{B}$ of $\Zb$ ($\%$) & $\mathcal{B}$ of 
$\Zbp$$(\%)$\tabularnewline
\hline
$\Upsilon(1S)\pi^{+}$ & $0.54^{+0.16+0.11}_{-0.13-0.08}$ & 
$0.17^{+0.07+0.03}_{-0.06-0.02}$\tabularnewline
$\Upsilon(2S)\pi^{+}$ & $3.62^{+0.76+0.79}_{-0.59-0.53}$ & 
$1.39^{+0.48+0.34}_{-0.38-0.23}$\tabularnewline
$\Upsilon(3S)\pi^{+}$ & $2.15^{+0.55+0.60}_{-0.42-0.43}$ & 
$1.63^{+0.53+0.39}_{-0.42-0.28}$\tabularnewline
$h_{b}(1P)\pi^{+}$ & $3.45^{+0.87+0.86}_{-0.71-0.63}$ & 
$8.41^{+2.43+1.49}_{-2.12-1.06}$\tabularnewline
$h_{b}(2P)\pi^{+}$ & $4.67^{+1.24+1.18}_{-1.00-0.89}$ & 
$14.7^{+3.2+2.8}_{-2.8-2.3}$\tabularnewline
$B^{+}\bar{B}^{*0}+\bar{B}^{0}B^{*+}$ & $85.6^{+1.5+1.5}_{-2.0-2.1}$ & 
--\tabularnewline
$B^{*+}\bar{B}^{*0}$ & -- & $73.7^{+3.4+2.7}_{-4.4-3.5}$\tabularnewline
\end{tabular}
\end{ruledtabular}
\label{tab:ZbBr}
\end{table}

Employing sums of BW functions for the resonance signals the experimental
analyses gave masses for both $Z_b$ states slightly above the corresponding open
flavor thresholds together with narrow widths. It seems in conflict with the
hadronic molecular picture, and was claimed to be consistent with the tetraquark
approach~\cite{Esposito:2016itg}. It is therefore important to note that a
recent analysis based on a formalism consistent with unitarity and analyticity
leads for both states to below-threshold pole
positions~\cite{Hanhart:2015cua,Guo:2016bjq}.\footnote{Notice that this,
however, does not exclude the possibility of above-threshold poles. In the used
parameterization, the contact terms are taken to be constants.
The possibility of getting an above-threshold pole is available once
energy-dependence is allowed in the contact terms. Nevertheless, the analyses at
least show that the below-threshold-pole scenario is consistent with the current
data.}

A few years after the discovery of $\Zb$ and $\Zbp$ in the Belle experiment, the
BESIII and Belle Collaborations almost simultaneously claimed the observation of
a charged state in the charmonium mass range,
$\Zc$~\cite{Ablikim:2013mio,Liu:2013dau}. It was shortly after confirmed by a
reanalysis of CLEO-c data~\cite{Xiao:2013iha}, and its neutral partner was also
reported in Refs.~\cite{Xiao:2013iha,Ablikim:2015tbp}. Soon after these
observations, the BESIII Collaboration reported the discovery of another charged
state $\Zcp$~\cite{Ablikim:2013wzq}, and its neutral partner was reported in
Ref.~\cite{Ablikim:2014dxl}.
These charmonium-like states show in many respects similar features as the
heavier bottomonium-like states discussed in the previous paragraphs, although
there are also some differences. On the one hand, while the $\Zc$ is seen in the
$J/\psi \pi$ channel and $\Zcp$ is seen in $h_c\pi$, there is no clear signal of
$\Zcp$ in $J/\psi\pi$ and $\Zc$ in $h_c\pi$, although in the latter case there
might be some indications of $\Zc\to h_c\pi$. This pattern might reflect a
strong mass dependence of the production mechanism~\cite{Wang:2013hga}.
On the other hand, in analogy to $\Zb$ and $\Zbp$, $\Zc$ and $\Zcp$ have masses
very close to the the $D\bar D^*$ and $D^*\bar D^*$ thresholds, respectively,
and they couple most prominently to these open-flavor channels regardless
of the significant phase space
suppression~\cite{Ablikim:2013xfr,Ablikim:2013emm,Ablikim:2015gda,
Ablikim:2015vvn}. The two $Z_c$ states are also widely regarded as hadronic
molecules~\cite{Wang:2013cya,Guo:2013sya,Voloshin:2013dpa,Cui:2013yva,
Wilbring:2013cha,Li:2013xia,
Zhang:2013aoa,Dong:2013iqa,Ke:2013gia,He:2013nwa,
Karliner:2015ina,Chen:2015ata,Gong:2016hlt}.

\begin{figure}[tb]
\begin{center}
  \includegraphics[width=0.586\linewidth]{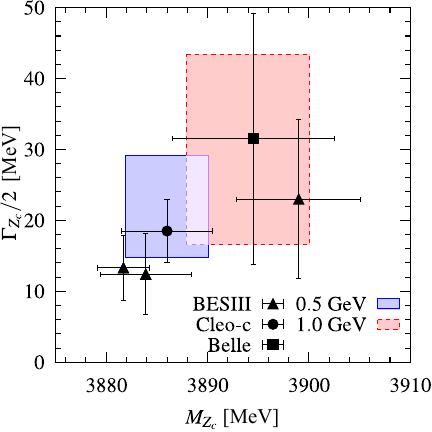}
 \caption{Poles determined in Ref.~\cite{Albaladejo:2015lob} (0.5~GeV and
 1.0~GeV refer to the cutoff values used therein) in comparison with the mass
 and width values for the $Z_c(3900)$ reported in Refs.~\cite{Ablikim:2013mio, 
 Ablikim:2013xfr, Ablikim:2015swa, Liu:2013dau, Xiao:2013iha}. Taken from
 Ref.~\cite{Albaladejo:2015lob}.
 }
 \label{fig:Zcpole}
\end{center}
\end{figure}

Analogous to the $Z_b$ case, the experimental analyses of the two $Z_c$ states
based on sums of BW distributions result in masses above the continuum thresholds as
well. However, this does not allow the correct extraction of the pole locations.
In order to obtain reliable pole locations  an analysis in the spirit of
Refs.~\cite{Hanhart:2015cua,Guo:2016bjq} is necessary for these charmonium-like
states. Such an analysis was done for the $\Zc$ in
Ref.~\cite{Albaladejo:2015lob}. By fitting to the available BESIII data in the
$Y(4260)\to J/\psi\pi^+\pi^-$~\cite{Ablikim:2013mio} and the $Y(4260)\to
J/\psi\pi^+\pi^-$~\cite{Ablikim:2015swa} modes, it is found that the current
data are consistent with either an above-threshold resonance pole or a
below-threshold virtual state pole. A comparison of the resonance pole obtained
therein with various determinations in experimental papers is shown in
Fig.~\ref{fig:Zcpole}.

\subsubsection{$Y(4260)$ and other vector states}\label{sec:Y(4260)}

At present, the vector channel with $J^{PC}=1^{--}$, in both the bottomonium and
the charmonium sector, is the best investigated one experimentally, since it can
be accessed directly in $e^+e^-$ annihilations.
Note that a pair of ground state open-flavor mesons, such as $D\bar{D}$,
$D\bar{D}^*+c.c.$, $D^*\bar{D}^*$, $D_s\bar{D}_s$, etc., carry positive parity
in the $S$ wave, and thus cannot be directly accessed in $e^+e^-$ 
annihilations. 
Accordingly, if the $S$-wave hadronic molecules exist in the vector channel,
they should be formed (predominantly) by constituents different from them. In 
particular,
it suggests that the $S$-wave molecular states in the vector channel should be
heavier than those thresholds opened by a pair of ground state $D^{(*)}$ mesons.

\begin{figure}[tb]
\begin{center}
  \includegraphics[width=0.45\textwidth]{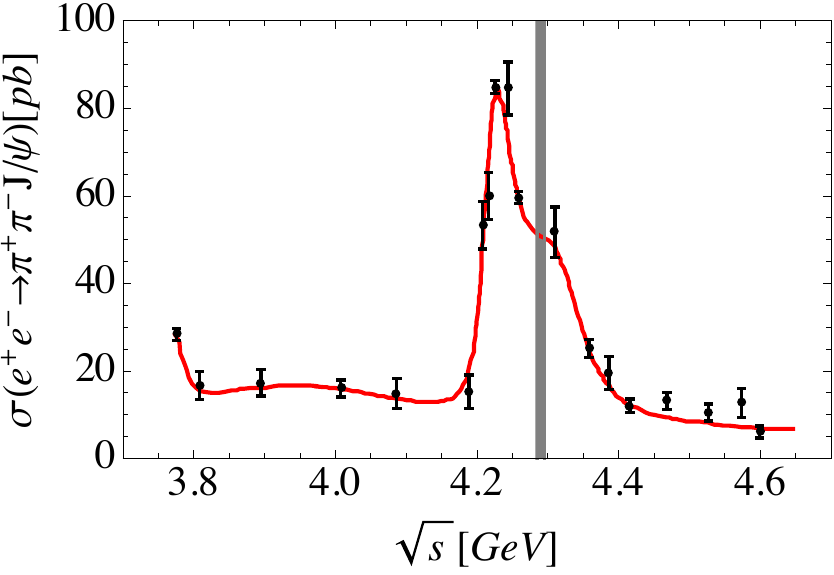}
\caption{ The cross section of $e^+e^-\to \pi^+\pi^-J/\psi$ for center-of-mass 
energies from $3.77~\gev$ to $4.6~\gev$~\cite{Ablikim:2016qzw}.
It shows a clear shoulder around the $D_1\bar D$ threshold (marked by the
vertical gray band) as predicted in Ref.~\cite{Cleven:2013mka}. The red solid curve is from the 
analysis of BESIII~\cite{Ablikim:2016qzw}. A comparison of these data with the
BESIII scan data can be found in~\cite{Gao:2017sqa}. }
\label{fig:Y4260-BESIII}
\end{center}
\end{figure}

As the first $Y$ state, the $Y(4260)$
was observed by  the BaBar Collaboration in the $J/\psi\pi^+\pi^-$ channel
in the initial state radiation (ISR) process
$e^+e^-\to\gamma_\text{ISR} J/\psi\pi^+\pi^-$~\cite{Aubert:2005rm}.
The fitted mass and width are $\left(4259\pm 8^{+2}_{-6}\right)\mev$ and 
$50\ldots 90~\mev$,
respectively. It was confirmed by
CLEO-c~\cite{He:2006kg}, Belle~\cite{Yuan:2007sj} and an additional analysis of
BaBar~\cite{Lees:2012cn}, with, however, mass values varying in different 
analyses. We notice that a recent combined analysis of the BESIII data in four 
different channels $e^+e^-\to \omega\chi_{c0}$~\cite{Ablikim:2015uix}, 
$\pi^+\pi^-h_c$~\cite{BESIII:2016adj}, 
$\pi^+\pi^- J/\psi$~\cite{Ablikim:2016qzw}, 
and $D^0D^{*-}\pi^++c.c.$~\cite{yuantalk}
gives a mass of $(4219.6\pm3.3\pm5.1)$~MeV and a width of 
$(56.0\pm3.6\pm6.9)$~MeV~\cite{Gao:2017sqa}.

\begin{figure*}[tb]
\parbox{5.5cm}{
  \includegraphics[width=0.31\textwidth]{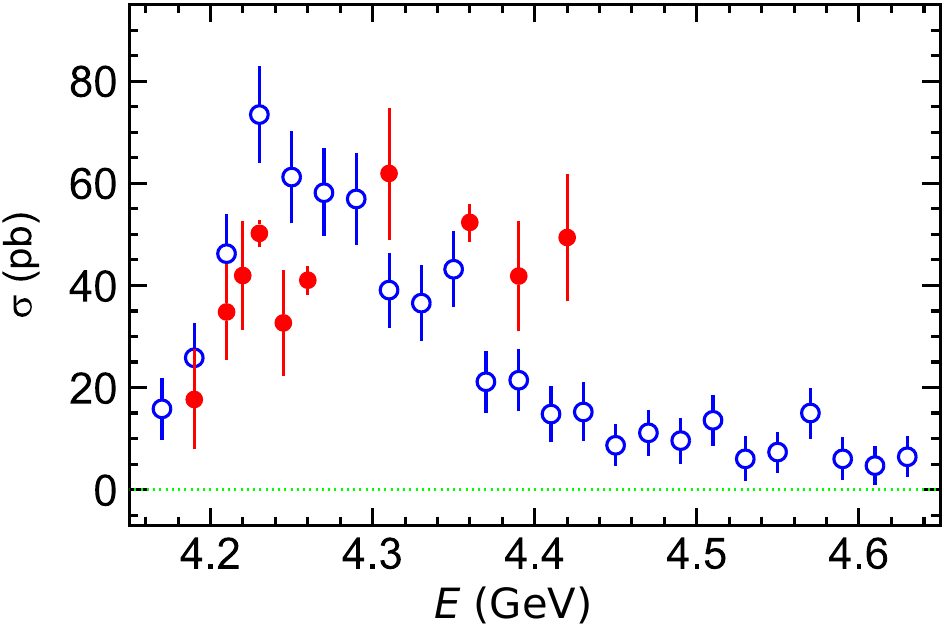}}\hfill
\parbox{5.5cm}{\vspace{-0.1cm}
   \includegraphics[width=0.33\textwidth]{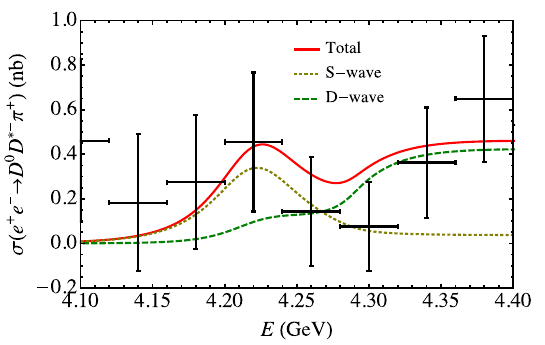}}
 \hfill
\parbox{5.5cm}{ 
     \includegraphics[width=0.29\textwidth]{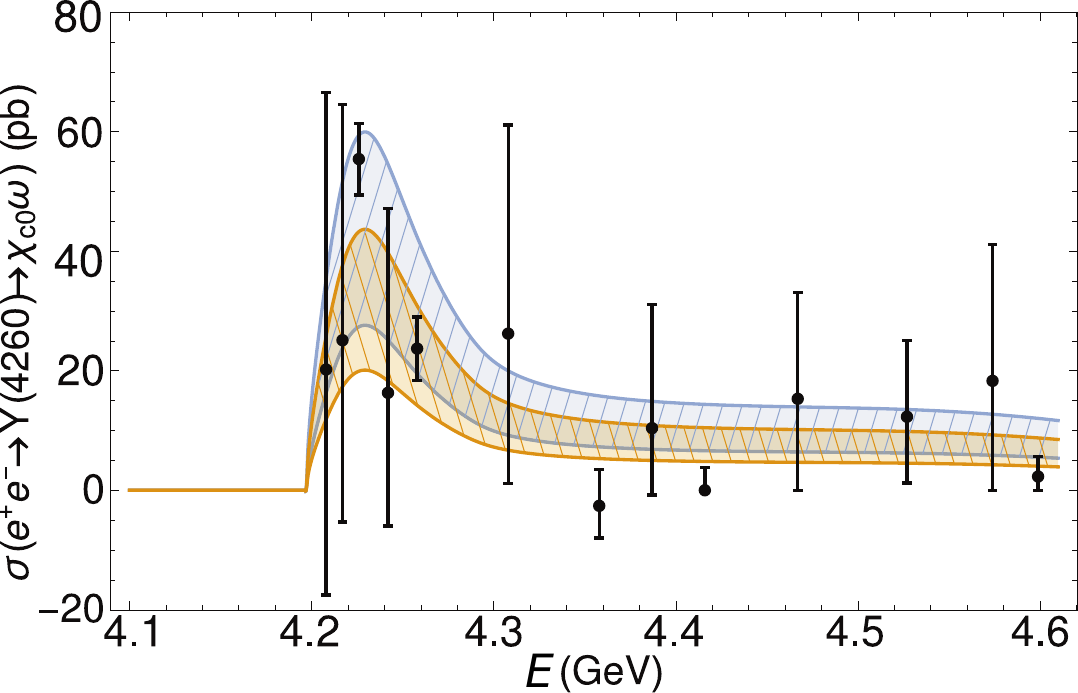}}
\caption{
The first plot shows the cross sections of the $e^+e^-\to h_c \pi\pi$ (red solid
circles) from BESIII~\cite{Ablikim:2013wzq} and the $e^+e^-\to J/\psi\pi\pi$
(blue hollow circles) from Belle~\cite{Liu:2013dau} (note that the recent BESIII
data for $e^+e^-\to J/\psi\pi\pi$ have much smaller errors as shown in
Fig.~\ref{fig:Y4260-BESIII}).
The first plot is taken from~\cite{Chang-Zheng:2014haa}. The second one is the
line shape for the $D\bar{D}^*\pi$ channel within the $D_1\bar D$ molecular
picture~\cite{Cleven:2013mka} compared to the Belle data~\cite{Pakhlova:2009jv}. 
The predicted line shape is similar to the solid 
line of the right panel of Fig.~\ref{fig:lineshapes2} with unstable constituent 
in Sec.~\ref{sec:3} (note 
that an updated analysis can be found in~\cite{Qin:2016spb} and the new data 
from BESIII can be found in~\cite{Gao:2017sqa}). The last plot is the line 
shape of $e^+e^-\to\chi_{c0}\omega$ measured by BESIII~\cite{Ablikim:2014qwy}
and the bands are theoretical calculations in the $D_1\bar D$ molecular 
picture~\cite{Cleven:2016qbn}.
}
\label{fig:Y4260LineShape}
\end{figure*}

The $Y(4260)$ was early recognized as a good candidate for an exotic state since
there are no quark-model states predicted around its mass.
Moreover, it does not show a strong coupling to $D\bar D$ as generally expected
for vector $c\bar c$ states, and it does not show up as a pronounced enhancement
in the inclusive cross sections for $e^+e^-\to$ hadrons (or the famous $R$ value
plot).
It is still believed to be a prime candidate, e.g., for a hybrid
state~\cite{Close:2005iz} (for a recent discussion see
Ref.~\cite{Kalashnikova:2016bta}) or a hadrocharmonium
state~\cite{Dubynskiy:2008mq,Li:2013ssa}.
However, it is also suggested to be a $D_1(2420)\bar D$ molecular
state~\cite{Ding:2008gr,Wang:2013cya,Cleven:2013mka,Li:2013bca,Li:2013yla,
Wu:2013onz} (the hadrocharmonium picture and the molecular picture are
contrasted in Ref.~\cite{Wang:2013kra}).
This picture is further supported by the fact that  the recent high-statistics
data from BESIII~\cite{Ablikim:2016qzw}  (see Fig.~\ref{fig:Y4260-BESIII}) shows
an enhancement at the $D_1(2420)\bar D$ threshold in the $J/\psi\pi\pi$
channel.\footnote{In this  context it is interesting to note that also the
hybrid picture predicts a large coupling of $Y(4260)$ to $D_1(2420)\bar
D$~\cite{Barnes:1995hc,Close:2005iz,Kou:2005gt}, which could be
interpreted as the necessity of considering $D_1\bar D$ as a component.} The
observations of $Z_c(3900)\pi$ (Sec.~\ref{sec:zc}) and
$X(3872)\gamma$~\cite{Ablikim:2013dyn} in the mass region of the $Y(4260)$
provide further support for a sizable $D_1(2420)\bar D$ component in its wave
function as will be discussed in Sec.~\ref{sec:6}.
The suppression of an $S$-wave production, in the heavy quark limit, of the
$1^{--}$ $D_1(2420)\bar D$ pair in $e^+e^-$
collisions~\cite{Eichten:1978tg,Eichten:1979ms,Li:2013yka} could be the reason
for the dip around the $Y(4260)$ mass in the inclusive cross section of
$e^+e^-\to$ hadrons~\cite{Wang:2013kra}.
In addition,  the data from Belle in $e^+e^-\to
\bar{D}D^*\pi$~\cite{Pakhlova:2009jv} and from BESIII on $e^+e^-\to
h_c\pi\pi$~\cite{BESIII:2016adj}, $\chi_{c0}\omega$~\cite{Ablikim:2014qwy}  are
highly nontrivial (Fig.~\ref{fig:Y4260LineShape}) and are claimed to be
consistent with the molecular picture~\cite{Cleven:2013mka,Cleven:2016qbn}.
A combined analysis of the BESIII data in different channels is presented in
Ref.~\cite{Gao:2017sqa}.

The absence of a signal for $Y(4260)$ in $J/\psi
K\bar{K}$~\cite{He:2006kg,Yuan:2007bt,Shen:2014gdm} questions the tetraquark
picture of $Y(4260)$ with a diquark-antidiquark $[cs][\bar c\bar s]$
configuration~\cite{Esposito:2014rxa}.
In addition, the ground state in the tetraquark picture~\cite{Esposito:2014rxa},
$Y(4008)$, is not confirmed by the recent high-statistics data from
BESIII~\cite{Ablikim:2016qzw}.
Meanwhile, the cross sections for $e^+e^-\to
\psi'\pi\pi$~\cite{Aubert:2007zz,Wang:2007ea}, $\eta'
J/\psi$~\cite{Ablikim:2016ymr} and $\eta J/\psi$ and $\pi^0
J/\psi$~\cite{Ablikim:2015xhk} do not show any structure around the $Y(4260)$
energy region. It remains to be seen if these findings allow for further conclusions on the nature of 
 the $Y(4260)$.

It is interesting to observe that some properties of the $Y(4260)$, like its
proximity and strong coupling to the $D_1\bar D$ threshold, are mirror imaged by
the $\Upsilon(11020)$ in the bottomonium sector~\cite{Bondar:2016pox}. Belle II
appears to be an ideal instrument to investigate this connection in more detail
in the future~\cite{Bondar:2016hva}.

Searching for new decay modes of the $Y(4260)$, BaBar scanned the line shapes of
$e^+e^-\to \psi(2S)\pi^+\pi^-$ and found a new structure, named $Y(4360)$ with a
mass of $(4324\pm 24)~\mev$ and a width of $(172\pm
33)~\mev$~\cite{Aubert:2007zz}. In the same year, Belle \cite{Wang:2007ea}
analyzed the same process and found two resonant structures: a lower one
consistent with $Y(4360)$ and a higher one, named $Y(4660)$, with a mass of
$(4664\pm 11\pm 5)~\mev$ and a width of $(48\pm 15\pm 3)~\mev$. A combined fit
~\cite{Liu:2008hja} to the cross sections of the process $e^+e^-\to
\psi(2S)\pi^+\pi^-$ from both BaBar and Belle gives the parameters for the two
resonances $M_{Y(4360)}=\left(4355^{+9}_{-10}\pm 9\right)\mev$,
$\Gamma_{Y(4360)}=\left(103^{+17}_{-15}\pm 11\right)\mev$ and
$M_{Y(4660)}=\left(4661^{+9}_{-8}\pm6\right)\mev$,
$\Gamma_{Y(4660)}=\left(42^{+17}_{-12}\pm6\right)\mev$ for $Y(4360)$ and
$Y(4660)$, respectively. The fit at the same time provides an upper limit for
$\br{Y(4260)\to \psi(2S)\pi^+\pi^-}\Gamma_{e^+e^-}$ as $4.3~\ev$.
Those measurements were updated in Ref.~ \cite{Wang:2014hta}.
Later on the Belle Collaboration found a structure in the
$\Lambda_c^+\Lambda_c^-$ channel with a peak position $30~\mev$ lower than that
of $Y(4660)$~\cite{Pakhlova:2008vn}, which might either point at an additional
state, called $Y(4630)$, or, be an additional  decay channel of the
$Y(4660)$~\cite{Cotugno:2009ys,Guo:2010tk}. The latter is the view taken in the
2016 Review of Particle Physics~\cite{Olive:2016xmw}.
Of particular interest to this review is the observation of
Ref.~\cite{Guo:2010tk} that within the $\psi^\prime f_0(980)$ hadronic molecular
picture~\cite{Guo:2008zg,Wang:2009hi} the line shape of $Y(4630)$ in the
$\Lambda_c^+\Lambda_c^-$ channel could be understood as the signal of $Y(4660)$
with the $\Lambda_c^+\Lambda_c^-$ final state interaction.
As a byproduct, Ref.~\cite{Guo:2009id} predicts the properties of its spin
partner, an $\eta_c^\prime f_0(980)$ hadronic molecule, at around 4.61~GeV.

As stressed in Refs.~\cite{Wang:2013hga,Bondar:2016pox} the 
production of $Z_c(3900)$ and $Z_c(4020)$ in the mass region of 
$Y(4260)$ and $Y(4360)$
as well as that of $\Zb$ and $\Zbp$ in the mass region of $\Upsilon(10860)$ and 
$\Upsilon(11020)$, respectively, are sensitive to the TS mechanism. A peculiar 
feature of such a mechanism is that whether peaks appear in certain invariant 
mass distributions depends strongly on the kinematics. The recent observation 
of a peak in the $\psi'\pi$ invariant mass distribution by the BESIII 
Collaboration~\cite{Ablikim:2017oaf} shows exactly this behavior.
The correlations between the initial $S$-wave thresholds and the final $S$-wave
thresholds could be a key for understanding the rich phenomena observed in this 
energy region~\cite{Liu:2015taa}.

\begin{table*}
\caption{Same as Table~\ref{tab:1} but in the  baryon
sector.}
\begin{ruledtabular}
 \begin{tabular}{l c c c c c}
state & $I(J^{P})$ & $M[\mathrm{MeV}]$ & $\Gamma[\mathrm{MeV}]$ & $S$-wave 
threshold(s) [$\mathrm{MeV}$] & Observed mode(s) (branching
ratios)\tabularnewline
\hline
$\Lambda(1405)$ & $0(\frac{1}{2}^{-})$ & $1405.1_{-1.0}^{+1.3}$ & $50.5\pm2.0$ & 
$N\bar{K}(-29.4_{-1.0}^{+1.3})$ &
$\Sigma\pi(100\%)$\tabularnewline
&&&&$\Sigma\pi(76.2^{+1.3}_{-1.0})$&\tabularnewline
\hline
$\Lambda(1520)$& $0(\frac 32^-)$ & $1519.5\pm 1.0$  & $15.6\pm 1.0$  
&$\Sigma(1385)^-\pi^+(-7.3\pm 1.1)$   & $N\bar{K} (45\pm 1)\%$
\tabularnewline
&&&&$\Sigma(1385)^+\pi^-(-2.9 \pm 1.1)$& $\Sigma\pi(42\pm 1)\%$\tabularnewline
&&&&$\Sigma(1385)^0\pi^0(0.8 \pm 1.4)$& $\Lambda\pi\pi(10\pm 
1)\%$\tabularnewline
\hline
$\Lambda(1670)$
& $0(\frac 12^-)$ &
$\approx1670$ & $\approx35$  &$\Lambda\eta\,(4)$   & $N\bar{K} (20\sim 30)\%$
\tabularnewline
&&&&& $\Sigma\pi(25\sim 55)\%$\tabularnewline
&&&&& $\Lambda\eta(10\sim 25)\%$\tabularnewline
\hline
$\Lambda_c(2595)$ & $0(\frac{1}{2}^{-})$ & $2592.25\pm 0.28$ & $2.6\pm0.6$ & 
$\Sigma_c(2455)^{++}\pi^- (-1.04\pm 0.31)$ &
$\Sigma_c(2455)^{++}\pi^-(24\pm 7)\%$\tabularnewline
 &  &  & & $\Sigma_c(2455)^0\pi^+ (-0.82\pm 0.31)$&
$\Sigma_c(2455)^{0}\pi^+(24\pm 7)\%$\tabularnewline
 &  &  & & $\Sigma_c(2455)^+\pi^0 (4.62 \pm 0.49)$&
$\Lambda_c^{+}\pi^+\pi^-~\text{3-body}~(18\pm 10)\%$\tabularnewline
\hline
$\Lambda_c(2625)$ & $0(\frac{3}{2}^{-})$ & $2628.11\pm 0.19$ & $<0.97$ & 
$\Sigma_c(2455)\pi (36.53\pm 0.24)$ &
$\Lambda_c^+\pi^+\pi^-(67\%)$\tabularnewline
 &  &  & & &
$\Sigma_c(2455)^{++}\pi^-(<5\%)$\tabularnewline
 &  &  & & &
$\Sigma_c(2455)^0\pi^+(<5\%)$\tabularnewline
\hline
$\Lambda_c(2880)$ & $0(\frac{5}{2}^{+})$ & $2881.53\pm 0.35$ & $5.8 \pm 1.1$ & 
$ND^*(-65.91\pm 0.35)$&
$\Lambda_c^+\pi^+\pi^-$\tabularnewline
 & &  & & &
$\Sigma_c(2455)^{0,++}\pi^\pm $\tabularnewline
 & &  & & &
$\Sigma_c(2520)^{0,++}\pi^\pm $\tabularnewline
 & &  & & &
$pD^0 $\tabularnewline
\hline
$\Lambda_c(2940)$ & $0(\frac 32^-)$ &
$2939.3_{-1.5}^{+1.4}$ & $17^{+8}_{-6}$ & 
$ND^*(-8.1_{-1.5}^{+1.4})$ &
$\Sigma_c(2455)^{0,++}\pi^\pm$\tabularnewline
 & \cite{Aaij:2017vbw} &  &  &  &
$pD^0$\tabularnewline
\hline
$\Sigma_c(2800)$ & $1(?^?)$ & $2800_{-4}^{+5}$ & $70^{+23}_{-15}$ & $ND(-6 \pm 
5)$ &
$\Lambda_c^+\pi$\tabularnewline
\hline
$\Xi_c(2970)$ & $\frac{1}{2}(?^?)$ & $2969.4 \pm 1.7$ & $19.0\pm 3.9$ & 
$\Sigma_c(2455)K (20.2 \pm 1.7)$ &
$\Lambda_c^+\bar{K}\pi$\tabularnewline
 &  & & &  &
$\Sigma_c(2455)\bar{K}$\tabularnewline
 &  & & &  &
$\Xi_c2\pi$\tabularnewline
 &  & & &  &
$\Xi_c(2645)\pi$\tabularnewline
\hline
$\Xi_c(3055)$ & $?(?^?)$ & $3055.1 \pm 1.7$ & $11\pm 4$ & $\Sigma_c(2520)K (41.1 
\pm 1.7)$ &
\tabularnewline
 &  & & & $\Xi_c(2970)\pi (-52.3 \pm 2.4)$ & \tabularnewline
 \hline
 $\Xi_c(3080)$ & $\frac{1}{2}(?^?)$ & $3078.4 \pm 0.7$ & $5.0\pm 1.3$ & 
$\Sigma_c(2520)K (64.4 \pm 0.7)$ & $\Lambda_c^+\bar{K}\pi$
\tabularnewline
 &  & & & $\Xi_c(2970)\pi (-29.0 \pm 1.8)$ & 
$\Sigma_c(2455)\bar{K}$\tabularnewline
  &  & & &  & $\Sigma_c(2520)\bar{K}$ \tabularnewline
\hline
$P_c(4380)$ & $\frac 12(\frac 32^?/\frac 52^?)$& 
$4380\pm8\pm29$ & $205\pm18\pm86$ &  $\Sigma_c(2520)\bar{D}$
($-6 \pm 30$)  & $J/\psi p$\tabularnewline
 \cite{Aaij:2015tga} & & &  &  $\Sigma_c(2455)\bar{D}^*$ $(-82 \pm 30)$ 
 &\tabularnewline
\hline
$P_c(4450)$ &$\frac 12(\frac 32^?/\frac 52^?)$ 
&$4449.8\pm1.7\pm2.5$  &  $39\pm5\pm19$&  $\chi_{c1}p$ $(0.9 \pm 3.0)$  &
$J/\psi p$\tabularnewline
\cite{Aaij:2015tga} & & &  &  $\Lambda_c(2595)\bar{D}$ $(-9.9 \pm 3.0)$  &
\tabularnewline & & &  &   $\Sigma_c(2520)\bar{D}^*$ ($-77.2 \pm 3.0$) & \tabularnewline
& & &  &   $\Sigma_c(2520)\bar{D}$ ($64.2 \pm 3.0$) & \tabularnewline
& & &  &   $\Sigma_c(2455)\bar{D}^*$ ($-12.3 \pm 3.0$) & \tabularnewline
\end{tabular}
\end{ruledtabular}
\label{tab:baryon}
\end{table*}

\subsection{Baryon candidates for hadronic molecules}

We now switch to the experimental evidences for hadronic molecules in the baryon
sector. In analogy to  the meson sector we will focus on states which are
located close to $S$-wave thresholds of narrow meson-baryon
pairs~\footnote{Note that also $P_{11}(1440)$ was proposed to have a prominent
$f_0(500)N$ substructure~\cite{Krehl:1999km}. However, the large width of the
$f_0(500)$ prohibits a model-independent study of this claim.}. In the light
baryon spectrum the $\Lambda(1405)$ has been broadly discussed as a $\bar{K}N$
molecular state. A few charm baryons discovered in recent years are close to
$S$-wave thresholds, and has been suggested to be hadronic molecules in the
literature. The recently observed very interesting $P_c(4450)$ and $P_c(4380)$
have also been proposed to be hadronic molecules with hidden charm.

\subsubsection{Candidates in the light baryon sector}
\label{sec:lam1405exp}

The $\Lambda(1405)$ was discovered  in the $\pi\Sigma$ subsystems of  $Kp\to
\Sigma\pi\pi\pi$~\cite{Alston:1961zzd} (see also
Ref.~\cite{Kim:1965zzd,Hemingway:1984pz}).
Further experimental information about this state comes from old scattering
data~\cite{Humphrey:1962zz,Sakitt:1965kh,Watson:1963zz,Ciborowski:1982et}
complemented by the recent  $\bar{K}N$ threshold amplitude extracted from data
on kaonic hydrogen~\cite{Bazzi:2011zj,Bazzi:2012eq} as well as the older
so-called threshold ratios~\cite{Tovee:1971ga,Nowak:1978au}.
There are further data on $\Sigma\pi$ distributions from $pp\to \Sigma^\pm
\pi^\mp K^+p$~\cite{Zychor:2007gf,Agakishiev:2012xk}, the photoproduction
$\gamma p\to K^+\Sigma\pi$~\cite{Moriya:2014kpv} and additional reactions.
It appears also feasible that high energy experiments  like BaBar, Belle,
BESIII, CDF, D0 and LHCb investigate the $\Lambda(1405)$, e.g. via the decays
of heavy hadrons such as $\Lambda_b\to J/\psi
\Lambda(1405)$~\cite{Roca:2015tea}.
Note that a signal of $\Lambda(1405)$ was clearly visible  in an analysis of
$\Lambda_b\to J/\psi Kp$ performed by the LHCb
Collaboration~\cite{Aaij:2015tga}.

The $\Lambda(1405)$ has strangeness $S=-1$ with $I(J^P)=0(1/2^-)$ and a mass
about 30~MeV (see Table~\ref{tab:baryon}) below the $\bar{K}N$ threshold.
Note that a direct experimental determination of the spin-parity quantum numbers
was only given recently by the CLAS Collaboration \cite{Moriya:2014kpv}.
Since its mass is smaller than that of the nucleon counterpart
$N^*(1535)\,1/2^-$ and the mass difference from its spin-splitting partner state
$\Lambda(1520)$ $I(J^P)=0(3/2^-)$ is larger than that between $N^*(1535)\,1/2^-$
and $N^*(1520)\,3/2^-$, it can hardly be accepted by the conventional
three-quark picture of the constituent quark model~\cite{Hyodo:2011ur}.
It is fair to say that the $\Lambda(1405)$ {was most probably} the
first exotic hadron observed~\cite{Alston:1961zzd}. The theoretical aspects of the
$\Lambda(1405)$ will be discussed in Sec.~\ref{sec:1405th}.

\subsubsection{Candidates in the charm baryon sector}

The two light quarks in a charm baryon can be either in the symmetric sextet or
antisymmetric anti-triplet representation of SU(3). Since the color wave
function is totally antisymmetric,
 the spin-flavor-space wave functions must be symmetric. Hence the light-quark
system in the $S$-wave flavor sextet (anti-triplet) has spin $1$ ($0$).
After combining with a heavy quark, the sextet and anti-triplet give the
$B_6(1/2^+)$, $B_6^*(3/2^+)$ and $B_{\bar{3}}(1/2^+)$ baryon multiplets,
respectively, as~\cite{Yan:1992gz}
\begin{eqnarray*}
B_{6}=\left(\begin{array}{ccc}
\Sigma_{c}(2455)^{++} & \frac{1}{\sqrt{2}}\Sigma_{c}(2455)^{+} & 
\frac{1}{\sqrt{2}}\Xi_{c}^{\prime+}\\
\frac{1}{\sqrt{2}}\Sigma_{c}(2455)^{+} & \Sigma_{c}(2455)^{0} & 
\frac{1}{\sqrt{2}}\Xi_{c}^{\prime0}\\
\frac{1}{\sqrt{2}}\Xi_{c}^{\prime+} & \frac{1}{\sqrt{2}}\Xi_{c}^{\prime0} & 
\Omega_{c}^{0}
\end{array}\right),
\end{eqnarray*}
 \begin{eqnarray*}
B_{6}^{*}=\left(\begin{array}{ccc}
\Sigma_{c}(2520)^{++} & \frac{1}{\sqrt{2}}\Sigma_{c}(2520)^{+} & 
\frac{1}{\sqrt{2}}\Xi_{c}(2645)^{+}\\
\frac{1}{\sqrt{2}}\Sigma_{c}(2520)^{+} & \Sigma_{c}(2520)^{0} & 
\frac{1}{\sqrt{2}}\Xi_{c}(2645)^{0}\\
\frac{1}{\sqrt{2}}\Xi_{c}(2645)^{+} & \frac{1}{\sqrt{2}}\Xi_{c}(2645)^{0} & 
\Omega_{c}(2770)^{0}
\end{array}\right),
\end{eqnarray*}
\begin{eqnarray*}
B_{\bar{3}}=\left(\begin{array}{ccc}
0 & \Lambda_{c}^{+} & \Xi_{c}^{+}\\
-\Lambda_{c}^{+} & 0 & \Xi_{c}^{0}\\
-\Xi_{c}^{+} & -\Xi_{c}^{0} & 0
\end{array}\right).
\end{eqnarray*}
All the ground state charm baryons within these three multiplets have been well
established in experiments~\cite{Olive:2016xmw}.
Among the other charm baryons, $\Lambda_c(2765)$, $\Xi_c(2815)$ and
$\Xi_c(3123)$ are not well-established from the experimental
analysis~\cite{Olive:2016xmw}.
The $P$-wave $1/2^-$ and $3/2^-$ antitriplet states are identified as
$[\Lambda_c(2595)^+, \Xi_c(2790)^+,
\Xi_c(2790)^0]$~\cite{Cheng:2006dk,Cheng:2015naa} and $[\Lambda_c(2625)^+,
\Xi_c(2815)^+, \Xi_c(2815)^0]$, respectively~\cite{Cheng:2015naa}. Among the
remaining charm baryons,  the $\Lambda_c(2880)^+$ has a definite spin of
$5/2$~\cite{Olive:2016xmw,Aaij:2017vbw}
and the quantum numbers of the $\Lambda_c(2940)^+$ were measured to be
$J^P=3/2^-$~\cite{Aaij:2017vbw}.
Besides these two charmed baryons, LHCb also measured another charm baryon
$\Lambda_c(2860)^+$ which is consistent with the predication of the orbital
$D$-wave $\Lambda_c^+$ excitation~\cite{Chen:2016phw,Chen:2016iyi,Chen:2017aqm}.
The only available information of other measured charm baryons are their masses
and some of their decay modes.
For recent reviews on the heavy baryons, we recommend~\cite{Klempt:2009pi,
Chen:2016spr}.

Although the $\Lambda_c(2595)^+$ may be accommodated as a regular three-quark
baryon in quark
models~\cite{Copley:1979wj,Pirjol:1997nh,Tawfiq:1998nk,Zhu:2000py,
Blechman:2003mq,Migura:2006ep,Zhong:2007gp}, one cannot neglect one striking
feature of it~\cite{Hyodo:2013iga} which could provide other potential
interpretations:
It lies between the $\Sigma_c(2455)^+\pi^0$ and $\Sigma_c(2455)^0\pi^+$,
$\Sigma_c(2455)^{++}\pi^-$ thresholds as shown in Table~\ref{tab:baryon}.
Thus, the $\Lambda_c(2595)$ is proposed as a dynamically generated state of the
nearby $\Sigma_c(2455)\pi$ coupling with other possible higher
channels~\cite{Lutz:2003jw,Hofmann:2005sw,Mizutani:2006vq,GarciaRecio:2008dp,
JimenezTejero:2009vq,Haidenbauer:2010ch,Romanets:2012hm,Lu:2014ina,
Liang:2014kra,Garcia-Recio:2015jsa,Lu:2016gev,Long:2016oog}, such as $ND$,
$ND^*$ and so on. The strong coupling between the $\Lambda_c(2595)$ and  the
$\Sigma_c(2455)\pi$ channels even leads to a prediction of the existence of a
three-body $\Sigma_c\pi\pi$ resonance in Refs.~\cite{Long:2016oog,Long:2017bgz}.
The analysis in~\cite{Guo:2016wpy}, however, indicates that the compositeness of
$\Sigma_c(2455)^+\pi^0$ is smaller than $10\%$ leaving $\Lambda_c(2595)$
dominated by either other heavier hadronic channels (such as $ND$ and $ND^*$) or
compact quark-gluon structures. Some other charm baryons, such as
$\Lambda_c(2880)$~\cite{Lutz:2003jw},
$\Lambda_c(2940)$~\cite{He:2006is,He:2010zq,Ortega:2012cx,Zhang:2012jk,
Zhao:2016zhf} and
$\Sigma_c(2800)$~\cite{JimenezTejero:2009vq,JimenezTejero:2011fc,Zhang:2012jk},
have also been considered as dynamically generated states from meson-baryon
interactions.  In particular, the $\Lambda_c(2940)^+$ is very close to the
$ND^*$ threshold --- it even overlaps with the threshold if using the recent
LHCb measurement~\cite{Aaij:2017vbw},  and it can 
couple to $ND^*$ in an $S$-wave.
Both are favorable features for treating it as an $ND^*$ hadronic 
molecule~\cite{Ortega:2012cx,Zhao:2016zhf}.

\subsubsection{Pentaquark-like structures with hidden-charm}

Recently, LHCb reported two pentaquark-like structures $P_c(4380)^+$ and
$P_c(4450)^+$ in the $J/\psi p$ invariant mass distribution of $\Lambda_b\to
J/\psi p K^-$~\cite{Aaij:2015tga}. Their masses (widths) are $(4380\pm 8 \pm
29)$~MeV ($(205\pm 18\pm 86)$~MeV) and $(4449.8\pm 1.7\pm 2.5)$~MeV ($(39\pm
5\pm 19)$~MeV), respectively. In this analysis the $\Lambda^*$ states that
appear in the crossed channel were parametrized via BW functions. The LHCb
analysis reported preference of the spin-parity combinations $(3/2^-, 5/2^+)$,
$(3/2^+, 5/2^-)$ or $(5/2^+, 3/2^-)$  for these two states, respectively. The
branching ratio for $\Lambda_b\to J/\psi p K^-$ was also
measured~\cite{Aaij:2015fea}.

The data for the Cabibbo suppressed process $\Lambda_b\to J/\psi p \pi^-$ are
consistent with the existence of these two $P_c$ structures~\cite{Aaij:2016ymb}.
The same experiment also published a measurement of $\Lambda_b^0\to \psi(2S) p
K^-$, but no signals for the $P_c$ states were observed due to either the low
statistics or their absence in the $\psi(2S)p$ channel~\cite{Aaij:2016wxd}.

The production mechanism and the decay pattern imply a five-quark content of
these two states with three light quarks and a hidden heavy $c\bar{c}$ component
if they are hadronic states. In fact, pentaquark-like states with hidden charm
have been predicted in the right mass region as dynamically generated in
meson-baryon interactions a few years before the LHCb
discovery~\cite{Wu:2010jy,Wu:2010vk}.
There are several thresholds in the mass region of the two $P_c$ structures,
namely $\chi_{c1} p$, $\Sigma_c(2520)\bar{D}$, $\Sigma_c(2455)\bar{D}^*$,
$\Lambda_c(2595)\bar{D}$, and $\Sigma_c(2520)\bar{D}^*$ (see
Table~\ref{tab:baryon}), though not all of them couple in $S$-waves to the
reported preferred quantum numbers, suggesting different interpretations of the
two $P_c$ states, such as $\Sigma_c(2455)\bar{D}^*$, $\Sigma_c(2520)\bar{D}$ or
$\Sigma_c(2520)\bar{D}^*$ hadronic
molecules~\cite{Chen:2015loa,Chen:2015moa,Roca:2015dva,Chen:2016otp,
He:2015cea,Karliner:2015ina,Ortega:2016syt,Shimizu:2016rrd}. It has been suggested that their
decay patterns could be used to distinguish among various hadronic molecular
options~\cite{Wang:2015qlf,Lu:2016nnt,Shen:2016tzq,Lin:2017mtz}.
There are also other dynamical studies with different channel
bases~\cite{Xiao:2016ogq,Yamaguchi:2016ote,Azizi:2016dhy,Geng:2017hxc}.

The extreme closeness of the $P_c(4450)$ to the $\chi_{c1} p$ threshold and to a
TS from a $\Lambda^*(1890)\chi_{c1}p$ triangle diagram was pointed out in
Ref.~\cite{Guo:2015umn}. In Ref.~\cite{Bayar:2016ftu}, it is stressed  that the
$\chi_{c1}p$ needs to be in an $S$-wave so as to produce a narrow observable
peak in the $J/\psi p$ invariant mass distribution, and correspondingly $J^P$
needs to be $1/2^+$ or $3/2^+$. This TS and other possible relevant TSs
are also discussed in~\cite{Liu:2015fea,Guo:2016bkl}. It is worthwhile to
emphasize that the existence of TSs in the $P_c$ region does not exclude a 
possible existence of pentaquark states whether or not they are hadronic
molecules. In Ref.~\cite{Meissner:2015mza}, the possibility that the $P_c(4450)$
could be a $\chi_{c1}p$ molecule was investigated.

In order to confirm the existence of the two $P_c$ states and distinguish them
from pure kinematic singularities, {\sl c.f.} discussions in
Sec.~\ref{sec:6-ts}, one can either search for them in other processes, such as
$\Lambda_b\to\chi_{c1}p K^-$~\cite{Guo:2015umn},
photoproduction~\cite{Karliner:2015voa,Kubarovsky:2015aaa,
Wang:2015jsa,Kubarovsky:2016whd,Gryniuk:2016mpk,Blin:2016dlf,Huang:2016tcr},
heavy ion collisions~\cite{Wang:2016vxa}, pion-nucleon
reactions~\cite{Kim:2016cxr,Liu:2016dli,Lin:2017mtz}, or search for their
strange~\cite{Feijoo:2015kts,Feijoo:2015cca,Ramos:2016odk,Chen:2016heh,
Lu:2016roh}, neutral~\cite{Lebed:2015tna,Lu:2015fva} and
bottomonium~\cite{Xiao:2015fia} partners. The $\Lambda_b\to\chi_{c1}p K^-$ decay
process has been observed at the LHCb experiment~\cite{Aaij:2017awb}.

\section{Identifying hadronic molecules }
\label{sec:3}

Hadronic molecules are analogues of  light nuclei, most notably the deuteron. 
They can be treated to a good approximation as composite systems made of two 
or more hadrons which are bound together via the strong interactions.
In this section the general notion of a molecular state is introduced. As will
be demonstrated for near-threshold bound states this picture can be put into a
formal definition that even allows one to relate observables directly to the
probability to find the molecular component in the bound state wave function.
However, it appears necessary to work with a more general notion of hadronic
molecules as also resonances can be of molecular nature in the sense formulated
above. Before we proceed it appears necessary to review some general properties
of the $S$-matrix. In this subsection also the terminology of a bound state, a
virtual state and a resonance are discussed for those notions will be heavily
used throughout this review.

\subsection{Properties of the {\boldmath$S$}-matrix} 
\label{sec:Sproperties}

\begin{figure*}
 \centering
   \includegraphics*[width=\linewidth]{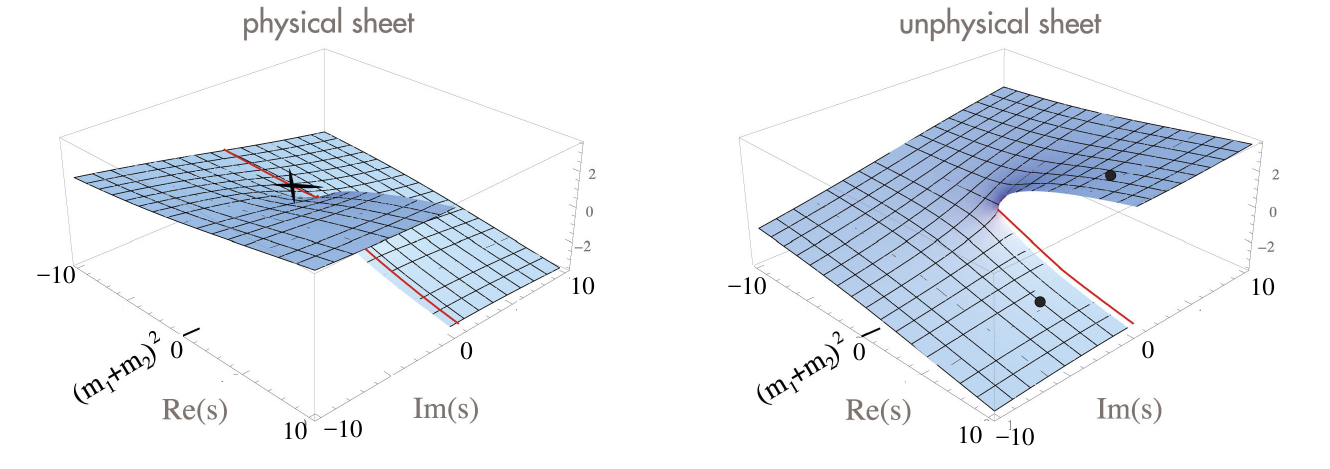}
   \caption{Sketch of the imaginary part of a typical single--channel amplitude in 
the complex $s$-plane. The solid dots indicate allowed positions for resonance poles,
the cross for a bound state.
The solid line is the physical axis (shifted by $i\epsilon$ into the physical sheet).
The two sheets are connected smoothly along their discontinuities.
\label{sheets}}
\end{figure*}

The unitary operator that connects asymptotic $in$ and $out$ states is called
the $S$-matrix.
It is an analytic function in the Mandelstam plane
 up to its branch points and poles.
 The $S$-matrix is the quantity that encodes all physics about a certain
 scattering or production reaction. In general it is
assumed that the $S$-matrix is analytic up to:
\begin{itemize}
 
\item {\it Branch points}, which occur at each threshold. On the one hand, there
are the so called right-hand cuts starting from the branch points at the
thresholds for an $s$-channel kinematically allowed process (e.g. at the $\bar
KK$ threshold in the $\pi \pi$ scattering amplitude). On the other hand, when
reactions in the crossed channel become possible one gets the left-hand cuts,
which are usually located in the unphysical region for the reaction studied but
may still influence significantly, e.g., the energy dependence of a reaction
cross section. Branch points can also be located inside the complex plane of 
the unphysical Riemann sheets: This is possible when the reaction goes via an 
intermediate state formed by one or more unstable states. It is clear that these
threshold branch points/cuts are determined kinematically and happen at the loop
level of Feynman diagrams.

In general, a loop Feynman diagram with more-than two intermediate particles has
more complicated kinematical singularities. They are called Landau
singularities~\cite{Landau:1959fi}, see,
e.g.,~\cite{Eden:1966,Chang:1983,Gribov:2009}.
For instance, in triangle diagrams the branch points of two intermediate pairs
can be very close to the physical region simultaneously, and such a situation
gives rise to the so-called {\it triangle singularity} already introduced
in Sec.~\ref{sec:2}. We come back to those in detail in
Sec.~\ref{sec:4-3ploop}.\footnote{The Landau singularity can even be a pole if
the one-loop Feynman diagram has at least five intermediate
particles~\cite{Gribov:2009}. However, this case is irrelevant for us and will
not be considered.}

\item {\it Poles}, which appear due to the interactions inherited in the
dynamics of the underlying theory. Depending on the locations, poles can be
further classified as follows:

-- {\it Bound states}, which appear as poles on the physical sheet. By causality
they are only allowed to occur on the real $s$-axis below the lowest threshold.
The deuteron in the isospin-0 and spin-1 proton-neutron system, which can be
regarded as the first established hadronic molecule, is a nice example.

-- {\it Virtual states}, which appear on the real $s$-axis, however, on the
unphysical Riemann sheet. A well-known example in nuclear physics is
the pole in the isospin-1 and spin-0 nucleon-nucleon scattering. It is within 1~MeV from the
threshold and drives the scattering length to a large value of about 24~fm.

-- {\it Resonances}, which appear as poles on an unphysical Riemann sheet close
to the physical one.  There is no restriction for the location of poles on the
unphysical sheets. Yet, Hermitian analyticity requires that, if there
is a pole at some complex value of $s$, there must be another pole at its
complex conjugate value, $s^*$.
Normally, the pole with a negative imaginary part is closer to the physical axis
and thus influences the observables in the vicinity of the resonance region more
strongly. However, at the threshold both poles are always equally important.
This is illustrated in Fig.~\ref{sheets}.

\end{itemize}

For a discussion of the analytic structure of the $S$-matrix with focus on
scattering experiments we refer to Ref.~\cite{Doring:2009yv} and references
therein.
Any of these singularities leads to some structure in the observables.
In a partial-wave decomposed amplitude additional singularities not related to
resonance physics may emerge as a result of the partial-wave projection. For a
discussion see, e.g., Ref.~\cite{hoehler}.

From the above classification, it is clear that the branch points are
kinematical so that they depend completely on the masses of the involved
particles in a certain physical process, while the poles are of dynamical origin
so that they should appear in many processes as long as they are allowed by
quantum numbers. 

We will call a structure observed experimentally a state if and only if the
origin of this structure is a pole in the $S$-matrix due to dynamics.
On the one hand this definition is quite general as it allows us to also call
the above mentioned pole in the isovector nucleon-nucleon scattering a state.
From the point of view of QCD, this definition appears to be quite natural since
it takes only a marginal change in the strength of the two-hadron potential
(e.g. via a small change in quark masses) to switch from a shallow bound state
to a near-threshold virtual state, and both leave a striking imprint in
observables (we come back to this in Sec.~\ref{sec:lineshapes}). 
{For example, various lattice QCD groups have observed the di-neutron to become a bound
state at quark masses heavier than the physical
value~\cite{Beane:2012vq,Yamazaki:2012hi,Yamazaki:2015asa,Berkowitz:2015eaa}.}
On the other hand, if a structure in the data finds its origin purely in a
kinematical singularity without a nearby pole, it would not be called a state.
There is currently a heated discussion going on in the literature whether some
of the $XYZ$ states are just threshold cusps or triangle
singularities~\cite{Bugg:2004rk,Chen:2011pv,Chen:2013coa,Swanson:2014tra,
Swanson:2015bsa,Pilloni:2016obd,Gong:2016jzb}.
It should be stressed, however, that pronounced near-threshold signals in the
continuum channel related to that threshold must find their origin in a nearby
pole~\cite{Guo:2014iya}.

 {In the physical world, }basically all candidates for hadronic molecules,
 {except for nuclei,} can decay strongly and thus can not be bound
 states in the rigorous sense of the word, since the lowest threshold is defined
 by the production threshold of the decay products.
 However, it still appears justified to e.g. call the $f_0(980)$ a $K\bar K$
 bound state, {or a quasi-bound state in a more rigorous sense,}
  if the corresponding pole is located on the physical sheet for
 the two-kaon system, or a virtual state if it is on the unphysical sheet for
 the two-kaon system, although the lowest threshold is the two-pion threshold.

\subsection{Definition of hadronic molecules}
\label{sec:wein}

 In order to proceed it is necessary to first of all define the notion of a
 molecular state. Naively one
 might be tempted to argue that if data can be described by a model 
where all interactions between continuum states come from 
  $s$-channel pole terms the resulting states have to be interpreted as
 ``elementary'' states. However, as we will discuss below, this is in
 general not correct.
 Analogously, a model that contains only non-pole interactions can still at the
 end lead to a pole structure of the $S$-matrix that needs to be interpreted as
 non-molecular. The origin of  the failure of intuition in these circumstances
 is the fact that a hadronic description of hadron dynamics can only be
 understood in the sense of an effective field theory with limited range of
 applicability. In particular the very short-ranged parts of the wave function
 as well as the interaction potential are model-dependent and can not be
 controlled within the hadronic prescription. 

 However, at least for near-threshold bound states (the term ``near'' will be
 quantified in the next subsection) there is a unique property of the wave
 function of a molecular state as long as it is formed by a (nearly) stable
 particle pair in an $S$-wave: The very fact that this particle pair can almost
 go on-shell leaves an imprint with observable consequences in the analytic
 structure of the corresponding amplitude, a feature absent to all other
 possible substructures. In fact, as a consequence of this feature, hadronic
 molecules can be very extended. To see this observe that a bound state wave
 function at large distances scales as $\exp(-\gamma r)/r$, where $r$ is the
 distance between the constituents and $\gamma$ denotes the typical momentum
 scale defined via
 \begin{equation}
 \gamma = \sqrt{2\mu E_B} \ ,
 \label{eq:gamdef}
 \end{equation}
 where $\mu=m_1m_2/(m_1+m_2)$ denotes the reduced mass of the two-hadron system
 and 
 \begin{equation}
  E_B = m_1+m_2-M
  \label{eq:Ebdef}
 \end{equation}
 the binding energy of the state with mass $M$ (note that we
 chose $E_B$ positive so that the bound state is located at $E=-E_B$ {with
 $E$ the energy relative to the threshold}).
 Thus, the size $R$ of a molecular state is given by $ R\sim 1/\gamma$.
  Accordingly, if $X(3872)$ with a binding energy of less than 200~keV with
  respect to the
 $D^0\bar D^{*\, 0}$ threshold were a molecule, it would be at least as large as
 10~fm.
 For a review of properties of systems with large scattering length we refer to
 Ref.~\cite{Braaten:2004rn}.

All these issues will be discussed in detail in the sections below. The
arguments will start in Sec.~\ref{sec:weinberg} from the classic definition
introduced by Weinberg long ago to model-independently capture  the nature of
the deuteron as a proton-neutron bound state. A detailed discussion of the
derivation will allow us to explain at the same time the limitations of this
definition.
Then in Sec.~\ref{sec:polecounting} it is demonstrated that the Weinberg
criterion is actually identical to the pole counting arguments by Morgan.
In Sec.~\ref{sec:poletrajectories} the generalization of the arguments to
resonances is prepared by a detailed discussion of pole trajectories that emerge
when some strength parameter that controls the location of the the $S$-matrix
poles varies. The compositeness criteria for resonances are very briefly
discussed in Sec.~\ref{sec:resonances}.

 \subsubsection{The Weinberg compositeness criterion}
 \label{sec:weinberg}

We start from the following ansatz for the physical wave function of a bound
state~\cite{Weinberg:1965zz} :
 \begin{equation}
|\Psi \rangle = \left(\lambda|\psi_0\rangle\atop
\chi (\bm{k}) |h_1h_2 \rangle \right),
\end{equation}
 where $|\psi_0\rangle$ denotes the compact component of the state and  $|h_1h_2
 \rangle$ its two-hadron component.
  Here compact denotes an object whose size is controlled by the
  confinement radius $R_{\rm conf.}< 1$ fm. Thus this component is assumed to be more
  compact than $R\sim 1/\gamma$, which denotes the characteristic size of a shallow bound
  state~\footnote{Actually, we may define the notion ``shallow'' by the request
  the $R>R_{\rm conf.}$, which translates into $E_B<1/(2\mu R_{\rm conf.}^2)$.}
  In addition, $\chi (\bm{k})$ is the wave
 function of the two-hadron part, where $\bm{k}$ denotes the relative 
 momentum of the two particles. 
 In this parameterization, by definition, $\lambda$ quantifies the contribution 
 of the compact component of the wave function to the physical wave function of 
 the state. Accordingly $\lambda^2$ denotes the probability to find the
 compact component of the wave function in the physical state, which
 corresponds to the wave function renormalization constant $Z$ in quantum field
 theory. Thus, the goal is to relate $\lambda$ to observables.

 In order to proceed one needs to define the interaction Hamiltonian. As shown 
 by Weinberg in Ref.~\cite{Weinberg:1963zza} under very general conditions one 
 may write
 \begin{equation}
 \hat {\mathcal H} |\Psi\rangle = E |\Psi\rangle,~~
\hat {\mathcal H} = \left(\begin{array}{cc}
\hat{H}_c&\hat{V}\\
\hat{V}&\hat{H}_{hh}^0
\end{array}
\right) .
 \end{equation}
 This expression exploits the observation that it is possible by a proper field 
 redefinition to remove all hadron-hadron interactions from the theory and to 
 cast them into $\psi_0$~\cite{Weinberg:1962hj,Weinberg:1963zza}. Then the 
 two-hadron Hamiltonian is given simply by the kinetic term $\hat{H}_{hh}^0=k^2/
 (2\mu)$, where $\mu = m_1m_2/(m_1+m_2)$
 denotes the reduced mass of the two-hadron system and $m_i$ the mass of hadron
 $h_i$.
 Introducing the transition form factor,
 \begin{equation}
 \langle \psi_0|\hat {V}|h_1 h_2 \rangle = f(\bm k),
 \end{equation}
 one finds the wave function in the momentum space as
 \begin{equation}
 \chi(\bm k)=\lambda \, \frac{f(\bm k)}{E-k^2/(2\mu)} \ .
 \end{equation}
 The wave function of a physical bound state needs to be normalized to have a
 probabilistic interpretation.
 We thus get \begin{eqnarray}\nonumber
1&=&\langle \Psi|\Psi\rangle = \lambda^2 \langle \psi_0|\psi_0\rangle {+}\!\!\! 
\int \!\! \frac{d^3k}{(2\pi)^3} |\chi(\bm k)|^2 \langle h_1h_2|h_1h_2\rangle \\
&=&  \lambda^2 \left\{1
+\int \frac{d^3k}{(2\pi)^3} \frac{f^2(\bm k)}{\left[E_B+k^2/(2\mu)\right]^2}
\right\}  .
\label{eq:norm}
 \end{eqnarray}
As mentioned above, $\lambda^2$ is in fact the wave function renormalization
constant $Z$, since the integral in the last line of Eq.~(\ref{eq:norm}) is 
nothing but the energy derivative of the self-energy. Because of the positivity 
of the integral, $\lambda^2$ is bound in the range between 0 and 1, and thus 
allows for a physical probabilistic interpretation for a bound state.

At this point a comment is necessary: in many textbooks on quantum field theory
it is written that the wave function renormalization constant $Z$ is scheme
dependent and is to be used to absorb the ultraviolet (UV) divergence of the
vertex corrections. Clearly this is correct. However, the scheme dependence and
UV divergence are only for the terms analytic in $E$.
What we find here is the LO piece of $Z$ in an energy expansion
around the threshold,\footnote{More discussion on this point can be found in
Sec.~\ref{sec:4-interactions}.} and as this
piece is proportional to $\sqrt{E}$ it can not be part of the Lagrangian.
Thus, the Weinberg criterion as outlined is based explicitly on the presence of
the two-particle cut which is responsible for the appearance of the square
root, whose presence is a distinct feature of the two-hadron component.

 The integral in Eq.~\eqref{eq:norm} converges if $f(\bm k)$ is a constant. The
 denominator contains solely model independent parameters, while the momentum
 dependence of the numerator is controlled by the relevant momentum range of the
 vertex function that may be estimated by $\beta$, the inverse range of forces.
 Thus, if $\beta \gg \gamma$,  the integral can be evaluated model-independently
 for the case of $S$-wave coupling which implies the constant $g_0=f(0)$ as the
 LO piece of $f(\bm k)$.\footnote{Note that in some works a model for
 the form factor $f(\bm k)$ is employed~\cite{Faessler:2007gv}.} Then one finds
  \begin{equation}
 1 = \lambda^2 \left[1 +\frac{\mu^2g_0^2}{2\pi\sqrt{2\mu E_B}}+ {\cal
 O}\left(\frac{\gamma}{\beta}\right)\right] .
 \end{equation}
From this
 we find the desired relation, namely
 \begin{equation}
 g_0^2 = \frac{2\pi\gamma}{\mu^2}  \left(\frac{1}{\lambda^2}-1\right)  ,
 \label{eq:gunrenorm}
 \end{equation}
which provides a relation between $\lambda^2$, the probability of finding the 
compact component of the wave function inside the physical wave function, and 
$g_0$, the bare coupling coupling constant of the physical state to the 
continuum, or $\lambda g_0$, the physical coupling constant.

The quantity $g_0$ appears also in the physical propagator of the bound state
since the self energy is given by 
\begin{eqnarray}\nonumber
\Sigma(E) &=& - \int \frac{d^3 k}{(2\pi)^3}
\frac{f^2(\bm k)}{E-k^2/(2\mu)+i\epsilon}
\\
&=& \Sigma(-E_B) + i\,g_0^2\frac{\mu}{2\pi}\sqrt{2\mu E+i\epsilon}  +
{\mathcal O}\left(\frac{\gamma}{\beta}\right)  .~~
\end{eqnarray}
We may therefore write for the $T$-matrix of the two continuum particles whose
threshold is close to the location of the bound state,
\begin{equation}
T_\text{NR}(E) = \frac{g_0^2}{E-E_0+\Sigma(E)} + \mbox{non-pole terms} \, ,
\end{equation}
where the subscript ``NR'' is a reminder of the nonrelativistic normalization
used in the above equation.
As long as the pole is close to the threshold, the amplitude near threshold 
should be dominated by the pole term (the non-pole terms are again controlled 
quantitatively by the range of forces). Using $ E_B = -E_0 + \Sigma(-E_B) - 
g_0^2\mu\gamma/(2\pi) $, which absorbs the (divergent) leading contribution of 
the real part into the bare pole energy and at the same
time takes care of the fact that the analytic continuation of the momentum term
also contributes at the pole,\footnote{When $E$ takes real values,
 the square root on the first sheet is defined by $\sqrt{2\mu E+i\epsilon} =
 +i\sqrt{-2\mu E}\, \theta(-E) + \sqrt{2\mu E}\,\theta(E) $.
 } we get
\begin{equation}
T_\text{NR}(E) = \frac{g_0^2}{E+E_B+g_0^2 \mu/(2\pi) (i k+\gamma)} \ ,
\label{eq:poleTmatrix}
\end{equation}
where we have introduced the two-hadron relative momentum $k=\sqrt{2\mu E}$.
Note that Eq.~\eqref{eq:poleTmatrix} is nothing but the one channel version of
the well-known Flatt\'e parametrization~\cite{Flatte:1976xu}.
  Thus, a measurement of near-threshold data allows one in principle to measure
 the composition of the bound state wave function, in line with the effective
 field theory analysis to be discussed later in Sec.~\ref{sec:6}, although in
 practice a reliable extraction of the coupling might be hindered by a scale invariance of the
 Flatt\'e parametrization that appears for large couplings~\cite{Baru:2004xg}.
 The phenomenological implications especially of Eq.~\eqref{eq:gunrenorm} on
 Eq.~\eqref{eq:poleTmatrix} and generalizations thereof will be discussed in
 Sec.~\ref{sec:lineshapes}.

 To make the last statement explicit we may match Eq.~(\ref{eq:poleTmatrix})
 onto the effective range expansion
 \begin{equation}
 T_\text{NR}(E)=- \frac{2\pi}{\mu} \frac{1}{1/a + (r/2)k^2-ik} \ ,
 \label{eq:ere}
 \end{equation}
 and find
 \begin{eqnarray} \nonumber
 a &=& - 2\, \frac{1-\lambda^2}{2-\lambda^2}\left(\frac1{\gamma}\right)+
 {\mathcal O}\left(\frac1\beta\right) , \\
  r&=& -\frac{\lambda^2}{1-\lambda^2} \left(\frac1{\gamma}\right)+ {\mathcal
  O}\left(\frac1\beta\right) .
  \label{eq:arwein}
 \end{eqnarray}
 Thus, for a pure molecule ($\lambda^2=0$) one finds that the scattering
 length gets maximal, $a=-1/\gamma$, and in addition $r={\mathcal O}(1/\beta)$,
 where the latter term is typically positive, while for a compact state 
 ($\lambda^2=1$) one gets $a=-{\mathcal O}(1/\beta)$ (in the presence of a bound
 state the scattering length is necessarily negative within the sign convention 
 chosen here) and $r\to -\infty$.
 These striking differences have severe implications on the line shapes of
 near-threshold states as will be discussed in detail in 
 Sec.~\ref{sec:lineshapes}.
 
 It is illustrative to apply the Weinberg criterion to the deuteron, 
 basically repeating the analysis presented already in Ref.~\cite{Weinberg:1965zz}.
 The scattering length and effective range
 extracted from proton-neutron scattering data in the deuteron channel
 are~\cite{Klarsfeld:1984es}
 \begin{equation}
a = -5.419(7) \ \mbox{fm \ and \ } r=1.764(8) \ \mbox{fm} \ ,
\label{arexp}
\end{equation}
where the sign of the scattering length was adapted to the convention employed here.
Furthermore, the deuteron binding energy reads~\cite{VanDerLeun:1982bhg}~\footnote{The
reference quotes $E_B=2.224575(9)$ MeV, however, for the analysis here such a high accuracy is
not necessary.}
  \begin{equation}
E_B= 2.22 \ \mbox{MeV} \ \Longrightarrow \ \gamma=45.7 \ \mbox{MeV} = 0.23 \ \mbox{fm}^{-1} \ .
\end{equation}
On the other hand, in case of the deuteron the range of forces is
provided by the pion mass --- accordingly the range corrections
that appear in Eqs.~(\ref{eq:arwein}) may in this case be estimated via
\begin{equation}
\frac1{\beta}\sim \frac1{M_\pi}\simeq 1.4 \ \mbox{fm} \ .
\end{equation}
Thus the effective range is of the order of the range corrections (and positive!)
 as required by the compositness criterion for a molecular state.
Using $\lambda^2=0$ in the expression for the scattering length we find
\begin{equation}
a_{\rm mol.} = -(4.3\pm 1.4) \ \mbox{fm} \ 
\end{equation}
also consistent with Eq.~(\ref{arexp}). Based on these considerations Weinberg
concluded that the deuteron is indeed composite.

 As mentioned previously, a location of a molecular state very near a threshold
 is quite natural, while a near-threshold compact state is
 difficult to accomplish~\cite{Jaffe:2007id,Hanhart:2014ssa}. This can now be 
 nicely illustrated on the basis of Eqs.~(\ref{eq:ere}) and (\ref{eq:arwein}). 
 By construction for $k=i\gamma$ the $T$ matrix develops a pole which may be
 read off from Eq.~(\ref{eq:ere})
 \begin{equation}
 \gamma = -\frac{1}{a} + \frac{\gamma^2 r}{2} \, .
 \label{eq:polecond}
 \end{equation}
 For a (nearly) molecular state $a\simeq -1/\gamma$ and $r\sim {\mathcal
 O}(1/\beta)$. Thus, for this case Eq.~(\ref{eq:polecond}) is largely saturated 
 by the scattering length term, and the range term provides only a small
 correction. However, for a predominantly genuine state we have $-1/a\simeq 
 \beta \gg \gamma$ and $r\to -\infty$. Thus in this case a subtle fine tuning 
 between the range term and the scattering length term appears necessary for the
 pole to be located very near threshold.

 While the low-energy scattering of the hadrons that form the bound state is
 controlled by scattering length and effective range, production reactions are
 sensitive to the residue of the bound state pole, which serves as the effective
 coupling constant,  to be called $g_{\rm eff}$, squared of the bound state to
 the continuum.  It is simply given by the bare coupling constant $g_0^2$
 introduced above multiplied by the wave function renormalization constant $Z$,
 which is $\lambda^2$  as explained above,
\begin{equation}
  g_\text{NR}^2\equiv{Z} g_0^2   = \frac{2\pi\gamma}{\mu^2} (1-\lambda^2) \, .
  \label{eq:residue_bs}
\end{equation}
 After switching to a relativistic normalization by multiplying
 with  $\left(\sqrt{2m_1}\sqrt{2m_2}\sqrt{2M}\right)^2$, and
 dropping terms of order $(E_B/M)$, we thus get
 \begin{equation}
 \frac{g_{\rm eff}^2}{4\pi}= 4 M^2\left(\frac{\gamma}{\mu}\right) 
 \left(1-\lambda^2\right)  .
 \label{eq:residue}
 \end{equation}
What is interesting about this equation is that it is bounded from above: The
effective coupling constant of a bound system to the continuum gets maximal for 
a pure two-hadron bound state. Since $1-\lambda^2$ is the probability of finding
the two-hadron composite state component in the physical wave function, it is
sometimes called ``compositeness''. Using Eq.~\eqref{eq:arwein}, the effective
coupling can be expressed in terms of the scattering length
\begin{equation}
  \frac{g_{\rm eff}^2}{4\pi}= \frac{4 M^2}{\mu}\frac{{-a\gamma}}{a+2/\gamma}\,,
  \label{eq:ga}
\end{equation}
which reduces to $-4M^2/(\mu a)$ in the limit of $\lambda^2=0$, reflecting the 
universality of an $S$-wave system with a large scattering 
length~\cite{Braaten:2004rn}.

Before closing this section some comments are necessary.
\begin{itemize}
\item The approach allows for model-independent statements only for $S$-waves,
since otherwise in the last integral of Eq.~(\ref{eq:norm}) there appears in the
numerator of the integrand an additional factor $k^{2L}$ from the centrifugal 
barrier. Accordingly, the integral can no longer be evaluated
model-independently without introducing additional parameters (regulator) to 
cope with the UV divergence.

\item For the same reason the continuum channel needs to be a two-body channel,
since otherwise the momentum dependence of the phase space calls for an 
additional suppression of the integrand. 

\item The binding momentum must be small compared to the inverse range of
forces, since otherwise the range corrections get larger than the terms that
contain the structure information.

\item For the applicability  of the formalism as outlined and an unambiguous
probabilistic interpretation, the state studied must be a bound state, since
otherwise the normalization condition of Eq.~(\ref{eq:norm}) is not applicable
which is at the very heart of the derivation.
However, nowadays there exist generalizations of the Weinberg approach also to
resonances which will be discussed in Sec.~\ref{sec:resonances}.

\item The constituents that form the bound state must be
narrow, since otherwise the bound system would also be
broad~\cite{Filin:2010se,Guo:2011dd}.
\end{itemize}

For long it seemed that the conditions are satisfied only by the deuteron and
Weinberg therefore closed his paper with the phrase~\cite{Weinberg:1965zz}:
``One begins to suspect that Nature is doing her best to keep us from learning
whether the `elementary' particles deserve that title.'' However, as outlined in
the introduction, there are now various near-threshold states confirmed
experimentally that appear to be consistent with those criteria, like $X(3872)$,
$D_{s0}^*(2317)$ and less rigorously $f_0(980)$ and others.

For illustration we would like to compare what is known about the effect of the
$D_{s0}^*(2317)$ on $DK$ scattering to the Weinberg criterion. Clearly, $DK$
scattering can not be measured directly in experiment, however, it can be
studied in lattice QCD using the so-called L\"uscher
method~\cite{Luscher:1990ux}.
A first study using this method for the $DK$ system is presented in
Ref.~\cite{Mohler:2013rwa}.
The scattering length and effective range extracted in this work for the lowest
pion mass ($M_\pi=156$ MeV) are $-(1.33\pm 0.20)$~fm and $(0.27 \pm 0.17)$~fm,
respectively. This number is to be compared to the Weinberg prediction for a
purely molecular state of $a=-(1\pm 0.3)$~fm and $r\sim 0.3$~fm, where the
inverse $\rho$-mass was assumed for the range of forces and we used that for
molecular states the effective range is positive and of the order of the range
of forces.
Scattering lengths of the same size were also extracted from a study of the
scattering of the light pseudoscalars off $D$-mesons using unitarized chiral
perturbation theory~\cite{Liu:2012zya}. Thus from both chiral dynamics on the
hadronic level as well as lattice QCD there are strong indications that
$D_{s0}^*(2317)$ indeed is a $DK$ molecule. The lattice aspects will be further
discussed in Sec.~\ref{sec:lattice}. Clearly, a direct experimental confirmation
of the molecular assignment for $D_{s0}^*(2317)$ is very desirable. A possible
observable could be the hadronic width of $D_{s0}^*(2317)$ as discussed in
Sec.~\ref{sec:isospinviol}.

So far we only focused on bound states. However, also very near-threshold poles
on the second sheet not accompanied by a first sheet pole, so-called virtual
states, leave a striking imprint in observables, {\sl c.f.}
Sec.~\ref{sec:Sproperties}.
A $T$-matrix that has its pole on the second, instead of on the first, sheet
reads, in distinction to Eq.~(\ref{eq:poleTmatrix}),
\begin{equation}
T_\text{NR}(E) = \frac{g_0^2}{E+E_v+ g_0^2 \mu/(2\pi) (i k-\gamma)} \, ,
\label{eq:poleTmatrix2}
\end{equation}
where now the virtual pole is located on the second sheet at $E=-E_v$, with
$E_v>0$ and we still use $\gamma$ to denote $\sqrt{2\mu E_v}$.
Here we use that on the second sheet below threshold the momentum is 
$-i|\sqrt{2\mu E}|$.

\subsubsection{The pole counting approach}
\label{sec:polecounting}

One of the classic approaches put forward to distinguish molecular states from
genuine ones is the so-called pole counting approach~\cite{Morgan:1992ge},
which may be summarized as: A bound state that is dominated by its compact
component (in the language of the previous section this implies $\lambda^2$
close to 1) manifests itself in two near--threshold singularities (one on the
first sheet, one on the second) while a predominantly molecular bound state
gives rise only to a single near-threshold pole on the first sheet.

To see that these criteria actually map perfectly on the Weinberg criterion it
is sufficient to observe that the poles of Eq.~(\ref{eq:ere}) are given by
\begin{equation}
k_{1/2} = \frac{i}{r} \pm \sqrt{-\frac{1}{r^2}-\frac{2}{ar}} \ .
\end{equation}
{Based on the sign convention employed in this work, cf. Eq.~(\ref{eq:ere}), in the presence of a bound state the scattering length is negative. In
addition, keeping only the leading terms for both the scattering length $a$ and the effective range $r$ as shown in 
Eqs.~(\ref{eq:arwein}), one obtains 
\begin{equation}
k_1 = i\gamma \ , \quad k_2=-i \gamma\left(\frac{2-\lambda^2}{\lambda^2}\right) 
\ .
\end{equation}
Thus
It is easy to see that $k_1$ and $k_2$ are positive and negative 
imaginary numbers, respectively, which implies that the former is a pole on the first Riemann
sheet (a bound state pole), while the latter is located on the second sheet.
When $\lambda$ approaches 0, which implies that the molecular component of the
state becomes increasingly important, the second pole disappears towards negative imaginary infinity,
which leaves $k_1$ as the only relevant pole.
} In particular one gets from this for the asymmetry of the pole locations
\begin{equation}
\frac{|k_1|-|k_2|}{|k_1|+|k_2|} = \lambda^2-1 \ .
\end{equation}
Thus the asymmetry of the pole locations is a direct measure of the amount of 
molecular admixture in the bound state wave function (as defined within the 
Weinberg approach) in line with the findings of Ref.~\cite{Morgan:1992ge}.
The close relation between the two approaches was first observed in 
Ref.~\cite{Baru:2003qq}.

\subsubsection{Remarks about pole trajectories}
\label{sec:poletrajectories}

QCD is characterized by a small number of parameters, namely the quark masses,
the number of colors ($N_c$) and $\Lambda_{\rm QCD}$ (the running coupling constant).
Accordingly those parameters determine completely the hadron spectrum. With
advanced theoretical tools it became possible recently to investigate the
movement of the QCD poles as QCD parameters are varied. There exist studies for
varying quark masses as well as varying numbers of colors, $N_c$,---both of them
allowing for deeper insights into the structure of the investigated states.

Studies that vary the number of colors are available mostly for light quark
systems. For a recent review, we refer to Ref.~\cite{Pelaez:2015qba}. In order
to connect the $N_c$ dependence of a given state in the spectrum to QCD in
Ref.~\cite{Pelaez:2015qba} a unitarized version of chiral perturbation theory,
the so-called inverse amplitude method, is employed, where the fact is exploited
that the leading $N_c$ behavior of the low-energy constants (LECs) is known.
Thus, once the unitarized amplitudes are fitted, e.g., to phase shifts it is
possible to investigate the impact of a varying $N_c$ by proper rescaling of
these LECs.
This kind of study was pioneered by the work reported in
Ref.~\cite{Pelaez:2003dy}, where it was demonstrated that the $N_c$ scaling of
the vector mesons $\rho$ and $K^*$ is in line with expectations for $\bar qq$
states, however,  that of $f_0(500)$  and $K_0^*(800)$ is completely at odds
with them. While the $N_c$ studies allow one to distinguish the quark content of
different states, they do not allow one to disentangle hadronic molecules from
other four-quark structures. In addition, both mentioned resonances are very
broad and as such do not allow one to straightforwardly quantify their molecular
component following the approach of the previous section.

Existing studies where quark masses are varied allow for a more direct contact
to the discussion of the previous section.
Understanding the quark mass dependence of hadrons with different composition is
not only interesting on its own sake, it is also important since lattice QCD
studies can be performed at arbitrary quark masses (and in fact are often
performed at enlarged quark masses {for practical reasons}).
Thus, as soon as we can relate certain pole trajectories to the structure of
the hadron it becomes feasible to ``measure'' the nature of the state using
lattice QCD.
More direct methods to use the lattice to determine the nature of certain
hadrons will be discussed in Sec.~\ref{sec:lattice}.

Let us consider pole trajectories of resonances as some generic strength
parameter is varied.
Here we follow the presentation in Ref.~\cite{Hanhart:2014ssa}. In this work it
is shown that in the presence of a pole the one-channel $S$-matrix can be
written as (for simplicity assuming the masses of the continuum particles to be
equal) \begin{eqnarray} \nonumber S &=&
\frac{(k-k_p-i\xi)(k+k_p-i\xi)}{(k-k_p+i\xi)(k+k_p+i\xi)} \\
&=&  \nonumber \frac{k^2-(k_p^2+\xi^2)-2ik\xi}{k^2-(k_p^2+\xi^2)+2ik\xi} \\
&=&\frac{s-s_0-4i(s-4m^2)^{1/2}\xi}{s-s_0+4i(s-4m^2)^{1/2}\xi} \ ,
\label{eq:Spole}
\end{eqnarray}
where $\xi\geq0$. {The unimodular form of Eq.~\eqref{eq:Spole} is because
of unitarity of the $S$-matrix.} The above parameterization accounts for the
facts that if the $S$-matrix has a pole at some complex momentum on the second
sheet, $k_p-i\xi$, it also has to have a pole $-k_p-i\xi$, which is the
realization of the Schwarz reflection principle in momentum space, and that any
pole on the second sheet is accompanied by a zero on the first.
As shown by the last equality, the momentum space expression can be
straightforwardly mapped onto the $s$-plane, where $s_0=4(k_p^2 + \xi^2 + m^2)$
was introduced. In the $s$-plane the Schwarz reflection principle calls for
poles at complex conjugate points.

\begin{figure} 
 \centering
   \includegraphics*[width=\linewidth]{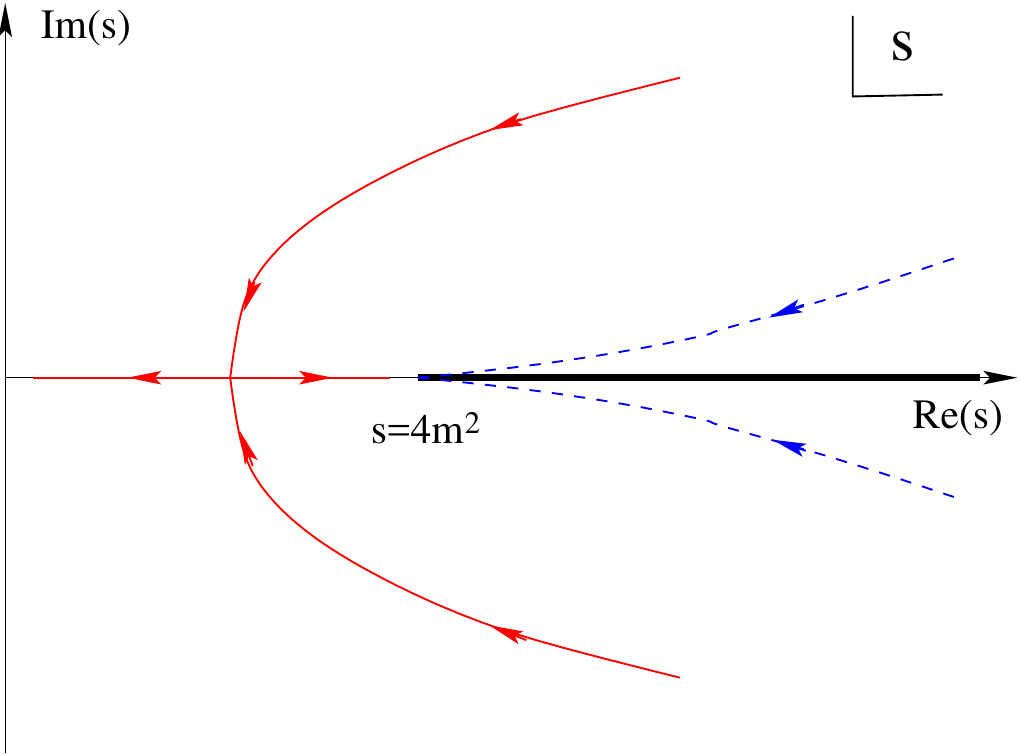}
   \caption{Typical pole trajectories for $S$-waves  (red, solid lines) and for 
higher partial waves (blue, dashed lines)
   in the second sheet of the complex $s$-plane. The thick line denotes the 
branch cut.}\label{fig:poletraj}
\end{figure}
To investigate the general behavior of the pole trajectories it is sufficient 
to vary the parameter $k_p^2$ from some finite positive
value to some finite negative value. Typical trajectories are shown in 
Fig.~\ref{fig:poletraj}.  The trajectories for $S$-waves
are depicted by the solid lines and for higher partial waves by the dashed 
lines. As long as $k_p^2$ is positive ($k_p$ is real),
Eq.~(\ref{eq:Spole}) develops two complex conjugate poles for all partial waves. 
When $k_p^2$ gets decreased, the poles approach each
other and eventually, for $k_p^2=0$, meet on the real axis. One of the poles 
switches to the first sheet at the
point were $k_p^2+\xi^2=0$, which is the threshold and at least for $S$-waves 
requires a negative value of $k_p^2$ ($k_p$ is imaginary).

The first nontrivial observation that can be read off Eq.~\eqref{eq:Spole} and
Fig.~\ref{fig:poletraj} straightforwardly is that $S$-waves and higher partial
waves behave very differently: The reason for this is the centrifugal barrier
that forces one to introduce a momentum dependence into $\xi$ according to
\begin{equation}
\xi(k) = \tilde \xi k^{2L} \ .
\end{equation}
This has a striking impact on the pole trajectories: For any $L>0$ the 
$\xi$-term is zero at $k=0$ and therefore
the point where the two pole trajectories  meet ($k_p^2=0$) coincides with 
$k=0$ which denotes the threshold. This is different for $S$-waves:
in this case the poles can meet somewhere below the threshold. Then, when 
$k_p^2$ is decreased further to negative 
values, both poles move away from the meeting point, such that one approaches 
the threshold while the other one goes away
from the threshold. This behavior can easily be interpreted via the pole 
counting approach: The further
away from the threshold the point is located where the two trajectories meet 
(this point is determined by the value of 
$\xi$ at the point where $k_p^2=0$), the more asymmetric are the two poles once 
one of them has switched to the first sheet, and thus the  molecular component of
 the state is more pronounced.

To make more explicit the above connection between the trajectories and the
molecular nature of the states we may study the scattering length and the
effective range that emerge from Eq.~(\ref{eq:Spole}):
\begin{equation}
a=-\frac{2\xi}{\xi^2+k_p^2} \ ; \quad r=-\frac{1}{\xi} \ .
\end{equation}
If we now use that the binding momentum $\gamma = \kappa_p-\xi$, where we 
introduced $\kappa_p=i k_p$, we can read off from the above equations and
Eq.~\eqref{eq:arwein}
\begin{equation}
\lambda^2 = 1-\frac{\xi}{\kappa_p} \ .
\end{equation}
To see the implications of this expression we may parameterize the relevant
quantities via
\begin{equation}
\gamma = \epsilon \delta \ ; \ \ \xi = \delta \ ; \ \ \kappa_p=(1+\epsilon)\delta \ \longrightarrow \lambda^2=\frac{\epsilon}{1+\epsilon} \ ,
\end{equation}
where $\delta>0$ and $\epsilon>0$.
Here it was already used that for a bound state to exist with a finite
binding energy $\kappa_p$ must exceed $\xi$. A vanishing binding momentum
($\gamma\to 0$) can be achieved either by $\epsilon\to 0$ for finite $\delta$, 
which immediately implies that $\lambda^2\to 0$, so that the state is purely
molecular, or $\delta\to 0$ for finite $\epsilon$. This case allows for a 
compact admixture of the very near-threshold pole, however, at the price of an 
extreme fine tuning of $\kappa_p$ and $\xi$ as both then need to go to zero 
simultaneously. This is the same kind of fine tuning already observed for 
non-molecular near-threshold states in Sec.~\ref{sec:wein} from a different
perspective.

As an example in Ref.~\cite{Hanhart:2014ssa} it was demonstrated explicitly that
the pole trajectories of the $f_0(500)$ meson as well as the $\rho$
meson that emerge when the quark masses are varied can be easily parameterized
in terms of the parameters $\xi$ and $\kappa_p$ introduced above.
In particular it was shown that while $\xi$ changes only mildly in the parameter
range studied $k_p^2$ changes a lot and in particular $\xi$ is sizable at the
point where $k_p^2$ is zero in the $f_0(500)$ channel. Accordingly the authors
of Ref.~\cite{Hanhart:2014ssa} conclude that, at least for unphysically large
quark masses, the $f_0(500)$ meson behaves like a hadronic molecule.

What should be clear from the considerations above is that states born off
hadron-hadron dynamics with poles above the relevant threshold are necessarily
broad; after all their coupling to this continuum channel is maximal.
This should also be clear from the pole trajectories illustrated in
Fig.~\ref{fig:poletraj}. An example for such a scenario is the very broad
$f_0(500)$ most probably generated by nonperturbative $\pi\pi$ interactions.
Such a property is in contrast to the tetraquark picture advocated in
Ref.~\cite{Esposito:2016itg}, where the authors argue that tetraquarks that are
visible in experiment must be narrow and slightly above threshold. It is
therefore of utmost importance that the pole locations of exotic candidates are
determined with high precision.

\subsubsection{Generalizations to resonances}
\label{sec:resonances}

The first work where the Weinberg approach was generalized to resonances is
Ref.~\cite{Baru:2003qq}, where the spectral density was employed to supplement
the parameter $\lambda^2$ introduced above for bound states.
The subject was later elaborated in various
papers~\cite{Hyodo:2011qc,Aceti:2012dd,Hyodo:2013iga,Hyodo:2013nka,
Sekihara:2014kya,Sekihara:2016xnq,
Guo:2015daa,Xiao:2016dsx,Xiao:2016wbs,Kang:2016ezb,Xiao:2016mon}; however, what
is common to all of them is that a quantitative, probabilistic extraction of the level of compositeness is not possible rigorously as soon as one moves to
resonances. The reason is that states that belong to poles on the second sheet
are not normalizable and as such one looses the condition of Eq.~(\ref{eq:norm})
that is crucial for the probabilistic interpretation.

However, it still appears reasonable to take over the key finding of
Sec.~\ref{sec:wein}, namely that the coupling of a state is larger for a larger
molecular component, also to resonance states.
As we will see in the following paragraphs, in certain situations this leads to
quite striking observable consequences. When being translated to an effective
field theory this observation implies that for molecular states loop
contributions appear always already at LO. The resulting power
counting will be detailed in Sec.~\ref{sec:4}.

When the state of interest is located below the production threshold of the
two hadrons that possibly form the molecular state, one can still distinguish
between quasi-bound states and virtual states, depending on whether the leading
pole is located on the first or the second sheet with respect to the mentioned
two-hadron system. The phenomenological implications of this will be discussed
in some detail in Sec.~\ref{sec:lineshapes}.

\subsection{Characteristic line shapes of hadronic molecules}
\label{sec:lineshapes}

\begin{figure} 
 \centering
   \includegraphics*[width=\linewidth]{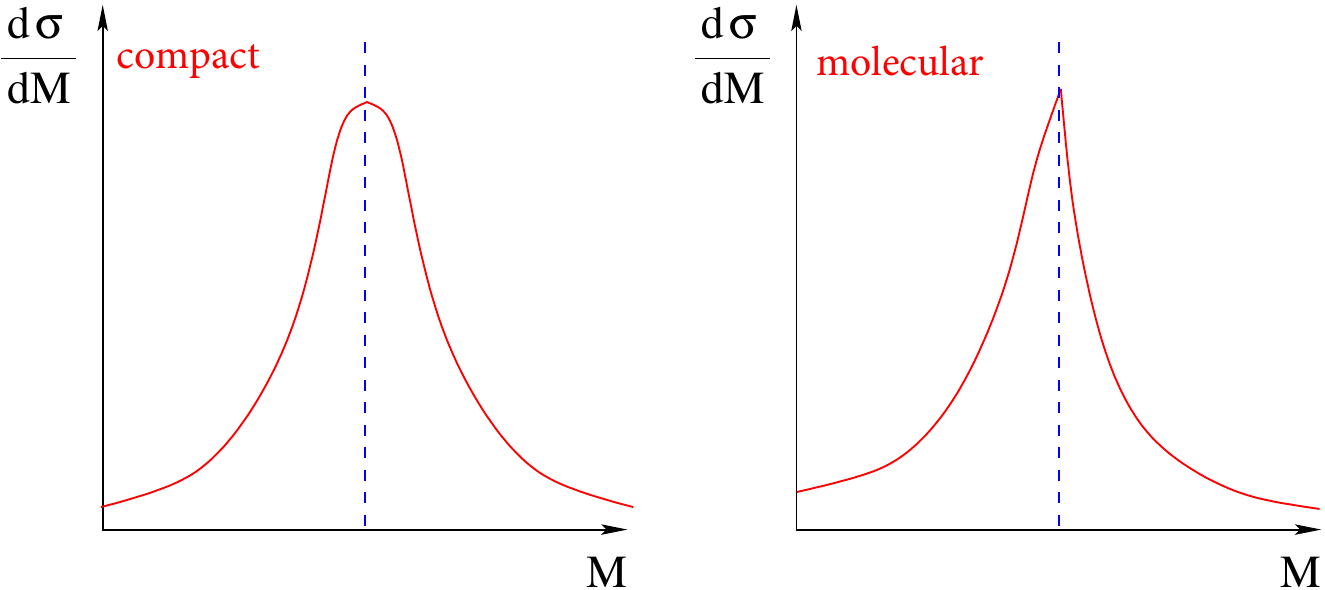}
   \caption{Sketch of typical near-threshold line shapes that emerge for compact
   (left panel) and molecular states (right panel). The dashed perpendicular line
   indicates the location of the threshold. The $x$--axis shows $M=m_1+m_2+E$.}\label{fig:lineshapes}
\end{figure}

Besides the deuteron all other (candidates for) hadronic molecules are unstable.
Then the scattering $T$-matrix needs to be modified compared to the form
discussed in Sec.~\ref{sec:wein}. In particular Eq.~(\ref{eq:poleTmatrix}) now
reads
\begin{equation}
T_{\rm in.}(E) = \frac{g^2/2}{E-E_r+(g^2/2)(i k+\gamma)+i\Gamma_0/2} \ ,
\label{eq:poleTmatrix_inel}
\end{equation}
where $E=k^2/(2\mu)$ and $\Gamma_0$ accounts for inelasticities not related
to the channel whose threshold is nearby. Those channels will be called
inelastic channels below.
We also changed the parameter that controls the pole location from $E_B$ to
$-E_r$ since it now refers to a resonance instead of a bound state.
Following the logic of the previous sections in the near-threshold regime the
dominant energy/momentum dependence for molecular states comes from the term
proportional to $g^2$ which is very large in this case, {\sl c.f.}
Eq.~(\ref{eq:gunrenorm}). On the contrary, for compact states the $k^2$ term
controls the momentum dependence. As a result of this in the former case the
line shape of the state that appears in any of the inelastic channels is very
asymmetric while in the latter it is symmetric. The two scenarios are sketched
in Fig.~\ref{fig:lineshapes}.
In addition, the line shape for the molecular state shows a clearly visible
nonanalyticity at the two-particle threshold which would be much weaker in the
other case.\footnote{It is completely absent only if the coupling between the
state with the two hadrons vanishes. However, in this case the $T$-matrix given
in Eq.~\eqref{eq:poleTmatrix_inel} vanishes as well.} Its presence follows 
directly from Eq.~(\ref{eq:poleTmatrix_inel}), since
\begin{equation}
\frac{\partial T_{\rm in.}(E)}{\partial E} \propto -\frac{1+(ig^2/2)(\partial k
/\partial E)}{(E-E_r+g^2/2(i k+\gamma)+i\Gamma_{0}/2)^2} \ ,
\end{equation}
with $\partial k/\partial E=\sqrt{\mu/(2E)}$. It is this derivative that is
not continuous when the energy crosses zero, the location of the threshold.
One might expect from this discussion that the coupling $g^2$ can be read off
from the line shape directly allowing for a direct interpretation of the
structure of the underlying state. However, a scale invariance of the expression
for the line shape appears as soon as the $g^2$-term dominates, and it
hinders a quantitative study in practice~\cite{Baru:2004xg}. In addition, even
for large values of  $g^2$ the nonanalyticity might well not show up clearly in 
the line shape since its visibility depends in addition on a subtle interplay 
of $E_r$, $\Gamma_0$ and $\gamma$.

It was already mentioned at the end of Sec.~\ref{sec:polecounting} that also
near-threshold virtual poles leave a striking impact on observables. In the
presence of inelastic channels a virtual state always leads to a peak of the
line shape exactly at the threshold while a near-threshold bound state still has
strength even below the threshold. For very near-threshold states like the
$X(3872)$ these two scenarios might be difficult to disentangle.

\begin{figure} %
 \centering
   \includegraphics*[width=\linewidth]{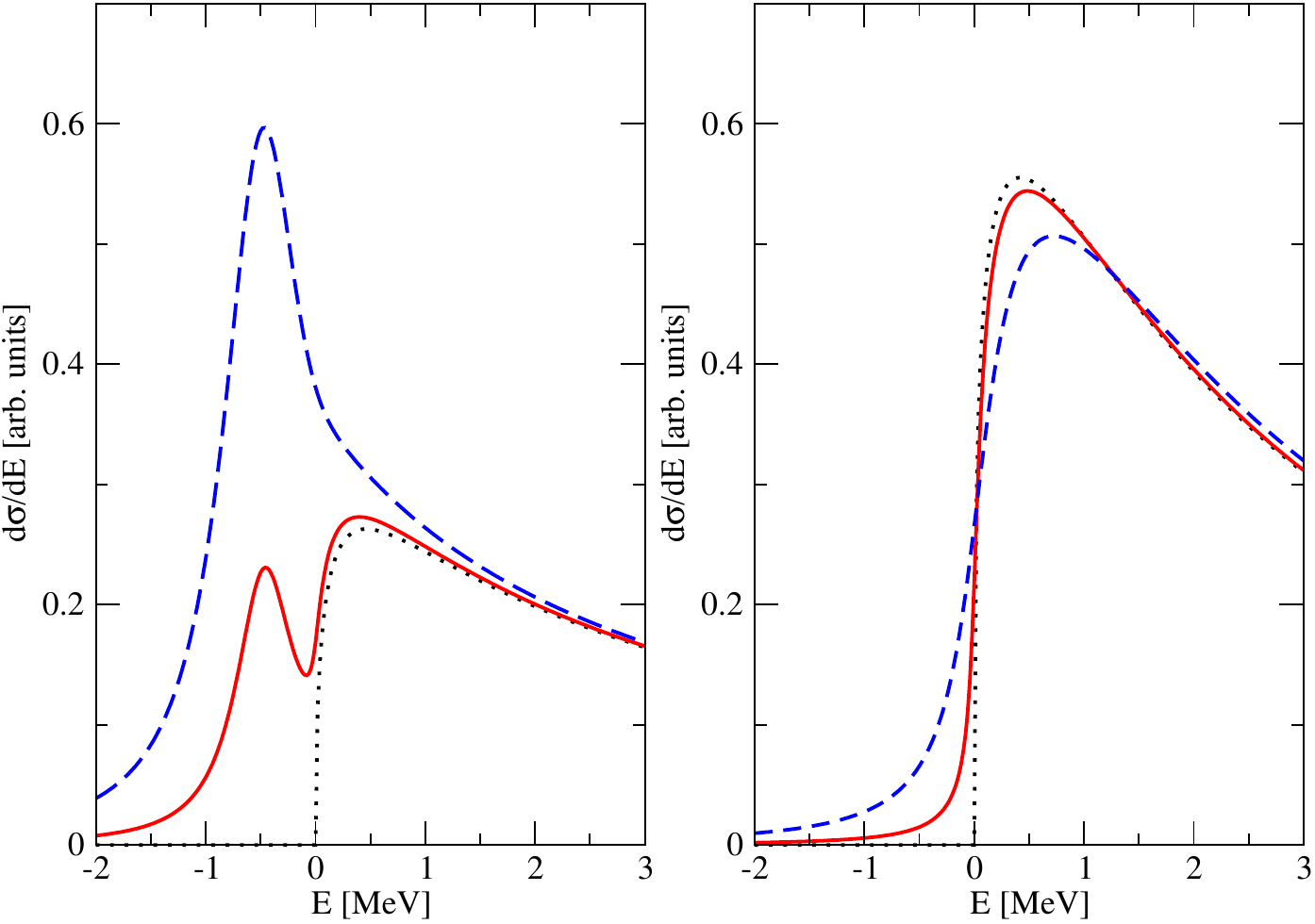}
   \caption{Line shapes that emerge for a bound state (left panel)
   and for a virtual state (right panel) once one of the constituents is 
   unstable.
   The dotted, solid and dashed line show the results for $\Gamma=0, 0.1$ and 
   $1$ MeV, respectively.
   The other parameters of the calculation are given in
Eq.~(\ref{eq:para}).}\label{fig:lineshapes2}
\end{figure}

As first stressed in Ref.~\cite{Braaten:2007dw} the line shapes in the elastic
channel might well be very interesting, if at least one of the constituents of
the system studied is unstable so that some strength of the amplitude leaks into
the region below the threshold~\footnote{At the same time the nonanalyticity
discussed above gets smeared out.}. To be concrete: If a state is located near
the threshold of particles  $A$ and $B$ and $A$ decays to $a$ and $b$, then  the
spectra in the  $abB$ channel need to be studied both below and above the
nominal $AB$ threshold.
To implement the necessary changes, in the formulae above for narrow
constituents one may simply replace the momentum $k$ in 
Eq.~(\ref{eq:poleTmatrix_inel}) by~\cite{Braaten:2007dw} 
\begin{eqnarray}\nonumber k_{\rm eff} &=&
\sqrt{\mu}\sqrt{\sqrt{E^2{+}\Gamma^2/4}{+}E} \\
& & \qquad + i\sqrt{\mu}\sqrt{\sqrt{E^2{+}\Gamma^2/4}{-}E}\ ,
\label{eq:unstableconst1}
\end{eqnarray}
where $\Gamma$ denotes the width of the unstable constituent, and $\mu$ is the
reduced mass of the two-hadron system evaluated using the mass of the unstable 
state. In addition, the subtraction term $\gamma$ needs to be replaced by
\begin{eqnarray}
\gamma_{\rm eff} = \pm\sqrt{\mu}\sqrt{\sqrt{E_r^2{+}\Gamma^2/4}{-}E_r} \ ,
\label{eq:unstableconst2}
\end{eqnarray}
where the upper sign leads to a (quasi-)bound state while the lower one to a
virtual state.
Clearly, for $\Gamma\to 0$ these expressions map nicely on those used for $k$
and $\gamma$ above for $E_r<0$.
Eqs.~(\ref{eq:unstableconst1}) and (\ref{eq:unstableconst2}) hold as long as the line shape for the unstable
constituent is well described by a BW distribution, namely for
$\Gamma/2\ll m_A-m_a-m_b$.
As soon as the energy dependence of $\Gamma$ starts to matter, more
sophisticated expressions need to be used, {\sl c.f.}
Ref.~\cite{Hanhart:2010wh}.
For simplicity we here use the expressions given above.
 The resulting line shapes in the elastic channel are shown for various values
of $\Gamma$ in Fig.~\ref{fig:lineshapes2},
where for illustration we used the parameters
 \begin{equation}
 E_r = -0.5 \ \mbox{MeV}, ~~ \Gamma_0= 1.5 \ \mbox{MeV},~~ g^2=0.1,~~ \mu =
 0.5\, .
 \label{eq:para} 
 \end{equation}
 The left panel shows the results for the (quasi-)bound state ($\gamma_{\rm
 eff}>0$), and the right one is for the virtual state ($\gamma_{\rm eff}<0$).
 Clearly, above the nominal two-hadron channel ($E=0$) the  spectra in the left
 and the right panels look very much alike, however, for $\Gamma>0$ drastic
 differences appear between the two cases for negative values of $E$.

 Following the Weinberg criterion, for the bound state case it is the relative
 height of the peak for $E<0$ and the bump for $E>0$, which are difficult to
 distinguish for the largest value of $\Gamma$, that is a measure of the
 molecular admixture of the studied state. Therefore a high resolution
 measurement of the line shape of $X(3872)$ would be very valuable to deduce its
 nature~\cite{Hanhart:2007yq,Braaten:2007dw,Hanhart:2010wh,Meng:2014ota,
 Kang:2016jxw}.

 In this context it is interesting to note that a line shape very similar to the
one shown in the left panel of Fig.~\ref{fig:lineshapes2}
 was also predicted for $Y(4260)\to D^*\pi \bar D$ under the assumption that
 $Y(4260)$ is a $D_1\bar D$ molecular state (note that $D^*\pi$ is the most
 prominent decay channel of $D_1(2420)$)~\cite{Wang:2013kra}, see the middle 
panel of Fig.~\ref{fig:Y4260LineShape}. A similar line shape also shows up in
the calculation in Ref.~\cite{Debastiani:2016xgg} for the $f_1(1285)$
strongly coupled to $K^*\bar K$.

Everything said so far had the implicit assumption that there are at most two
near-threshold poles on the two relevant sheets. The possible line shapes change
dramatically as soon as additional poles are located in the near-threshold
regime as is discussed in detail in
Refs.~\cite{Baru:2010ww,Artoisenet:2010va,Hanhart:2011jz}.

\subsection{Heavy quark spin symmetry}
\label{sec:3_HQSS}

 In the limit of infinitely heavy quarks, the spin of heavy quarks decouples
 from the system and is conserved individually. As a result, the total angular
 momentum of the light degrees of freedom becomes a good quantum number as well.
 This gives rise to the so-called heavy quark spin symmetry
 (HQSS)~\cite{Isgur:1989vq}.
 In the real world quarks  are not infinitely heavy, however, heavy quark
 effective field theory allows one to systematically include corrections that
 emerge from finite quark masses
in a systematic expansion in $\Lambda_{\rm QCD}/M_Q$, where $M_Q$ denotes the
heavy quark mass. For an extensive review we refer to
Ref.~\cite{Neubert:1993mb}. HQSS is the origin for the near degeneracy of, e.g.,
$D^*$ and $D$ as well as $B^*$ and $B$. Similarly, it also predicts
straightforwardly multiplets of hadronic molecules made of a heavy hadron or
heavy quarkonium and light hadrons~\cite{Guo:2009id,Yamaguchi:2014era}, and of
hadroquarkonium as well~\cite{Cleven:2015era}.
In addition, HQSS allows one to predict ratios of different transitions
involving heavy hadrons in the same spin multiplet, and in particular for your
interest transitions of hadronic molecules.
Examples for those predictions can be found in
Refs.~\cite{Fleming:2008yn,Fleming:2011xa} where the decays of the $\X$ into the
final states $\chi_{cJ}\pi$ and $\chi_{cJ}\pi\pi$ are discussed in the XEFT
framework {which will be discussed in Sec.~\ref{sec:4-pc}}. The ratios
among various decays of the $Z_b(10610)$ and $Z_b(10650)$ into $h_b(mP)\pi$ and
$\chi_{bJ}(mP)\gamma$ (from the neutral $Z_b$ states) were computed in both
XEFT~\cite{Mehen:2011yh} and \nreft~\cite{Cleven:2013sq} frameworks to be
discussed in the next section.
They are consistent with the result solely based on the
HQSS~\cite{Ohkoda:2012rj}.
For other predictions based on HQSS on the radiative and strong decays of
hadronic molecules in the heavy-quarkonium sector, we refer
to~\cite{Ma:2014ofa,Ma:2014zva}.

In special cases HQSS also allows one to make predictions for bound systems of
two or more heavy
mesons~\cite{Voloshin:2011qa,Mehen:2011yh,Nieves:2012tt,Guo:2013sya,Liu:2013rxa,
Baru:2016iwj} since certain potentials get linked to each
other. This will be discussed in details in Sec.~\ref{sec:4-interactions}.

\section{Nonrelativistic effective field theories}
\label{sec:4}

All the candidates for hadron resonances, and in particular the candidates of
hadronic molecules, which are the focus of this review, were discovered via
their strong decays into other hadrons. Therefore, to understand these
structures requires  also a study of their decays. Because of the
nonperturbative nature of QCD at hadronic energy scales, a first-principle
calculation of the spectrum of hadronic resonances at the level of quarks and
gluons can only be done using lattice QCD.
Although there has been tremendous progress in lattice QCD, a reliable
calculation of the full hadronic resonance spectrum for physical quark masses is
still out of reach. In addition, even if such calculations were available, the
interpretation of the emerging spectra still requires
 additional theoretical analyses.

Only in the special case discussed in Sec.~\ref{sec:3}, {\sl i.e.} for shallow bound
states coupling in an $S$-wave to a nearby continuum channel comprised of two
stable or at least narrow hadrons, one finds a direct and physical
interpretation for the leading and nonanalytic contribution of the wave function
renormalization constant $Z$ as the (normalizable) probability to find the
continuum contribution in the physical state.
 Because of the closeness of the
threshold to the mass of the physical composite state, such systems are ideal
objects to apply the concept of effective field theories (EFTs), which makes use
of the separation of scales and which per definition include  a
cutoff~\cite{Lepage:1989hf}. Of particular relevance here are the
nonrelativistic EFTs~(NREFTs). Note that the
general principles underlying any EFT are formulated in Weinberg's
paper on phenomenological Lagrangians~\cite{Weinberg:1978kz}.

As mentioned in Sec.~\ref{sec:3}, hadronic molecules are located close to some
strongly coupled thresholds.
We denote the low-energy (low-momentum) scale characterizing such a system,
given by the binding energy (binding momentum) defined in Eq.~(\ref{eq:Ebdef})
(Eq.~(\ref{eq:gamdef})), generically by $Q$.
All other hadronic scales that we may collectively label as $\Lambda$ are thus
regarded as hard. This enables one to construct a perturbation theory in
$Q/\Lambda$, which for near-threshold states should be a small number.
As will become clear below, it depends on the system which scale is appropriate
for $\Lambda$. For example when investigating the $f_0(980)$ as a candidate for a
$\bar KK$ molecular state, the inverse range of forces, the natural candidate
for $\Lambda$, is given by the mass of the allowed lightest exchange meson, the
rho meson. A phenomenologically adequate value for the binding energy is 10~MeV.
It corresponds to a binding momentum of 70~MeV, and thus $Q/\Lambda \sim
1/10$ is a good expansion parameter.\footnote{The subtle interplay of scales in
molecular transitions is  is discussed in detail in Ref.~\cite{Hanhart:2007wa}
on the example of decays of the $f_0(980)$.} Furthermore, the closeness to
threshold also means that the constituent hadrons can be treated
nonrelativistically.

As discussed in the preceding section, the most interesting information about
the structure of a near-threshold state is contained in its coupling strength to
the threshold channel, which measures the probability for finding the two-body
bound state component in the physical state. This is consistent with the
intuition that a state is the more composite the larger its coupling to the
continuum. As shown in Eq.~\eqref{eq:residue}, for a bound state the coupling
reaches its maximal value, if the physical state is purely an $S$-wave bound
state, $\lambda^2=0$. Hence, it is important to extract the value of the
coupling constant for understanding the nature of near-threshold structures.
In addition, a large coupling implies the prominence of hadronic loops not only
in the formation of the state but also in transitions and decays.
In this section, we will discuss the NREFT formalism which is a natural
framework for studying the transitions involving hadronic molecules with a small
energy release. It can also be used to compute the universal long-distance part
of the production/decay processes of hadronic molecules, which will be discussed
in Sec.~\ref{sec:6}.

The analytic structure of the three-point scalar loop integral (including the
TS) will be discussed in Sec.~\ref{sec:4-3ploop}. The power counting rules
for the NREFT treating all intermediate particles on the same footing will be
detailed in Sec.~\ref{sec:nreft1}. We denote such a theory  as \nreft. When
one of the intermediate particles is much more off-shell than the others, it can
be integrated out from \nreft~and one gets another effective field theory, here
called \nreftii, which  was originally introduced as XEFT to study the
properties of the $X(3872)$. The XEFT and its relation to \nreft~will be
discussed in Secs.~\ref{sec:4-XEFT} and \ref{sec:4-nreft2xeft}. 
Sec.~\ref{sec:4-interactions} is devoted to a brief description of the
formation of hadronic molecules.

\begin{figure}[t!]
  \begin{center}
   \vglue2mm
   \includegraphics[width=0.65\linewidth]{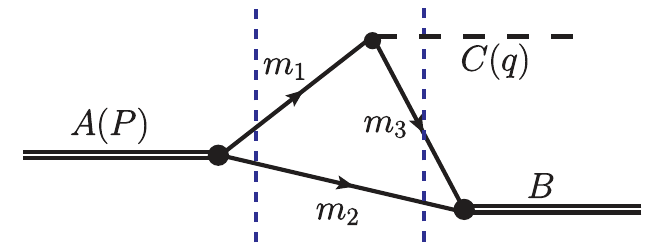}
   \caption{ A triangle diagram illustrating the long-distance contribution to
the transition between two heavy particles $A$ and $B$ with the emission of a
light particle $C$. The two vertical dashed lines denote the two relevant cuts.
   \label{fig:triangle}}
  \end{center}
\end{figure}

The formation of hadronic molecules may be viewed as a result of
nonperturbative hadron-hadron interactions. It is therefore natural to ask if
there is also an impact of hadron loops on the properties of more regular
excited hadrons.  Indeed, for certain transitions the effective field theory
\nreft~mentioned above  predicts prominent loop effects.
As examples, we briefly discuss single-pion/eta transitions and hindered M1
transitions between heavy quarkonia in Sec.~\ref{sec:4-NREFT_ccbar}. It will
become clear that whether the hadron-loop effects are important for
properties of an excited hadron is process-dependent. In particular, the
location of an excited hadron close to a threshold is a necessary but not
a sufficient condition.

\subsection{Power counting schemes}
\label{sec:4-pc}

As demonstrated in Sec.~\ref{sec:3}, the decisive feature of molecular states as
compared to more compact structures is the prominence of a two-hadron cut.
In some decays the cuts induced by intermediate particles might also matter.
To illustrate this point, we start this section by a discussion of the analytic
structure of three-point loop functions. This will shed light on the NREFT
power counting as well.

\subsubsection{Analytic structure of the three-point loop integral}
\label{sec:4-3ploop}

If a hadronic molecule has at least one unstable constituent, it can decay
directly through the decays of that unstable particle when phase space allows.
It can also decay into another heavy particle with a mass of the same order by
emitting light particles such as pions or photons from its constituents. The
mechanism for a transition accompanied by the emission of a single light
particle is depicted in Fig.~\ref{fig:triangle}.
In the figure the two vertical dashed lines show the relevant branch cuts:
They correspond to the time slices at which the intermediate particles can go
onto their mass shells.

We denote the intermediate particles as 
$M_{1,2,3}$ with masses $m_{1,2,3}$, and the external particles as $A,B,C$ 
with masses $m_{A,B,C}$, as shown in Fig.~\ref{fig:triangle}.  If 
all intermediate particles are nonrelativistic we can formulate a power 
counting based on the velocities of the intermediate particles. 
Let us start from the scalar triangle loop integral
\begin{widetext}
\begin{eqnarray}
    I(q) &=& i\int\!\frac{d^4l}{(2\pi)^4}
\frac{1}{\left(l^2-m_1^2+i\epsilon\right) \left[(P-l)^2-m_2^2+i\epsilon\right]
    \left[(l-q)^2-m_3^2+i\epsilon\right]}
    \nonumber \\
    &\simeq& \frac{i}{N_m} \int\!\frac{dl^0 d^3\bm l}{(2\pi)^4}
\frac1{\left[l^0{-}T_1(|\bm l|)+i\epsilon\right] \left[P^0{-}l^0{-}
T_2(|\bm{l}|)+i\epsilon\right] \left[l^0{-}E_C{-}T_3(|\bm
l{-}\bm{q}|)+i\epsilon\right] },
    \label{eq:scalarI}
\end{eqnarray}
where $\epsilon=0^+$, $N_m=8m_1m_2m_3$, $T_i(p)=p^2/2m_i$ denotes the kinetic 
energy for a 
heavy meson
with mass $m_i$, and $E_C$ the energy of the particle $C$ in the rest frame 
of the initial particle $A$. The second line is obtained by treating all 
the intermediate states 
nonrelativistically in the rest frame of the initial particle. Performing the 
contour integration over $l^0$, one gets a
convergent integral over the three-momentum. Defining
$\mu_{ij}=m_im_j/(m_i+m_j)$, $b_{12} = m_1+m_2-m_A$ and 
$b_{23}=m_2+m_3+E_C-m_A$, one has
\begin{equation}
  I(q) \simeq \frac{4\mu_{12}\mu_{23} }{N_m} \int\! \frac{d^3\bm 
l}{(2\pi)^3}\left[
  \left(\bm{l}^{\,2} + c_1 -i\epsilon\right) \left(\bm{l}^{\,2} + c_2 - 
\frac{2\mu_{23}}{m_3} \bm{l}\cdot\bm{q} - i\epsilon
  \right) \right]^{-1},
  \label{eq:loopinter}
\end{equation}
where $c_1= 2\mu_{12}b_{12}$, and
$c_2=2\mu_{23}b_{23}+\left(\mu_{23}/m_3\right){ q}^2$ with $q\equiv|\bm q|$.
The two terms in the denominator of the integrand
 contain a unitary cut each, as indicated by the
vertical dashed lines in Fig.~\ref{fig:triangle}. The other
two-body cut crossing the lines of $M_1$ and $M_3$ corresponds to the case that
the particle $M_3$ is propagating back in time (we assume implicitly that
it is $M_1$ and not $M_2$ that can decay to $M_3$ and $C$ in near on-shell 
kinematics). This is a relativistic effect which is
neglected here. The intermediate particles $M_1$ 
and $M_2$ are on shell when $\bm l^{\,2} + c_1 = 0$; $M_2$ and $M_3$ (as well 
as $C$) are on shell for $\bm{l}^{\,2} + c_2 - 
{2\mu_{23}} \bm{l}\cdot\bm{q}/m_3 = 0$. Accordingly,
 $\sqrt{|c_1|}$ and $\sqrt{|c_2|}$ define  two
different momentum scales where the corresponding intermediate states go on
shell.
Their values depend on all of the masses involved and may be very different 
from each other. For the nonrelativistic approximation to hold both must be 
small compared
to $m_i (i=1,2,3)$.

The integral of Eq.~(\ref{eq:loopinter}) can be presented in closed 
form~\cite{Guo:2010ak,Mehen:2015efa}
\begin{eqnarray}
    I(q) &=& {\cal N} \frac{1}{\sqrt{a}} \left[
\arctan\left(\frac{c_2-c_1}{2\sqrt{a(c_1-i\epsilon)}}\right) -
\arctan\left(\frac{c_2-c_1-2a}{2\sqrt{a(c_2-a-i\epsilon)}}\right) \right],
    \label{eq:Iexp} \\
    &=& {\cal N} \frac{1}{\sqrt{a}} \left[
\arcsin\left(\frac{c_2-c_1}{\sqrt{(c_2-c_1)^2+4ac-i\epsilon}}\right)
    - 
\arcsin\left(\frac{c_2-c_1-2a}{\sqrt{(c_2-c_1)^2+4ac-i\epsilon}}\right)
\right].
    \label{eq:Iexp2}
\end{eqnarray}
\end{widetext}
where ${\cal N}=\mu_{12}\mu_{23}/(2\pi m_1m_2m_3)$, and
$a = \left(\mu_{23}/m_3\right)^2 { q}^2$.
Especially Eq.~(\ref{eq:Iexp2}) highlights the presence of a special singularity
at
\begin{equation}
  (c_2-c_1)^2+4ac_1 = 0 \, .
  \label{eq:nrtrising}
\end{equation}
When rewriting the inverse trigonometric functions in terms of 
logarithms, one finds that this is a logarithmic divergence.
The solution of this equation gives the leading Landau 
singularity~\cite{Landau:1959fi} (for early
and recent reviews, see~\cite{Eden:1966,Chang:1983,Aitchison:2015jxa})
for a triangle diagram, also called triangle singularity, evaluated 
in nonrelativistic kinematics~\cite{Guo:2014qra}. The singularity location is 
slightly 
shifted from that found by solving the relativistic Landau equation. A 
comparison for a specific example can be found in the appendix of 
Ref.~\cite{Guo:2014qra}.  

Being nonlinear in all of the involved masses, Eq.~\eqref{eq:nrtrising} as well
as the Landau equation allow for different solutions. However, a direct
evaluation of the loop integral reveals that only in a very restricted
kinematics, one of the solutions produces an observable effect, namely
 when this solution is located on the physical
boundary, {\sl i.e.}, the upper edge of the branch cut in the first Riemann
sheet or alternatively the lower edge of the branch cut in the
second~\cite{Schmid:1967ojm}, see Fig.~\ref{sheets}. In this case, the TS can
produce a narrow peak in the invariant mass distribution, which may even mimic a
resonance. This effect
 was already indicated in Sec.~\ref{sec:TS} and will be further  illustrated  in
 Sec.~\ref{sec:6-ts}.
We therefore discuss now under which circumstances the singularity appears on
the physical boundary. This case is contained in the Coleman--Norton
theorem~\cite{Coleman:1965xm} (for triangle diagrams see
Ref.~\cite{Bronzan:1964zz}). The physical picture becomes most transparent using
the simple triangle singularity equation derived in Ref.~\cite{Bayar:2016ftu}:
\begin{equation}
  q_{\rm on+} = q_{a-}  \, ,
  \label{eq:trianglesing}
\end{equation}
where $q_{\rm on+}$
is the center-of-mass (CM) momentum of particles $M_1$ and $M_2$ when they are
on shell, and $q_{a-}$ is the momentum of particle $M_2$ in the rest frame
of A when $M_2$ and $M_3$ are on shell (being on shell is necessary but not
sufficient to define $q_{a-}$ as will be discussed immediately).
One finds
\begin{eqnarray}
  q_{{\rm on}+} &=& \frac1{2 m_A} \sqrt{\lambda(m_A^2,m_1^2,m_2^2)}\,,
  \nonumber\\
  q_{a-} &=& \gamma \left( \beta \, E_2^* - p_2^* \right) , 
\label{eq:qon}
\end{eqnarray}
where
\begin{equation}
  E_2^* = \frac{m_{B}^2+m_2^2-m_3^2}{2 m_{B}},\quad
  p_2^* = \frac{\sqrt{\lambda(m_{B}^2,m_2^2,m_3^2)}}{2 m_{B}},
\end{equation}
are the energy and the magnitude of the three-momentum of particle $M_2$ in the 
rest frame of particle B, {\sl i.e.} the CM frame of the ($M_2$, $M_3$) 
system, respectively, $\beta=q/E_B$ is the magnitude of the velocity of 
particle B in the rest frame of A, and $\gamma= 1/{\sqrt{1-\beta^2}} =
{E_{B}}/{m_{B}}$ is the Lorentz boost factor. 
Eq.~\eqref{eq:trianglesing} is the condition for the 
amplitude $I(q)$ to have a TS on the physical 
boundary. Note that if particle $M_1$ can go on shell
simultaneously with $M_3$ and $C$, it must be 
unstable. Consequently, its width 
moves the logarithmic divergence into the complex 
plane and the physical amplitude becomes finite.

Let us consider the kinematical region where the momentum of particle $M_2$ 
is positive so that Eq.~\eqref{eq:trianglesing} can be satisfied: $p_2 =q_{a-} 
= \gamma (\beta\,E_2^*-p_2^*) >0$. Then  $p_3=\gamma (\beta\,E_3^*+p_2^*)$ 
(where $E_3^*$ is the
energy of particle $M_3$ in the rest frame of particle B), the momentum of 
particle $M_3$ in the rest frame
of the initial particle, is positive as well.
This means that particles $M_2$ and $M_3$ move in the same direction in that 
frame. The corresponding velocities are given by
\begin{eqnarray}
  \beta_2 = \beta\, \frac{E_2^*-p_2^*/\beta}{E_2^*-\beta\, p_2^*}\,,\quad
  \beta_3 = \beta\, \frac{E_3^*+p_2^*/\beta}{E_3^*+\beta\, p_2^*}\,,
\end{eqnarray}
respectively. It is easy to see that $p_2>0$ leads to 
\begin{equation}
  \beta_3>\beta>\beta_2\,,
\end{equation}
which means that particle $M_3$ moves faster than $M_2$ and in the same 
direction in
the rest frame of the initial particle A.
This, together with the requirement that all intermediate particles are on their
mass shells, gives
the condition for having a TS on the physical boundary.
This is in fact the Coleman--Norton theorem~\cite{Coleman:1965xm} applied to the
triangle diagram:
the singularity is on the physical boundary if and only if the diagram can be 
interpreted as a classical process in spacetime. For other discussions about
TSs using the Mandelstam variables, we refer to two recent
works~\cite{Szczepaniak:2015eza,Liu:2015taa} and references therein.

To finish this section, we point out again that the TS mechanism has been around
for more than half a century, but only in recent years has become a viable tool
in hadron physics phenomenology due to the data discussed in this review. In
fact, many of the calculations outlined in which the TS plays a dominant role
can be and often are done without  recourse to an EFT. Still, in a broader view
it can nicely be embedded in the framework outlined here. In any case, whenever
the TS can play a role, it has to be included.

\subsubsection{\nreft}
\label{sec:nreft1}

A key component for any EFT is the power counting in terms of some dimensionless
small quantity, which allows for a systematic expansion and an estimate for the
uncertainty of the calculation caused by the truncation of the series at some
finite order. The natural small quantity in nonrelativistic systems is the
velocity $v$ (measured in units of the speed of light) which is much smaller
than one by assumption.

As mentioned in Sec.~\ref{sec:4-3ploop},  triangle diagrams with all
three intermediate particles being nonrelativistic in fact have two momentum
scales given by $\sqrt{|c_1|}$ and $\sqrt{|c_2|}$. Accordingly, one may define 
$v_1 = \sqrt{|c_1|}/(2\mu_{12})$ and $v_2=\sqrt{|c_2-a|}/(2\mu_{23})$ for the 
velocities of the intermediate mesons. 

From the previous analysis, three-point loop diagrams have two kinds of
singularities: two-body threshold cusps and TSs.
The two-body threshold singularities are encoded in the two velocities defined
above. When the TS, with its location implicitly defined via
Eq.~\eqref{eq:nrtrising}, is not in the considered kinematic region, the loop
function of Eq.~\eqref{eq:Iexp2} can be expanded in a power series as
\begin{eqnarray}
 I(q) &=& \frac{\mathcal{N}}{\sqrt{a} } \Bigg[ \left(\frac{\pi}{2} - 
\frac{2\sqrt{ac_1}}{c_2-c_1} \right) - \left(\frac{\pi}{2} - 
\frac{2\sqrt{ac_2}}{c_2-c_1} \right)  \nonumber\\ 
&& + \order{ \frac{\left(4 a c_1\right)^{3/2}}{(c_2-c_1)^3} } \Bigg] 
\nonumber\\ 
&=& {\cal N} \frac{2}{\sqrt{c_2}+\sqrt{c_1}} + \ldots\,.
  \label{eq:simplifiedI}
\end{eqnarray}
When the masses of all three intermediate particles are similar, 
$m_i\sim m$, the LO term in the above equation may be written as 
\begin{eqnarray}
    I(q) \sim \frac{{\cal N}}{m} \frac{2}{v_1+v_2} \, .
  \label{eq:Ipc}
\end{eqnarray}
Thus the arithmetic mean of the two velocities characterizes the size of the 
triangle loop. It is therefore the relevant parameter to estimate the leading 
loop contribution 
for the transition of a heavy state into a light state and another heavy state.

The power counting in nonrelativistic velocities for a given loop diagram can be
obtained by applying the following rules: The three-momentum of the intermediate
nonrelativistic particles counts as $\order{v}$, the nonrelativistic energy
counts as $\order{v^2}$, and each nonrelativistic propagator is of
$\order{v^{-2}}$. Thus, $I(q)$ scales as $\order{ v^5/(v^2)^3 }=\order{v^{-1}}$.
Comparing with Eq.~\eqref{eq:Ipc}, one sees that the velocity in the power
counting should be understood as the average of $v_1$ and
$v_2$~\cite{Guo:2012tg}.
In addition to the parts discussed the amplitude for a given process can have
factors of the external momentum $q$. To be general we do not count the external
momentum $q$ in powers of $v$, but keep it explicitly.
This defines the power counting as detailed in
Refs.~\cite{Guo:2009wr,Guo:2010zk,Guo:2010ak}. We denote this theory as \nreft.

\begin{figure}[tb]
  \begin{center}
   \includegraphics[width=0.48\linewidth]{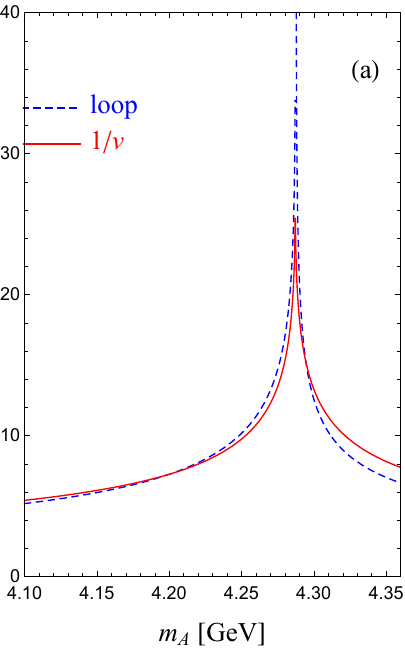}\hfill
   \includegraphics[width=0.48\linewidth]{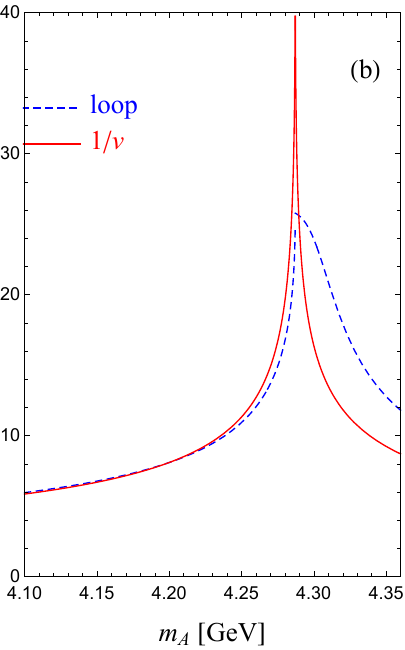}
   \caption{
   Comparison of the power counting rule for the scalar three-point loop 
integral, $1/v$, with the numerical result evaluated using
Eq.~\eqref{eq:Iexp}. The numerical result is normalized to $1/v$ at 
$m_A=4.22$~GeV. The involved masses are given in Eq.~\eqref{eq:masses}, and 
the mass for the final heavy particle takes the value of 3.886~GeV for (a) and 
3.872~GeV for (b).
   \label{fig:pc}}
  \end{center}
\end{figure}
In order to demonstrate how the power counting rules work, we compare in 
Fig.~\ref{fig:pc} the values of $1/v$ with $v=(v_1+v_2)/2$ and an explicit 
calculation of the loop function as given in Eq.~\eqref{eq:Iexp}. The curves 
are  normalized at $m_A=4.22$~GeV. The values used for the calculation are:
\begin{equation}
  m_1 = 2.420~\text{GeV},~~ m_2= 1.867~\text{GeV}, ~~
 m_3 = 2.009~\text{GeV}. 
\label{eq:masses}
\end{equation}
For the external light particle we take $m_C=0.140$~GeV for 
Fig.~\ref{fig:pc}~(a) and $m_C=0$ for Fig.~\ref{fig:pc}~(b). 
In addition, we take two values for $m_B$, 3.886~GeV and $3.872$~GeV, and the 
results are shown as (a) and (b), respectively. 
Then (a) and (b) correspond to the loop integrals in the amplitudes for the 
$\Y\to\Z\pi$ and $\Y\to\X\gamma$, respectively, which will be discussed later 
in Sec.~\ref{sec:6-long}.
Using 
Eq.~\eqref{eq:trianglesing} or \eqref{eq:nrtrising}, we find that for 
$m_B=3.886$~GeV, there is a TS at $m_A=4.288$~GeV, which is 
the reason for the sharp peak in the dashed line in Fig.~\ref{fig:pc}~(a).
Note that in
the plots the widths of the intermediate mesons were neglected. For 
$m_B=3.872$~GeV, which is smaller than $m_2+m_3$, and $m_C=0$, the TS
moves to the complex plane at $m_A=(4.301-i\,0.018)$~GeV with a clearly
visible effect on the line shape, see the dashed line in 
Fig.~\ref{fig:pc}~(b). 
One sees from the figure that the simple power counting rule of
Eq.~(\ref{eq:Ipc}) agrees remarkably well with the explicit
calculation except for energies very close to the TS.

\begin{figure}[tb]
  \begin{center}
   \includegraphics[width=0.6\linewidth]{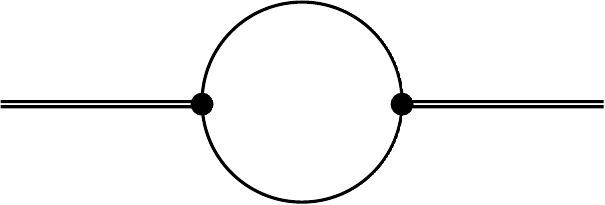}\\
   \caption{A one-loop two-point self-energy diagram.
   \label{fig:selfenergy}}
  \end{center}
\end{figure}

In addition to counting the loop integral as discussed above, one also needs to
take into account the vertices in order to obtain a proper estimate for a given
loop amplitude.
To illustrate the method let us start from the simplest two-point self-energy
diagram shown in Fig.~\ref{fig:selfenergy}. We assume that the mass of the state
is close to the threshold of the internal particles that can therefore be
treated nonrelativistically. If the coupling is in an $S$-wave, then the loop
scales as $\order{v^5/(v^2)^2}=\order{v}$.\footnote{Here we only focus on the
velocity scaling and neglect the geometric factor of $1/(4\pi)$.} If the
coupling is in a $P$-wave, each vertex contributes an additional factor of $v$
and the loop scales as $\order{v^3}$. Of course, the real part of the loop
integral is divergent, and the resulting correction to the mass is
scale-dependent.
However, since the scale dependence can be formally absorbed into the bare mass
of the state this discussion is not of relevance here. Thus, we find that the
effect of the two-hadron continuum on the self-energy of heavy quarkonia is
parametrically suppressed, if the state is close to the threshold which implies
a small value of $v$, and that this suppression increases for increasing orbital
angular momentum of the two-hadron state.

Next we consider the one loop diagram for the decay process $\text{A}\to 
\text{B\,C}$, with A and B heavy and C  light, as depicted in 
Fig.~\ref{fig:triangle}. To be concrete, 
we assume that C couples to the intermediate states $M_1$ and $M_3$ in a
$P$-wave (such as the pion couples to the ground state heavy mesons).
This coupling structure leads to  a 
factor of  $q$ (in the rest frame of A). The generalization to 
other situations is easy. The power counting rules for a few typical cases are 
then
as follows:
\begin{itemize}
  \item[(1)] Both A and B couple to the intermediate states in an $S$-wave.
  As a result the final state particles $B$ and $C$ must be in a $P$-wave. 
Therefore the expression of  Eq.~\eqref{eq:Ipc} needs to be multiplied by $q$. 
Still, the $1/v$ enhancement factor quantifies the relative importance of the 
triangle diagram for the transition: the closer both A and B to the 
corresponding thresholds, the more important the intermediate states. On top of 
this may come an additional enhancement
driven, e.g., by large couplings characteristic for molecular states as derived 
in Sec.~\ref{sec:3}.

  \item[(2)] Either A or B couples to the  intermediate states in an 
$S$-wave with the other one in a $P$-wave. In this case, because there is 
only one 
possible linearly independent external momentum for two-body decays, the 
internal momentum at the $P$-wave vertex must be turned into an external 
momentum. The amplitude scales as $\order{q^2/(m^2v)}$. Since the decay 
should be in an $S$-wave in this case, we have introduced a factor $1/m^2$ to 
balance the dimension of the $q^2$ factor~\cite{Guo:2010ak} as in this case
the loop contribution needs to be compared to a constant tree-level 
contribution.
 
  \item[(3)] Both A and B couple to the intermediate states in a $P$-wave. Each 
$P$-wave vertex contributes a factor of the internal momentum. In the power 
counting of \nreft, the external momentum is kept explicitly. As a result, 
there are two possibilities for the scaling of the $P$-wave vertices: Each 
$P$-wave vertex scales either as $m\,v$ or as the external momentum $q$. More 
insights can be obtained if we take a closer look at the relevant tensor loop 
integral:
\begin{equation}
  I^{ij}(\bm q\,) =  i\int\!\frac{d^4l}{(2\pi)^4} l^il^j\times 
\left[\text{integrand of }I(q)\right].
\end{equation}
In the rest frame of the initial particle, it can be decomposed into an 
$S$-wave part and a $D$-wave part as
\begin{equation}
  I^{ij}(\bm q\,) = P_S^{ij} I_S(q) + P_D^{ij} I_D(q),
\end{equation}
where 
\begin{equation}
  P_S^{ij}=\frac{\delta^{ij}}{\sqrt{3} }\,,\quad  
  P_D^{ij}= \frac1{\sqrt{6}} \left( 3\frac{q^iq^j}{q^2} -\delta^{ij} \right), 
\end{equation}
are the $S$- and $D$-wave projectors, respectively, which satisfy 
$P_S^{ij}P_S^{ij} = 1$, $P_D^{ij}P_D^{ij} = 1$, and $P_S^{ij}P_D^{ij} = 0$. 
Then in the $S$-wave part $I_S(q)$, the internal momentum scales as 
$\order{v}$, and $I_S(q)\sim\order{v}$. In the $D$-wave part $I_D(q)$, the 
internal momentum turns external, and one gets $I_D(q)\sim\order{q^2/(m^2 
v)}$, which would have the same scaling as $I_S(q)$ if 
$q/m\sim v$.\footnote{Noticing that $P_S^{ij} l^il^j=l^2/\sqrt{3}$ and 
$P_D^{ij} 
l^il^j=2l^2 P_2(\cos\theta)/\sqrt{6}$ with $P_2(\cos\theta)$ the second 
Legendre polynomial, it can be shown that $I_S(q)$ is UV divergent while 
$I_D(q)$ is UV convergent~\cite{Albaladejo:2015dsa,Shen:2016tzq}. The power 
counting of the $D$-wave part was not discussed in Ref.~\cite{Guo:2010ak}.} For 
the 
decay amplitude, the factor of $q$ from the vertex coupling C to intermediate 
states needs to be taken into account additionally.
 
\end{itemize}

This nonrelativistic power counting scheme was proposed in
Ref.~\cite{Guo:2009wr} to study the coupled-channel effects of charm meson
loops in charmonium transitions, and studied in detail later~\cite{Guo:2010ak}.
Applications to transitions between two heavy quarkonium states can be found in
Refs.~\cite{Guo:2010zk,Guo:2010ca,Guo:2011dv,Mehen:2011tp,Guo:2012tg,
Guo:2014qra}, and to transitions involving one or two $XYZ$ states in
Refs.~\cite{Cleven:2011gp,Cleven:2013sq,Guo:2013nza,
Esposito:2014hsa,Mehen:2015efa,Huo:2015uka,Wu:2016dws,Abreu:2016xlr,
Chen:2016mjn}. In particular in Ref.~\cite{Cleven:2013sq} the implications of
items (1) and (2) are demonstrated.
It is shown that, while the transitions of the $Z_b$ states to $\Upsilon(nS)\pi$
potentially suffer from large higher order corrections, the transitions to 
$h_b(mP)\pi$ and $\chi_{bJ}(mP)\pi$ should be dominated by the triangle
topology. The near-threshold cross section for $e^+e^-\to D\bar D$ was studied
in~\cite{Chen:2012qq} using NREFT as well.

It is clear that the power counting can only be applied to processes where the
intermediate hadrons are nonrelativistic and especially close to their mass
shells. Otherwise the loop diagrams receive contributions from large momenta and
cannot be treated in a simple EFT including only the hadronic degrees of freedom
of A, B, C, $M_1$, $M_2$ and $M_3$.

\subsubsection{\nreftii~and XEFT}
\label{sec:4-XEFT}

Because the $\X$ is arguably the most important and interesting candidate for a
hadronic molecule, here we will discuss in some detail one NREFT
designed specifically for studying the properties of the $\X$. It is called XEFT
and was proposed in Ref.~\cite{Fleming:2007rp} following the
Kaplan--Savage--Wise approach to describe the nucleon-nucleon
system~\cite{Kaplan:1998tg,Kaplan:1998we}. It can be regarded as a special
realization of \nreftii. Similar effective theories can be constructed for other
possible hadronic molecules which are located very close to thresholds. For
instance, in the framework of a similar theory, the $Z_b(10610)$ and
$Z_b(10650)$ were studied in Refs.~\cite{Mehen:2011yh,Mehen:2013mva} and the
$Z_c(3900)$ in Ref.~\cite{Wilbring:2013cha}.

The XEFT assumes the $\X$ to be a hadronic molecule of $D^0\bar D^{*0}+c.c.$.
The tiny binding energy~\cite{Olive:2016xmw}
\begin{equation}
  B_{X}=M_{D^0} +M_{D^{*0}}-M_X=(0.00\pm0.18)~\text{MeV},
  \label{eq:Xbe}
\end{equation}  
implies that the long-distance part of the $\X$ wave function is universal
and is insensitive to the binding mechanism which takes place at a much shorter
distance.
The long-distance degrees of freedom are $D^0$, $D^{*0}$, $\bar
D^0$, $\bar D^{*0}$ and $\pi^0$. All of them are treated nonrelativistically.
For processes dominated by the long-distance scales such as the decays $\X\to
D^0\bar D^0\pi^0$ and $\X\to D^0\bar D^0\gamma$ which can occur via the decay of
the vector charm meson directly, the XEFT at LO can reproduce the results
from the effective range theory which makes use of the universal two-body wave
function of the $\X$ at asymptotically long 
distances~\cite{Voloshin:2003nt,Voloshin:2005rt}
\begin{equation}
  \psi_X(r) \propto \frac{e^{-\gamma_0 r}}{r},
\end{equation}
where the $\X$ is assumed to be below the $D^0\bar D^{*0}$ threshold, and
$\gamma_0=\sqrt{2\mu_0 B_{X}}\leq20$~MeV with $\mu_0$ the reduced mass of 
$D^0$
and $\bar D^{*0}$.
Yet, it has the merit of being improvable order by order by including local
operators and pion exchanges although unknown short-distance coefficients will 
be involved. For processes involving shorter-distance scales such as the decays 
of the $\X$ into a charmonium and light particles, the XEFT can still be used by
parameterizing the short-distance physics in terms of local operators employing
factorization theorems and the operator product 
expansion~\cite{Braaten:2005jj,Braaten:2006sy}. The XEFT can also be used even
if the $\X$ is a virtual state with a nonnormalizable wave
function~\cite{Hanhart:2007yq} or a resonance above threshold.

The power counting and the NLO corrections to the decay $\X\to D^0\bar D^{0}
\pi^0$ were studied in Ref.~\cite{Fleming:2007rp}. The XEFT was also used to
study the decays of the $\X$ to the $\chi_{cJ}$ with one and two
pions~\cite{Fleming:2008yn,Fleming:2011xa}, the radiative transitions
$\X\to\psi(2S)\gamma$, $\psi(4040)\to\X\gamma$~\cite{Mehen:2011ds} and
$\psi(4160)\to \X\gamma$~\cite{Margaryan:2013tta}, the scattering of an
ultrasoft pion~\cite{Braaten:2010mg} or $D$ and $D^*$~\cite{Canham:2009zq} off
the $\X$, and the quark mass dependence and finite volume corrections of the
$\X$ binding energy~\cite{Jansen:2013cba,Jansen:2015lha}.
The relation between the XEFT and the formalism of \nreft~was clarified in
Ref.~\cite{Mehen:2015efa}.
As an extension of the XEFT in Ref.~\cite{Alhakami:2015uea} a modified power
counting was suggested to take into account also an expansion in the ratio
between the pion mass and the charm meson masses. The need for such an
expansion is removed, however, as soon as Galilean invariance is imposed on the
interactions~\cite{Braaten:2015tga}.

In the following, we use the decay $\X\to D^0\bar D^0\pi^0$ as an example to
illustrate the power counting of the XEFT. The binding momentum
$\gamma_0\leq20$~MeV sets the long-distance momentum scale in this theory.
The typical momenta for the $D^0$ and $D^{*0}$ are of the order $p_D\sim
p_{D^*}\sim\gamma_0$. The pion kinetic energy is less than 7~MeV, and thus the
momentum for either an internal or external pion is also counted as
$p_\pi\sim\gamma_0$. Furthermore, the pion exchange introduces another small
scale $\mu=\sqrt{\Delta^2-M_{\pi^0}^2}\simeq44$~MeV,  with
$\Delta=M_{D^{*0}}-M_{D^0}$. Denoting all the small momentum scales by $Q$, we
have
\begin{equation}
  \{p_D, p_{D^*}, p_\pi, \mu, \gamma_0\} =\order{Q}.
  \label{eq:XEFTpc}
\end{equation}
Thus, the measure for one-loop integral is of $\order{Q^5}$, and each 
nonrelativistic propagator is of $\order{Q^{-2}}$.
All Feynman diagrams can then be assigned a power of $Q$.

The XEFT keeps as the degrees of freedom only those modes with a very
low-momentum $\sim\gamma_0$.
The binding momentum for the $D^+D^{*-}+c.c.$ channel at the $\X$ mass is
$\gamma_c\simeq126$~MeV. It is treated as a hard scale, and the charged charm
mesons are integrated out from the XEFT.

Denoting the field annihilating the 
$D^0, \bar D^0$, $D^{*0}$ and $\bar D^{*0}$ by $D,\bar D$, $\bd$ and $\bdbar$, 
respectively, and taking the phase 
convention that $\X$ is $\left( D^0\bar D^{*0} +\bar D D^{*0} 
\right)/\sqrt{2}$, the relevant Lagrangian for the calculation up to the NLO is 
written as~\cite{Fleming:2007rp}
\begin{widetext}
\begin{eqnarray}
 {\cal L}_{\rm XEFT} 
  &=& 
 \sum_{\bm{\phi}=\bd,\bdbar } \bm{\phi}^{\dagger} \bigg(i\partial_0 + 
\frac{\bm{\nabla}^2}{2 M_{D^{*0}}
		    }\bigg)\bm{\phi} 
 +  \sum_{{\phi}=D,\bar D } D^\dagger \bigg(i\partial_0 + 
\frac{\bm{\nabla}^2}{2 M_{D^0} } \bigg) D 
 + \pi^\dagger \bigg(i\partial_0 + \frac{\bm{\nabla}^2}{2 M_{\pi^0}}
   + \delta\bigg) \pi
\nonumber \\
&&+ \left[  
\frac{g}{2 F_\pi} \frac{1}{\sqrt{ 2M_{\pi^0} } }
 \left( D \bd^\dagger \cdot \bm{\nabla}\pi  
   + \bar D^\dagger \bdbar \cdot \bm{\nabla}\pi^\dagger \right) + {\rm
 H.c.} \right]
\nonumber \\
 && 
- \,  
\frac{C_0}{2} \, \left(\bdbar D + \bd \bar D \right)^\dagger 
\cdot \left(\bdbar D + \bd \bar D \right) 
+  \left[   \frac{C_2 }{16} \, 
\left(\bdbar D + \bd \bar D \right)^\dagger 
\cdot \left(\bdbar (\overleftrightarrow \nabla)^2 D 
        + \bd (\overleftrightarrow \nabla)^2 \bar D \right) + {\rm H.c.} \right]
\nonumber \\
&&+ \left[  \frac{B_1 }{\sqrt{2}}\frac{1}{\sqrt{2 M_{\pi^0}}} \left(\bdbar D + 
\bd \bar D \right)^\dagger \cdot D \bar{D} \bm{\nabla} \pi + {\rm 
H.c.}\right] \nonumber\\
&& + \frac{C_\pi}{2 M_{\pi^0}} \left( D^\dagger \pi^\dagger D \pi + 
\bar D^\dagger 
\pi^\dagger \bar D \pi \right) + C_{0D} D\,^\dagger\bar{D}^\dagger D\bar D \,,
\label{eq:XEFTlag}
\end{eqnarray}
\end{widetext}
where $\delta=\Delta-M_{\pi^0}\simeq7$~MeV and
$F_\pi=92.2$~MeV is the pion decay constant. The first line 
contains the kinetic terms for the pseudoscalar and vector charm mesons as 
well as for the nonrelativistic pion, the second is for the axial coupling of 
the pion to charm mesons with $g\simeq0.6$ determined from the $D^{*}$ width, 
the third line contains the LO and NLO contact interaction terms, and the 
fourth line contains the terms for a  short-distance emission of a pion. The contact terms in 
the last line were not considered in Ref.~\cite{Fleming:2007rp}, but,
as is argued below, also contribute to $\X\to D^0\bar D^0\pi^0$ at NLO. In 
particular, the $C_{0D}$ term may have a significant impact on the line shapes
as will become clear in the  discussion below.

\begin{figure}[tb]
  \begin{center}
   \includegraphics[width=\linewidth]{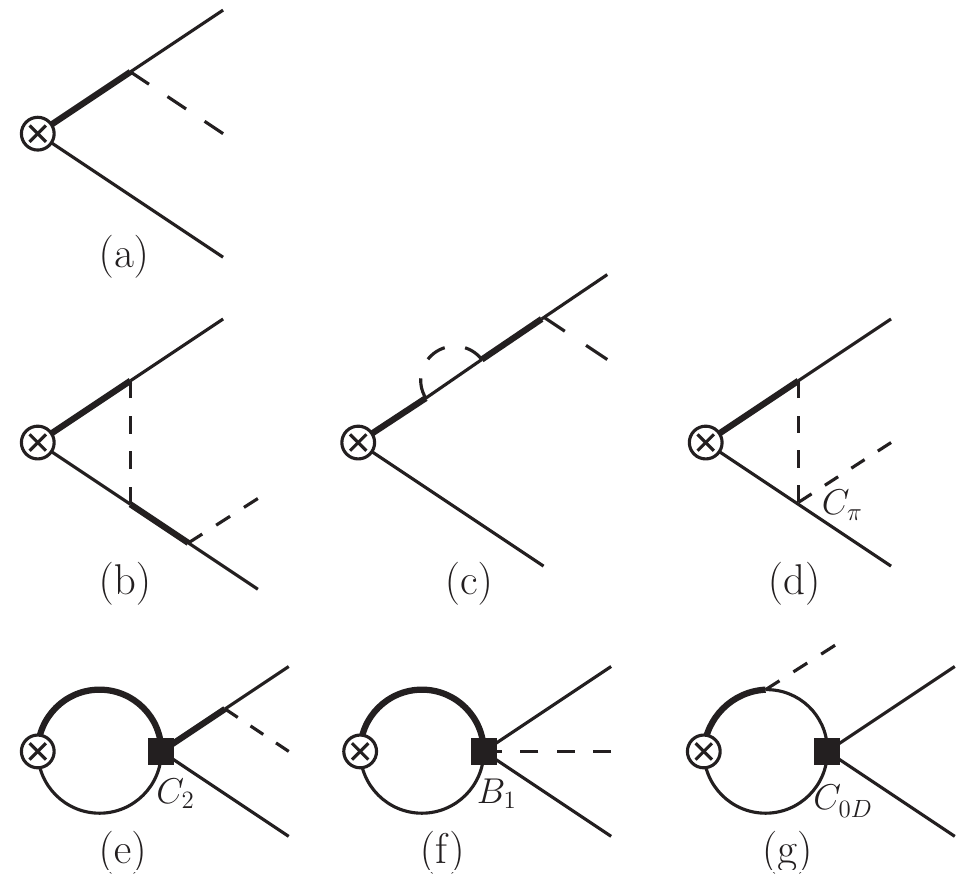}\\
   \caption{LO (a) and NLO (b,\ldots, g) diagrams for the calculation of the 
$\X\to D^0\bar D^0\pi^0$ decay width. The circled cross denotes an insertion of 
the $\X$, the thin and thick solid lines represent the pseudoscalar and vector 
charm mesons, respectively, and the dashed lines denote the pions.
   \label{fig:XDDpi}}
  \end{center}
\end{figure}

The Feynman diagrams relevant for the calculation of the $\X\to D^0\bar
D^0\pi^0$ decay width up to NLO are shown in Fig.~\ref{fig:XDDpi}. Diagram (a)
contributes at LO, (b,c) and (e,f) are the NLO diagrams calculated in
Ref.~\cite{Fleming:2007rp}, and (d,g) are two new diagrams from the new terms in
the last line of the Lagrangian in Eq.~\eqref{eq:XEFTlag}. Here we only discuss
the power counting for each diagram and the contributions missing in the
original work~\cite{Fleming:2007rp}, and refer  to Ref.~\cite{Fleming:2007rp}
for details of the calculation. One essential point of the XEFT is that the
pion-exchange is treated perturbatively based on the observation that the
two-pion exchange contribution is suppressed relative to the one-pion exchange
by
\begin{equation}
  \frac{g^2\mu_0\mu }{8\pi F_\pi^2} \simeq \frac1{20} \cdots \frac1{10} .
\end{equation}
Then the $\X$ is generated through a resummation of the $D\bar D^*$ contact 
terms (the 
charge conjugated $\bar D D^*$ channel is always implied). The 
pole of the $\X$ is at $E=-B_{X}$, and thus at LO 
\begin{equation}
  1+C_0\Sigma_0(-B_{X})=0 \ ,
  \label{eq:Xpole}
\end{equation}
where 
\begin{eqnarray}
  \Sigma_0(E) &=& - \left( \frac{\Lambda_{\rm PDS}}{2\pi} \right)^{4-D}\! 
\int\! \frac{d^{D-1} l}{(2\pi)^{D-1} } \frac1{E-l^2/(2\mu_0)+i\epsilon} 
\nonumber\\
  &=& \frac{\mu_0}{2\pi} \left(\Lambda_{\rm PDS} -\sqrt{-2\mu_0 E 
-i\epsilon} \right)
\label{eq:Sigma}
\end{eqnarray}
is the two-point one-loop integral containing nonrelativistic $D^0$ and $\bar 
D^{*0}$ propagators in the power divergence subtraction (PDS) 
scheme~\cite{Kaplan:1998tg,Kaplan:1998we}, where $E$ is the energy defined 
relative to the threshold and $\Lambda_{\rm PDS}$ is the PDS scale. 
For Eq.~\eqref{eq:Xpole} to be renormalization 
group invariant, $C_0$ needs to absorb the scale dependence of the loop 
integral:
\begin{equation}
  C_0(\Lambda_{\rm PDS}) = \frac{2\pi}{\mu_0\left(\gamma_0 - \Lambda_{\rm PDS} 
\right) }.
\label{eq:C0PDS}
\end{equation}
Keeping only momentum modes of order $Q$, the power counting for the loop 
integral is  $\Sigma_0(E)=\order{Q^5/(Q^2)^2}=\order{Q}$. One sees that 
the scale-independent part of $C_0$, $\bar C_0 = [\Lambda_{\rm PDS} + 
1/C_0( \Lambda_{\rm PDS}) ]^{-1} = 2\pi/(\mu_0\gamma_0)$, indeed scales as 
$Q^{-1}$.

Now we consider the power counting of the decay amplitudes from the 
diagrams 
in Fig.~\ref{fig:XDDpi}. The decay rate can be obtained from these amplitudes 
taking into account properly the wave function renormalization 
$Z$~\cite{Fleming:2007rp} which accounts for the insertion of the $\X$ 
interpolating field shown as circled crosses in Fig.~\ref{fig:XDDpi}. Notice 
that for the calculation of the decay rate up to the NLO, one needs $Z$ up to 
NLO (LO) for the LO (NLO) amplitude. The amplitude from diagram (a) scales as 
$\order{Q/Q^2}=\order{Q^{-1}}$ since there is one nonrelativistic propagator 
and one $P$-wave vertex which gives a factor of $p_\pi\sim Q$. Both one-loop 
diagrams (b) and (c) have four nonrelativistic propagators and three $P$-wave 
vertices, and thus scale as $\order{Q^0}$, one order higher 
than the LO diagram (a). The coefficients $C_2$ and $B_1$ scale as 
$Q^{-2}$~\cite{Fleming:2007rp}. Noticing that there are two derivatives in the 
$C_2$ term and one derivative in the $B_1$ term in the Lagrangian, the 
amplitudes from diagrams (e) and (f) should be counted as $\order{Q^0}$ as well.

Let us discuss diagrams (d) and (g) which were missing in the original 
calculation in Ref.~\cite{Fleming:2007rp}. The $C_\pi$ contact term can be 
matched to the chiral Lagrangian for the interaction between heavy and light 
mesons~\cite{Burdman:1992gh,Wise:1992hn,Yan:1992gz,Guo:2008gp}. At LO of the 
chiral expansion the interaction between pions and pseudoscalar heavy 
mesons receives contributions from the Born term from the exchange of $D^*$, 
which constitutes a subdiagram to (b) and (c), and the Weinberg--Tomozawa term. 
It turns out that the amplitude for $D^0\pi^0\to D^0\pi^0$ vanishes at LO. At 
NLO of the chiral expansion, there are several operators, see 
Refs.~\cite{Guo:2008gp,Guo:2009ct}.  In particular, it is easy to see that the 
$h_0$ and $h_1$ terms therein are proportional to the light quark mass or 
equivalently to $M_\pi^2$. The Feynman rule for the $D^0\pi^0\to D^0\pi^0$ vertex 
from these two terms (using relativistic normalization for all the fields) is
\begin{equation}
  i\,\Amp_{h_0,h_1} = i \frac{2}{3} \left(6h_0+h_1\right) 
\frac{M_\pi^2}{F_\pi^2} .
\end{equation}
The value of $h_1$ is fixed to be 0.42 from the mass splitting between the 
$D_s$ and $D$ mesons, and the $1/N_c$ suppressed parameter $h_0\simeq0.01$ from 
fitting to the lattice data for the pion mass dependence of charm meson 
masses~\cite{Liu:2012zya}. One sees $\Amp_{h_0,h_1}\simeq 0.65$. Hence, by 
matching to the chiral Lagrangian, $C_\pi$ should scale as $Q^0$, which leads 
to the scaling of $\order{Q^0}$ for diagram (d). 

\begin{figure}[tb]
  \begin{center}
   \includegraphics[width=\linewidth]{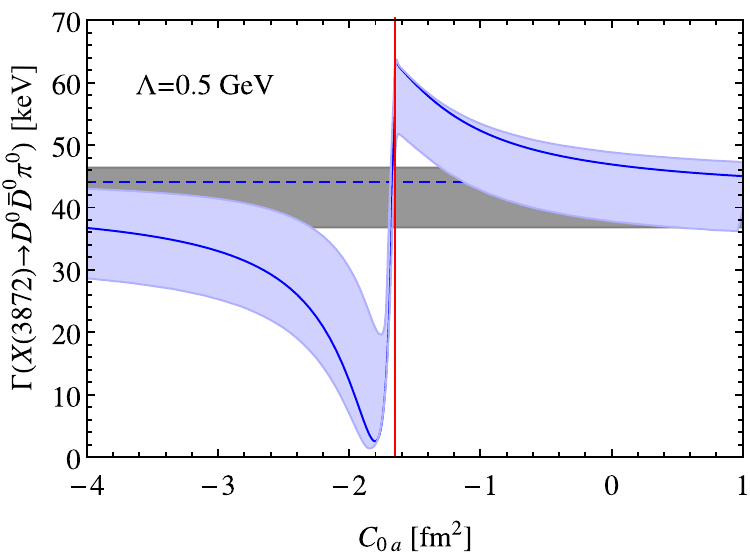}\\
   \caption{Decay width of the $\X\to D^0\bar D^0\pi^0$ calculated in 
Ref.~\cite{Guo:2014hqa} taking into account the $D\bar D$ final state 
interaction in the framework of Lippmann--Schwinger equation regularized by a 
Gaussian form factor. Here the cutoff in the Gaussian regulator is taken to be 
$\Lambda=0.5$~GeV, and $C_{0a}$ is the unknown isoscalar part of the $D\bar D$ 
contact term. The gray and blue bands correspond to the uncertainty bands 
without and with the $D\bar D$ final state interaction, respectively. The 
vertical line denotes the $D^0\bar D^0$ threshold. Adapted from 
Ref.~\cite{Guo:2014hqa}. 
   \label{fig:XDDpi_FSI}}
  \end{center}
\end{figure}

Diagram (g) involves a short-distance contact interaction between $D^0$ and
$\bar D^0$. If the vertex $C_{0D}$ scales as $Q^0$, then diagram (g)
$=\order{Q^0}$. However, the situation could be more complicated. From the HQSS
analysis of the $\X$ in Sec.~\ref{sec:3_HQSS}, the $\X$ as a $D\bar D^*$
hadronic molecule should have three spin partners in the strict heavy quark
limit. One of them has quantum numbers $J^{PC}=0^{++}$ and couples to $D\bar D$
and $D^*\bar D^*$. Therefore, there is the possibility that the $D\bar D$
interaction needs to be resummed to generate a near-threshold pole. In this
case, $C_{0D}$ needs to be promoted to be $\order{Q^{-1}}$, analogous to $C_0$.
Then diagram (g) appears at $\order{Q^{-1} }$ making it a LO contribution.
Clearly this can cause a large correction to the $\X\to D^0\bar D^0\pi^0$  decay
rate. This effect can be seen clearly in Fig.~\ref{fig:XDDpi_FSI}, which is the
result obtained in Ref.~\cite{Guo:2014hqa} using \nreft~in combination with the
framework to be discussed in Sec.~\ref{sec:4-interactions}. The unknown
parameter $C_{0a}$ in the figure parameterizes the isoscalar part of $D^\dag\bar
D^\dag D\bar D$ contact interaction, see Eq.~\eqref{C0++} below, playing a role
similar to $C_{0D}$ introduced in Eq.~\eqref{eq:XEFTlag}.

Since $\X\to D^0\bar D^0\pi^0$ is an important process sensitive to 
the long-distance structure of the $\X$, it would be interesting to revisit it 
considering the missing diagrams in XEFT. In particular, it was found that the 
nonanalytic corrections from the pion-exchange diagrams (b) and (c) of Fig,~\ref{fig:XDDpi} only 
contribute to $\sim1\%$ of the decay rate~\cite{Fleming:2007rp}. Whether this 
remains true after considering diagram (d) remains to be seen.

It should be stressed that the role of nonperturbative pions on the $\X$ 
properties is studied in various
papers~\cite{Baru:2011rs,Baru:2013rta,Baru:2016iwj} which in many cases 
confirm the results of XEFT. However, also in these studies diagrams of the 
types shown in diagrams (d) and (g) of Fig.~\ref{fig:XDDpi} were not included.

\subsubsection{From \nreft~to XEFT}
\label{sec:4-nreft2xeft}

From the discussions above, we see that all momentum scales much larger than 
$\gamma_0\leq20$~MeV have been integrated out from the XEFT. This is different 
from \nreft, where all nonrelativistic modes are kept as effective degrees 
of freedom including those with a momentum of the order of a few hundreds of 
MeV. \nreft~when applied to the $\X$ can be regarded as the high-energy theory 
for the XEFT. The short-distance operators in XEFT at the scale of a few 
hundreds of MeV can be matched to \nreft. This is
discussed in some detail in Ref.~\cite{Mehen:2015efa} in the context of 
calculations of the reactions $\X\to\chi_{cJ}\pi^0$.

To show the relation between \nreft~and XEFT explicitly let us consider the case 
$c_2\gg c_1$.
The quantities $c_2$ and $c_1$  introduced in Eq.~\eqref{eq:loopinter} define
 the locations of the two-body cuts of the triangle diagram. In the low-momentum region $l\sim \sqrt{c_1}$, 
 the second factor in the integrand of Eq.~\eqref{eq:loopinter} can 
be expanded in powers of $l^2/c_2$ and one gets
\begin{eqnarray}
  I(q) &=& \frac{4\mu_{12}\mu_{23}}{N_m c_2} \int^\Lambda \frac{d^3
l}{(2\pi)^3}\frac1{{l}^{\,2} + c_1 -i\epsilon} \left[1+ \order{\frac{c_1}{c_2}} 
\right] \nonumber\\
&\simeq& \frac{\mu_{12}}{2\pi N_m  \left[b_{23} + q^2/(2 m_3) \right] }  \left(\Lambda_{\rm PDS} -\sqrt{c_1 {-}i\epsilon} \right)\!.
~~~~
\end{eqnarray}
The resulting momentum integral in the first line is divergent
 and  needs to be regularized. The 
natural UV cutoff of the new effective theory
is set by $\Lambda<\sqrt{c_2}$. We denote such a theory as 
\nreftii. It reduces to the XEFT when applied to the $\X$. In order to compare 
with the XEFT, in the second line of the above equation we evaluate the 
integral in the PDS scheme which is equivalent to the sharp cutoff 
regularization by letting $\Lambda_{\rm PDS}= 2 \Lambda/\pi$ and dropping the 
terms of $\order{1/\Lambda}$. For a detailed comparison of 
dimensional versus cutoff regularization we refer to
Ref.~\cite{Phillips:1998uy}.

For $m_{1}=M_{D^{*0}}$, $m_2=M_{D^0}$, $E_C=E_\pi$, and $m_A=M_X$, the second 
line of the above equation reduces to 
\begin{equation}
  -\frac1{N_m \left(E_\pi + \Delta_H \right)} \frac1{ 
C_{0}(\Lambda_{\rm PDS} ) },
\end{equation}
where $\Delta_H=M_{D^0}+m_3-M_X$, and the term $q^2/(2m_3)$ has been neglected. 
Terms of the above form appear in the XEFT amplitudes for transitions between 
the $\X$ and a charmonium with the emission of a light 
particle~\cite{Fleming:2008yn,Fleming:2011xa,Mehen:2011ds,Margaryan:2013tta}.

The different power countings of XEFT and \nreft~has various implications 
that we now illustrate by two examples:

   Since \nreft~keeps all nonrelativistic modes explicitly, the charged
  $D\bar D^*$ channel which has a momentum of $\gamma_c\simeq126$~MeV needs to
  be kept as soft degrees of freedom. On the contrary, the XEFT only keeps the
  ultrasoft neutral charm mesons dynamically and the charged ones are integrated
  out.
  It was pointed out in Ref.~\cite{Mehen:2015efa} that it is crucial to take
  into account the charged charm mesons for the calculation of the
  $\X\to\chi_{cJ}\pi^0$ decay rate in \nreft~because their contribution
cancels to a large extent the one
  from the neutral charm mesons as usual in isospin violating transitions
({\sl c.f.}
the discussion in Sec.~\ref{sec:isospinviol}).\footnote{The role of the
charged
 charm mesons for certain decays of the $\X$ was already stressed in
 \cite{Gamermann:2009fv}.}
  The situation for decays into an isoscalar pion pair, $\X\to\chi_{cJ}\pi\pi$,
  is different. We expect that the charged and neutral channels are still of
  similar order, but add up constructively.
  
\begin{figure}[tb]
    \begin{center}
     \includegraphics[width=\linewidth]{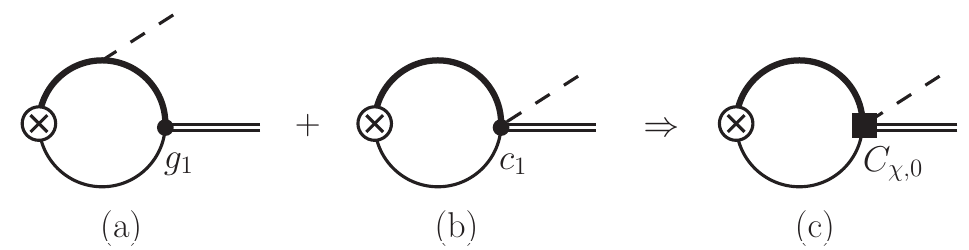}\\
     \caption{Diagrams for calculating the decay rate for the process
     $\X\to\chi_{c1}\pi^0$.
     The circled cross denotes an insertion of the $\X$, the thin and thick 
     solid lines represent the pseudoscalar and vector charm mesons, 
     respectively, the dashed lines present the pions, and the double lines
     correspond to the $\chi_{c1}$.
     \label{fig:Xchic1pi}}
    \end{center}
  \end{figure}

  Furthermore, in the XEFT calculation for  $\X\to\chi_{cJ}\pi^0$, there appears a new,
  reaction specific
  short-distance operator, labeled by $C_{\chi,0}$ in
  Fig.~\ref{fig:Xchic1pi}~(c). 
  To estimate its size it is matched onto two contributions in heavy meson
  chiral perturbation theory in Refs.~\cite{Fleming:2008yn,Fleming:2011xa}. Those
  are given by the exchange of a charm meson, which is proportional to the
  $\chi_{cJ}H\bar H$ coupling constant $g_1$, and a contact term accompanied
  by a low energy constant $c_1$, shown as diagram (a)
  and (b) in Fig.~\ref{fig:Xchic1pi}, respectively. The final result in XEFT
  then depends on the unknown ratio $g_1/c_1$. In \nreft, however, the two
  contributions appear at different orders, since the amplitude from diagram (b) is
  suppressed by $v^2$ compared with that from diagram (a).
  
\subsection{Formation of hadronic molecules}
\label{sec:4-interactions}

While so far the focus was on transitions of molecular candidates, we now 
turn to their formation through two-hadron scattering. 
For illustration we focus in this chapter on the scattering of open-flavor heavy
mesons off their antiparticles in a framework of  
NREFT similar to the EFT for nucleon-nucleon 
interactions~\cite{Epelbaum:2008ga}. 
The example of the formation of $\Lambda(1405)$ from
similar dynamics is discussed in Sec.~\ref{sec:1405th}.
In this section we mainly discuss the method used in 
Refs.~\cite{Nieves:2011vw,Nieves:2012tt,Valderrama:2012jv,
HidalgoDuque:2012pq,Guo:2013sya,Guo:2013xga}. It is based on the 
Lippmann--Schwinger equation (LSE) regularized using a Gaussian vertex form 
factor. The coupled-channel LSE 
reads 
\begin{eqnarray}
      T_{ij}(E;\bm k',\bm k) &=& V_{ij}(\bm k', \bm k) \\
    &+&  \sum_n \int\! \frac{d^3l}{(2\pi)^3} 
\frac{ V_{in}(\bm{k}',\bm{l})\, T_{nj}(E;\bm l, \bm k)  }{  
E-l^2/(2\mu_{n}) - \Delta_{n1} + i\epsilon  }, \nonumber
 \label{eq:lse}
\end{eqnarray}
where $\mu_{n}$ is the reduced mass in the $n$-th channel, $E$ is the 
energy defined relative to threshold of the first channel, and $\Delta_{n1}$ is 
the difference between the $n^{\rm th}$ threshold and the first one. When the 
potential takes a separable form $V_{ij}(\bm k', 
\bm k)=\xi_{i}{(\bm k')}V_{ij}\varphi_{j}{(\bm k)}$, where the $V_{ij}$ are 
constants, the equation can be  
simplified greatly. In addition, for very near-threshold states one should 
expect a momentum 
expansion for the potential to converge fast and a dominance of $S$-waves. 
Both the separability as well as the absence of higher partial waves will be 
spoiled as soon as the one-pion exchange is included on the potential level;
this case will be discussed briefly later in this section.

With a UV regulator such as of the Gaussian form, see, 
e.g., Ref.~\cite{Epelbaum:2008ga}, 
\begin{equation}
  V_{ij}(\bm k',\bm k) =  e^{-\bm k^{\prime2}/\Lambda^2} V_{ij} e^{-\bm 
k^{2}/\Lambda^2} \ ,
\label{eq:potdef}
\end{equation}
the LSE can be solved straightforwardly.  
If the $T$-matrix has a near-threshold bound state pole, the effective coupling 
of this composite state to the constituents can be obtained by calculating the 
residue of the $T$-matrix element at the pole. For simplicity, we consider a 
single-channel problem with the LO contact term: $V(\bm k',\bm k) 
= C_0 e^{-\bm k^{\prime2}/\Lambda^2} e^{-\bm 
k^{2}/\Lambda^2}$. The nonrelativistic $T$-matrix element for the scattering of 
the two hadrons is then given by 
\begin{equation}
T_\text{NR}^{}(E) = \left[ C_0^{-1} + \Sigma_\text{NR}^{}(E) \right]^{-1}, 
\label{eq:T1c}
\end{equation}
where 
\begin{equation}
\Sigma_\text{NR}^{}(E)= \frac{\mu}{2\pi} \left[\frac{\Lambda}{\sqrt{2\pi}} 
- \sqrt{-2\mu E-i\epsilon} \right] +\order{ {\Lambda}^{-1} }
\label{eq:SigmaGaussian}
\end{equation}
is the nonrelativistic two-point scalar loop function 
defined in Eq.~\eqref{eq:Sigma} but evaluated with a Gaussian regulator. After 
renormalization by absorbing the cutoff dependence into $C_0$, we obtain
\begin{equation}
  T_\text{NR}^{}(E)=\frac{2\pi/\mu}{\gamma -\sqrt{-2\mu E-i\epsilon} } +\order{ 
{\Lambda}^{-1} } .
  \label{eq:TNR}
\end{equation}
The binding momentum $\gamma$ was defined in Eq.~(\ref{eq:gamdef}).
The 
effective coupling is obtained by taking the residue at the pole $E=-E_B$:
\begin{eqnarray}
  g_\text{NR}^2 &=& \lim_{E\to -E_B} (E+E_B) T_\text{NR}^{}(E) = \left[  
\Sigma_\text{NR}'(-E_B) \right]^{-1} \nonumber\\
  &=& \frac{2\pi\gamma}{\mu^2} .
  \label{eq:gNR}
\end{eqnarray}
It does not depend on $C_0$, and is scale independent up to terms suppressed by 
$1/\Lambda$.
Multiplying $g_\text{NR}^2$ by the factor $(8 m_1 m_2 M)$ to get the 
relativistic  
normalization, we recover the expression for $g_\text{eff}$ derived in 
Eq.~\eqref{eq:residue} for  
 $\lambda^2=0$. Thus we find that a potential of the
kind given in Eq.~(\ref{eq:potdef}) generates hadronic molecules.
Deviations of this result behavior can be induced, e.g., by momentum
dependent interactions (or terms of order $\gamma/\Lambda$). This
observation formed the basis for the generalization of the Weinberg 
compositeness criterion presented in 
Refs.~\cite{Aceti:2012dd,Hyodo:2011qc,Hyodo:2013nka,Sekihara:2014kya}.

To proceed we first need to say a few words about the scattering of heavy
mesons.
For infinitely heavy quarks the spin of the heavy quark decouples, and
accordingly in a reaction not only the total angular momentum is conserved but
also the spin of the heavy quark and thus the total angular momentum of the
light quark system as well. Therefore, a heavy-light quark system can be labeled
by the total angular momentum of the light quark system $j_\ell$. Accordingly
the ground state mesons $D$ and $D^*$  ($\bar B$ and $\bar B^*$) form a doublet
with $j^P_{\ell}=1/2^-$, where we on purpose deviate from the standard notation
$s_\ell^P$ to remind the reader that the light quark part can well be a lot more
complicated than just a single quark.
Candidates of the next doublets of excited states are $D^*_0(2400)$\footnote{The
$D\pi$ $S$-wave resonant structure is probably more complicated than a single
broad resonance, as demonstrated by a two-pole structure in
Ref.~\cite{Albaladejo:2016lbb}.} and $D_1(2430)$ (the corresponding $B$-mesons
are still to be found), characterized by $j^P_{\ell}=1/2^+$ and a width of about
300~MeV, and $D_1(2420)$ and $D_2^*(2460)$ ($\bar B_1(5721)$ and $\bar
B^*_2(5747)$) with $j^P_{\ell}=3/2^+$ and a width of about 30~MeV. Since the
states with $j^P_{\ell}=1/2^+$ are too broad to support hadronic
molecules~\cite{Filin:2010se,Guo:2011dd}, in what follows we  focus on the
scattering of the ground state mesons off their anti-particles as well as on
that of the $j^P_{\ell}=3/2^+$ mesons off the ground state ones with one of them
containing a heavy quark and the other a heavy anti-quark.

We start with the former system. To be concrete, we take the charm mesons. In
the particle basis, there are six $S$-wave meson pairs with given 
$J^{PC}$~\cite{Nieves:2012tt}:
\begin{eqnarray}
&&0^{++}:\quad\left\{D\bar{D}({^1S_0}),D^*\bar{D}^*({^1S_0})\right\},\nonumber\\
&&1^{+-}:\quad\left\{D\bar{D}^*({^3S_1},-),D^*\bar{D}^*({^3S_1})\right\},
\nonumber\\[-3mm]
\label{eq:basis}  \\[-3mm]
&&1^{++}:\quad\left\{D\bar{D}^*({^3S_1},+)\right\},\nonumber\\
&&2^{++}:\quad\left\{D^*\bar{D}^*({^5S_2})\right\},\nonumber
\end{eqnarray}
where the individual partial waves are labelled as $^{2S+1}L_J$, with $S$, $L$, 
and $J$ denoting the total spin, the angular momentum, and the total momentum 
of the two-meson system, respectively. We define the $C$-parity eigenstates as
\begin{equation}
D\bar{D}^*(\pm)=\frac{1}{\sqrt{2}}\left(D\bar{D}^*\pm D^*\bar{D}\right), \label{eq:DDstar}
\end{equation}
which comply with the convention\footnote{
Notice that a different convention for the $C$-parity operator was used in
Ref.~\cite{Nieves:2012tt}. As a consequence, the off-diagonal transitions
of $V_{\rm LO}^{(0{++})}$ in Ref.~\cite{Nieves:2012tt} have a different sign as 
compared to Eq.~(\ref{C0++}), see also
Sec.~VI~A in Ref.~\cite{Guo:2016bjq} for further
 details of our convention.}
 for the $C$-parity transformation $\hat{C}{\cal M}=\bar{\cal M}$.
Because of HQSS, the interaction at LO is independent of the heavy
quark spin, and thus can be described by the matrix elements  $\langle j_{1\, 
\ell}',
j_{2\, \ell}',{j_\ell}| \hat\Ham_I |  j_{1\, \ell}, j_{2\, \ell},{j_\ell} 
\rangle$ where
the light quark systems get coupled to a total light-quark angular momentum of 
the two-meson system, $j_\ell$.
 Thus, for the systems under
consideration, we have two independent terms for each isospin ($I=0$ or $1$):
  $\langle 1/2,1/2,0 | \hat\Ham_I | 1/2,1/2,0 \rangle$ and $
\langle 1/2,1/2,1 | \hat\Ham_I | 1/2,1/2,1 \rangle $.
This simple observation leads to the conclusion that in the strict heavy quark
limit the six pairs in Eq.~\eqref{eq:basis} are grouped into two
multiplets with $j_\ell=0$ and 1, respectively.
In the heavy quark limit, it is convenient to
use a basis of states characterized via ${j_\ell^{PC}}\otimes {s_{c\bar c}^{PC} 
}$, where $s_{c\bar
c}$ refers to the total spin of the $c$ and $\bar c$ pair.
For the case of $S$-wave interactions only, both ${j_\ell^{PC}}$ and ${s_{c\bar
c}^{PC}}$ can only be in $0^{-+}$ or $1^{--}$. Therefore, the spin multiplet 
with
${j_\ell=0}$ contains two states with quantum numbers:
\begin{equation}
  {0_\ell^{-+}}\otimes {0_{c\bar c}^{-+} } = 0^{++}, \qquad
    {0_\ell^{-+}}\otimes {1_{c\bar c}^{--} } = 1^{+-} ,
    \label{eq:sl0}
\end{equation}
and the spin multiplet for ${j_\ell=1}$ has the following four states:
\begin{equation}
  {1_\ell^{--}}\otimes {0_{c\bar c}^{-+} } = 1^{+-}, \qquad
    {1_\ell^{--}}\otimes {1_{c\bar c}^{--} } = 0^{++} \oplus
    {1^{++}} \oplus 2^{++} .
    \label{eq:sl1}
\end{equation}
It becomes clear that if the ${1^{++}}$ state $X(3872)$ is a $D\bar D^*$
molecule, then it is in the multiplet with $j_\ell=1$~\cite{Voloshin:2004mh}, 
and has three spin partners with $J^{PC}=0^{++}$, $2^{++}$ and
$1^{+-}$ in the strict heavy quark limit as pointed out in
Refs.~\cite{Hidalgo-Duque:2013pva,Baru:2016iwj}. Based on an
analogous reasoning it 
was suggested already earlier that  $Z_b(10610)$ and $Z_b(10650)$ might have four more 
isovector partners $W_{b0}^{(\prime)}$, $W^{}_{b1}$ and 
$W^{}_{b2}$~\cite{Bondar:2011ev,Voloshin:2011qa,Mehen:2011yh}. A detailed and
quantitative analysis of these $W_{bJ}$ states can be found in
Ref.~\cite{Baru:2017gwo}.

It is worthwhile to notice that the two $1^{+-}$ states are in different
multiplets with $j_\ell=0$ and $1$, respectively, and thus cannot be related to
each other via HQSS. However, the isovector $Z_b(10610)$ and $Z_b(10650)$ are
located with similar distances to the $B\bar B^*$ and $B^*\bar B^*$ thresholds,
respectively. Such an approximate degeneracy suggests that the isovector
interactions in the $j_\ell=0$ and $j_\ell=1$ sectors are approximately the
same, and the off-diagonal transition strength in the isovector channel between
the two meson pairs with $J^{PC}=1^{+-}$ in Eq.~\eqref{eq:basis} approximately
vanishes. A fit to the Belle data of the $Z_b$ line shapes with HQSS constraints
implemented also leads to nearly vanishing channel coupling~\cite{Guo:2016bjq}.
This points towards an additional ``light quark spin symmetry'' as proposed by
Voloshin very recently~\cite{Voloshin:2016cgm}. While a deeper understanding for
such a phenomenon is still missing, it seems to be realized in the charm sector as
well for the charged $Z_c(3900)$~\cite{Ablikim:2013mio,Liu:2013dau} and
$Z_c(4020)$~\cite{Ablikim:2013wzq} observed by the BESIII and Belle
collaborations.
Note that in Ref.~\cite{Valderrama:2012jv} 
it is argued that channel couplings are suppressed while in Ref.~\cite{Baru:2016iwj}
they were claimed to be important to keep a well defined spin symmetry limit.
We come back to this controversy briefly later in this section.

When the physical nondegenerate masses for the heavy mesons are used, one needs 
to switch to the basis in terms of physical states in Eq.~\eqref{eq:basis}.
 In this basis and for a given set of quantum numbers $\{JPC\}$, the LO EFT
 potentials $V^{(JPC)}_{\rm LO}$, which respect HQSS,
 read~\cite{AlFiky:2005jd,Nieves:2012tt,Valderrama:2012jv}
\begin{eqnarray}\label{C0++}
&&V_{\rm LO}^{(0{++})}=
\begin{pmatrix}
C_{0a} & -\sqrt{3}C_{0b} \\
-\sqrt{3}C_{0b} & C_{0a}-2C_{0b}
\end{pmatrix},
\label{Vct0++}\\
&&V_{\rm LO}^{(1{+-})}=
\begin{pmatrix}
C_{0a}-C_{0b} & 2C_{0b} \\
2C_{0b} & C_{0a}-C_{0b}
\end{pmatrix},
\label{Vct1+-}\\
&&V_{\rm LO}^{(1{++})}=C_{0a}+C_{0b} \label{eq:contact2-a},\label{Vct1++} \\
&&V_{\rm LO}^{(2{++})}=C_{0a}+C_{0b} \label{eq:contact2-b},\label{Vct2++}
\end{eqnarray}
where $C_{0a}$ and $C_{0b}$ are two independent low-energy constants.
Thus, since in the spin symmetry limit $D$ and $D^*$ are degenerate, implying
that $D\bar D^*$ and $D^*\bar D^*$ loops are equal, the above equality of the
potentials in the $1^{++}$ and $2^{++}$ channels immediately predicts equal
binding energies for the two states in this limit.

Once HQSS violation is introduced into the system by the use of the
physical masses, the
two-multiplet pattern gets changed, however, the close connection between the
$1^{++}$ and $2^{++}$ states persists.
An inclusion of the one-pion exchange necessitates an extension of the basis, 
since now also $D$-waves need to be included. In fact, HQSS is preserved
only if all allowed $D$-waves are kept in the system, even if the
masses of the open flavor states are still kept degenerate~\cite{Baru:2016iwj}.
The probably most striking effect of
the $D$-waves, once the $D^*$-$D$ mass difference is included, is
that now transitions of the $2^{++}$ $D^*\bar D^*$ $S$-wave state to the
$D\bar D$ and $D\bar D^*$ $D$-wave become possible. It allows for a width of
this state of up to several tens of MeV~\cite{Albaladejo:2015dsa,Baru:2016iwj}, 
which might be accompanied by a sizeable shift in mass.
In addition, spin symmetry relations might get modified via the
coupling of the molecular states with regular charmonia as discussed recently
in Ref.~\cite{Cincioglu:2016fkm}.
 
For near-threshold states it is natural to assume that the contact terms are
independent of the heavy quark mass --- phenomenologically they can be viewed as
parameterizing the exchange of light meson resonances. Then one can also predict
the heavy quark flavor partners of the $\X$. The heavy quark spin and flavor
partners of the $\X$ predicted in Ref.~\cite{Guo:2013sya} with $\Lambda=0.5$~GeV
are listed in Table~\ref{tab:predictions}.
 
\begin{table}[tb]
\caption{\label{tab:predictions} Predictions of the partners of the $\X$ for 
$\Lambda=0.5$~GeV in Ref.~\cite{Guo:2013sya}.
}
\begin{ruledtabular}
\begin{tabular}{l c c c }
        $J^{PC}$ & States & Thresholds (MeV) & Masses (MeV)
       \\\hline
       $1^{++}$ & $\frac1{\sqrt{2}}(D\bar D^*+D^*\bar D)$ &
       3875.87 & 3871.68 (input)  \\
                       $2^{++}$ & $D^*\bar D^*$ &
       4017.3  & $4012^{+4}_{-5}$  \\
       $1^{++}$ & $\frac1{\sqrt{2}}(B\bar B^*+B^*\bar B)$ &
       10604.4 & $10580^{+9}_{-8}$  \\
                       $2^{++}$ & $B^*\bar B^*$ &
       10650.2 & $10626^{+8}_{-9}$  \\
                       $2^{+}$ & $D^*B^*$ &
       7333.7 & $7322^{+6}_{-7}$  \\ 
   \end{tabular}
\end{ruledtabular}
\end{table}
The $Z_b(10610)$ can be related to the $Z_b(10650)$ when the off-diagonal 
interaction is neglected as discussed above.  Their hidden-charm partners
are found to be virtual states in this formalism~\cite{Guo:2013sya}, which may 
correspond to the $Z_c(3900)$ and $Z_c(4020)$. In fact, it is shown in 
Ref.~\cite{Albaladejo:2015lob} that the BES\-III data for the $Z_c(3900)$ in
both the $J/\psi\pi$~\cite{Ablikim:2013mio} and  $D\bar 
D^*$~\cite{Ablikim:2015swa} 
modes can be well fitted with either a resonance above the $D\bar D^*$
threshold or a virtual state below.

The number of the LO contact terms is larger for the interaction between a pair 
of $j_\ell=1/2$ and $j_\ell=3/2$ heavy and anti-heavy mesons.
For each isospin, 0 or 1, in the heavy quark limit, there are four independent 
interactions
denoted as $\langle
j_{1\, \ell},j_{2\, \ell},j_\ell|\hat\Ham_I |j_{1\, \ell}',j_{2\, \ell}',j_\ell
\rangle$, where now $j_\ell$ can take values 1 or 2
\begin{eqnarray}
  F_{Ij_\ell}^d &\equiv& \left\langle \frac12,\frac32,j_\ell \left|\hat\Ham_I 
\right|\frac12,\frac32,j_\ell \right\rangle , \nonumber\\
  F_{Ij_\ell}^c &\equiv& \left\langle \frac12,\frac32,j_\ell \left|\hat\Ham_I 
  \right|\frac32,\frac12,j_\ell \right\rangle .
  \label{eq:VHT}
\end{eqnarray}
The relevant combinations of these constants for a given heavy meson pair can 
be worked out by changing the basis by means of a unitary transformation
(see, e.g.,~\cite{Ohkoda:2012rj,Xiao:2013yca}):
\begin{eqnarray}
  &&| s_{1\,c},j_{1\, \ell},j_1; s_{2\,c},j_{2\, \ell},j_2;J\rangle \nonumber\\
  &=& \sum_{s_{c\bar c}, j_\ell }  
 \sqrt{ (2j_1+1)(2j_2+1) (2s_{c\bar c}+1)(2j_\ell+1) } \nonumber\\
 && \times 
 \begin{Bmatrix}
   s_{1\,c} & s_{2\,c} & s_{c\bar c} \\
   j_{1\, \ell} & j_{2\, \ell} & j_\ell \\
   j_1 & j_2 & J
 \end{Bmatrix}
 |s_{1\,c},s_{2\,c},s_{c\bar c};j_{1\, \ell},j_{2\, \ell},j_{\ell}; J\rangle ,
 ~~~
\end{eqnarray}
where $j_1$ and $j_2$ are the spins of the two heavy mesons, $J$ is the 
total angular momentum of the whole system, and
$s_{1\,c}$ and $s_{2\,c}$ are the spins of the heavy quark. 

Consider two mesons $A$ and $B$; each of them is not a $C$ parity eigenstate, but their linear combination can form $C$ parity eigenstates. With the phase convention specified below \eqref{eq:DDstar},
the $C=\pm$ eigenstates of a flavor-neutral system consisting of a pair of mesons are given by 
\begin{equation}
    |C=\pm \rangle=\frac{1}{\sqrt{2}}\left[A B \pm (-1)^{J-J_{A}-J_{B}} \bar B \bar A\right],
\end{equation}
where $J_A$ and $J_B$ are the spins of the mesons $A$ and $B$, and $J$ is the total spin of the two-body system.

Noticing that the
total spin of the heavy quark and anti-quark $s_{c\bar c}$ is conserved in the
heavy quark limit, and combining the meson pairs into eigenstates of 
$C$-parity, one can obtain the contact terms for the $S$-wave interaction 
between a pair of $j_\ell=\frac12$ and $\frac32$ heavy and anti-heavy mesons. 
The diagonal ones are listed in Table~\ref{tab:HTcontact}.
\begin{table}
  \caption{ The diagonal contact terms for the $S$-wave interaction between a 
pair of $j_\ell^P=1/2^-$ and $3/2^+$ heavy and anti-heavy mesons.
\label{tab:HTcontact}
}
\begin{ruledtabular}
  \centering\begin{tabular}{L C C}
    J^{PC} & \text{Meson pairs} & \text{Contact terms}\\\hline
    1^{{--}} & \frac{1}{\sqrt{2}} \left(D\bar{D}_1-D_1\bar{D}\right) & \frac{1}{8} \left(-F_{I1}^c-5
F_{I2}^c+3 F_{I1}^d+5 F_{I2}^d\right) \\
& \frac{1}{\sqrt{2}}\left(D^*\bar{D}_1+D_1\bar{D}^*\right) & \frac{1}{16} \left(7 F_{I1}^c-5
F_{I2}^c+11 F_{I1}^d+5 F_{I2}^d\right) \\
& \frac{1}{\sqrt{2}}\left(D^*\bar{D}_2-D_2\bar{D}^*\right) & \frac{1}{16} \left(-5
F_{I1}^c-F_{I2}^c+15 F_{I1}^d+F_{I2}^d\right) \\[2mm]

0^{--} & \frac{1}{\sqrt{2}}\left(D^*\bar{D}_1-D_1\bar{D}^*\right) & F_{I1}^c+F_{I1}^d \\[2mm]

2^{--} & \frac{1}{\sqrt{2}}\left(D\bar{D}_2-D_2\bar{D}\right) & \frac{1}{8} \left(3
F_{I1}^c-F_{I2}^c+3 F_{I1}^d+5 F_{I2}^d\right) \\
& \frac{1}{\sqrt{2}}\left(D^*\bar{D}_1-D_1\bar{D}^*\right) & \frac{1}{16} \left(F_{I1}^c-3
F_{I2}^c+F_{I1}^d+15 F_{I2}^d\right) \\
& \frac{1}{\sqrt{2}}\left(D^*\bar{D}_2{+}D_2\bar{D}^*\right) & \frac{1}{16} \left(9 F_{I1}^c+5 F_{I2}^c+9
F_{I1}^d+7 F_{I2}^d\right)
\\[2mm]

3^{--} & \frac{1}{\sqrt{2}}\left(D^*\bar{D}_2-D_2\bar{D}^*\right) & F_{I2}^d-F_{I2}^c \\[2mm]

0^{-+} & \frac{1}{\sqrt{2}}\left(D^*\bar{D}_1+D_1\bar{D}^*\right) & F_{I1}^d-F_{I1}^c \\ [2mm]

1^{-+} & \frac{1}{\sqrt{2}} \left(D\bar{D}_1+D_1\bar{D}\right) & \frac{1}{8} \left[5
\left(F_{I2}^c+F_{I2}^d\right)+F_{I1}^c+3 F_{I1}^d\right] \\
& \frac{1}{\sqrt{2}}\left(D^*\bar{D}_1-D_1\bar{D}^*\right) & \frac{1}{16} \left[5
\left(F_{I2}^c+F_{I2}^d\right)-7 F_{I1}^c+11 F_{I1}^d\right] \\
& \frac{1}{\sqrt{2}}\left(D^*\bar{D}_2+D_2\bar{D}^*\right) & \frac{1}{16} \left(5 F_{I1}^c+F_{I2}^c+15
F_{I1}^d+F_{I2}^d\right) \\ [2mm]

2^{-+} & \frac{1}{\sqrt{2}}\left(D\bar{D}_2+D_2\bar{D}\right) & \frac{1}{8} \left(-3 F_{I1}^c+F_{I2}^c+3
F_{I1}^d+5 F_{I2}^d\right) \\
& \frac{1}{\sqrt{2}}\left(D^*\bar{D}_1+D_1\bar{D}^*\right) & \frac{1}{16} \left[3 \left(F_{I2}^c+5
F_{I2}^d\right)-F_{I1}^c+F_{I1}^d\right] \\
& \frac{1}{\sqrt{2}}\left(D^*\bar{D}_2{-}D_2\bar{D}^*\right) & \frac{1}{16} \left(-9 F_{I1}^c-5
F_{I2}^c+9 F_{I1}^d+7 F_{I2}^d\right) \\ [2mm]
3^{-+} & \frac{1}{\sqrt{2}}\left(D^*\bar{D}_2+D_2\bar{D}^*\right) & F_{I2}^c+F_{I2}^d \\
\end{tabular}
\end{ruledtabular}
\end{table}
One sees that the linear combinations are different for all channels, and 
it is not as easy as in case of  the $\X$ to predict spin partners for the $\Y$
based on the assumption that it is predominantly a $D_1\bar D$ state. The
possibility of $S$-wave hadronic molecules with exotic quantum numbers $1^{-+}$
was discussed in~\cite{Wang:2014wga}. 
Here we also give the off-diagonal contact terms.
There are three channels in each of $1^{--}$, $2^{--}$, $1^{-+}$ and $2^{-+}$ sectors, 
we label them in each sector listed in Table~\ref{tab:HTcontact} from top to bottom as 1, 2, and 3. Then the off-diagonal contact terms for the $1^{--}$ sector are 
\begin{equation}
    \begin{aligned}
    V_{12} &= -\frac{1}{8 \sqrt{2}}\left[5\left(F_{I 2}^{c}+F_{I 1}^{d}-F_{I 2}^{d}\right)+F_{I 1}^{c}\right], \\
    V_{13}&= \frac{1}{8} \sqrt{\frac{5}{2}}\left(-3 F_{I 1}^{c}+F_{I 2}^{c}+F_{I 1}^{d}-F_{I 2}^{d}\right), \\
    V_{23}&= \frac{\sqrt{5}}{16} \left(5 F_{I 1}^{c}+F_{I 2}^{c}+F_{I 1}^{d}-F_{I 2}^{d}\right) . 
    \end{aligned}
\end{equation}
The off-diagonal contact terms for the $2^{--}$ sector are 
\begin{equation}
    \begin{aligned}
        V_{12} &= -\frac{1}{8} \sqrt{\frac{3}{2}}\left(F_{I 1}^{c}+5 F_{I 2}^{c}+F_{I 1}^{d}-F_{I 2}^{d}\right), \\
        V_{13} &= \frac{1}{8} \sqrt{\frac{3}{2}}\left(-3 F_{I 1}^{c}+ F_{I 2}^{c}-3 F_{I 1}^{d}+3 F_{I 2}^{d}\right),\\
        V_{23}& =\frac{3}{16}\left(F_{I 1}^{c}-3 F_{I 2}^{c}+F_{I 1}^{d}-F_{I 2}^{d}\right).
    \end{aligned}
\end{equation}
The off-diagonal contact terms for the $1^{-+}$ sector are 
\begin{equation}
\begin{aligned}
    V_{12}&= \frac{1}{8 \sqrt{2}}\left[5\left(F_{I 2}^{c}-F_{I 1}^{d}+F_{I 2}^{d}\right)+F_{I 1}^{c} \right], \\
    V_{13}&= \frac{1}{8} \sqrt{\frac{5}{2}}\left(3 F_{I 1}^{c}-F_{I 2}^{c}+F_{I 1}^{d}-F_{I 2}^{d}\right), \\
    V_{23}&= -\frac{1}{16} \sqrt{5}\left(5 F_{I 1}^{c}+F_{I 2}^{c}-F_{I 1}^{d}+F_{I 2}^{d}\right).
\end{aligned}
\end{equation}
The off-diagonal contact terms for the $2^{-+}$ sector are 
\begin{equation}
    \begin{aligned}
        V_{12} &= \frac{1}{8} \sqrt{\frac{3}{2}}\left(F_{I 1}^{c}+5 F_{I 2}^{c}-F_{I 1}^{d}+F_{I 2}^{d}\right), \\
        V_{13} &= \frac{1}{8} \sqrt{\frac{3}{2}}\left(3 F_{I 1}^{c}- F_{I 2}^{c}-3 F_{I 1}^{d}+3 F_{I 2}^{d}\right),\\
        V_{23} &= -\frac{3}{16} \left(F_{I 1}^{c}-3 F_{I 2}^{c}-F_{I 1}^{d}+F_{I 2}^{d}\right).
    \end{aligned}
\end{equation}

However, one non-trivial prediction for the spectrum of molecular states in the
heavy quarkonium spectrum becomes apparent immediately from the discussion
above: Since the most bound states appear in $S$-waves  the lightest negative
parity vector state can be formed only from $j_\ell^P=1/2^-$ and $3/2^+$
heavy and anti-heavy mesons.
Therefore the mass difference of $X(3872)$ as bound state of two ground state
$j_\ell^P=\frac12^-$ mesons ($D$ and $D^*$) and the lightest exotic vector state
$Y(4260)$ (388 MeV) should be of the order of the mass difference of the
lightest $3/2^+$ state and the $D^*$ (410 MeV). Clearly this prediction is
nicely realized in nature. Note that from this reasoning it also follows that
$if$ the $Y(4008)$ indeed were to exist it could not be a hadronic molecule. In
this context it is interesting to note that the most resent data from BESIII on
$e^+e^-\to J/\psi\pi\pi$~\cite{Ablikim:2016qzw} does not seem to show evidence
for the $Y(4008)$, {\sl c.f.} Fig.~\ref{fig:Y4260-BESIII}.

\subsection{Impact of hadron loops on regular quarkonia}
\label{sec:4-NREFT_ccbar}

In the previous sections we argued that meson loops play a prominent role in
both the formation and the decays of hadronic molecules. One may wonder if they
also have an impact on the properties of regular charmonia.
In this section we demonstrate that certain processes for regular hadrons,
largely well described by the quark model, can also be influenced by significant
meson loop effects, since reaction rates can receive an enhancement due to the
nearly on-shell intermediate heavy mesons.
The origin of this mechanism is that for most heavy quarkonium transitions
$M_{Q\bar Q}-2 M_{Q\bar q}\ll M_{Q\bar q}$, where $M_{Q\bar Q}$ and $M_{Q\bar
q}$ are the masses of the heavy quarkonium and an open-flavor heavy meson,
respectively. As a result, the intermediate heavy mesons are nonrelativistic
with a small velocity
\begin{equation}
    v\sim \sqrt{|{M_{Q\bar Q}}-2m_{Q\bar q}|/m_{Q\bar q}}\ll 1\,,
\end{equation}
and the meson loops in the transitions can be investigated 
using \nreft.
We will highlight this effect on two examples in what follows.\footnote{The
effects of meson loops in heavy quarkonium spectrum are investigated in,
e.g., Refs.~\cite{Eichten:1978tg,Eichten:1979ms,Ono:1983rd,Kalashnikova:2005ui,
Eichten:2005ga,Pennington:2007xr,Barnes:2007xu, Li:2009ad,Danilkin:2009hr,
Ortega:2010qq,Danilkin:2010cc,Liu:2011yp,Bali:2011rd,Zhou:2013ada,
Ferretti:2013faa,Ferretti:2013vua,Ferretti:2014xqa, Hammer:2016prh,Du:2016qcr,
Lu:2016mbb, Lu:2017hma,Zhou:2017dwj}, and in heavy quarkonium transitions in
Refs.~\cite{Ono:1985jt, Lipkin:1988tg,Moxhay:1988ri,Zhou:1990ik,
Li:2007xr,Meng:2007tk,Meng:2008bq,
Liu:2009dr,Zhang:2009kr,Zhang:2010zv,Wang:2011yh, Li:2011ssa,Wang:2012mf,
Guo:2012tj,Li:2013zcr,Cao:2016xqo}.}

We start with the hindered M1 transitions between two $P$-wave heavy quarkonia
with different radial excitations, such as the $h_c(2P)\to \gamma
\chi_{cJ}(1P)$. It was proposed in Ref.~\cite{Guo:2011dv,Guo:2016yxl} that such
transitions are very sensitive to meson-loop effects, and the pertinent partial
widths provide a way to extract the coupling constants between the $P$-wave
heavy quarkonia and heavy open flavor mesons.
\begin{figure}[t]
    \centering \includegraphics[width=\linewidth]{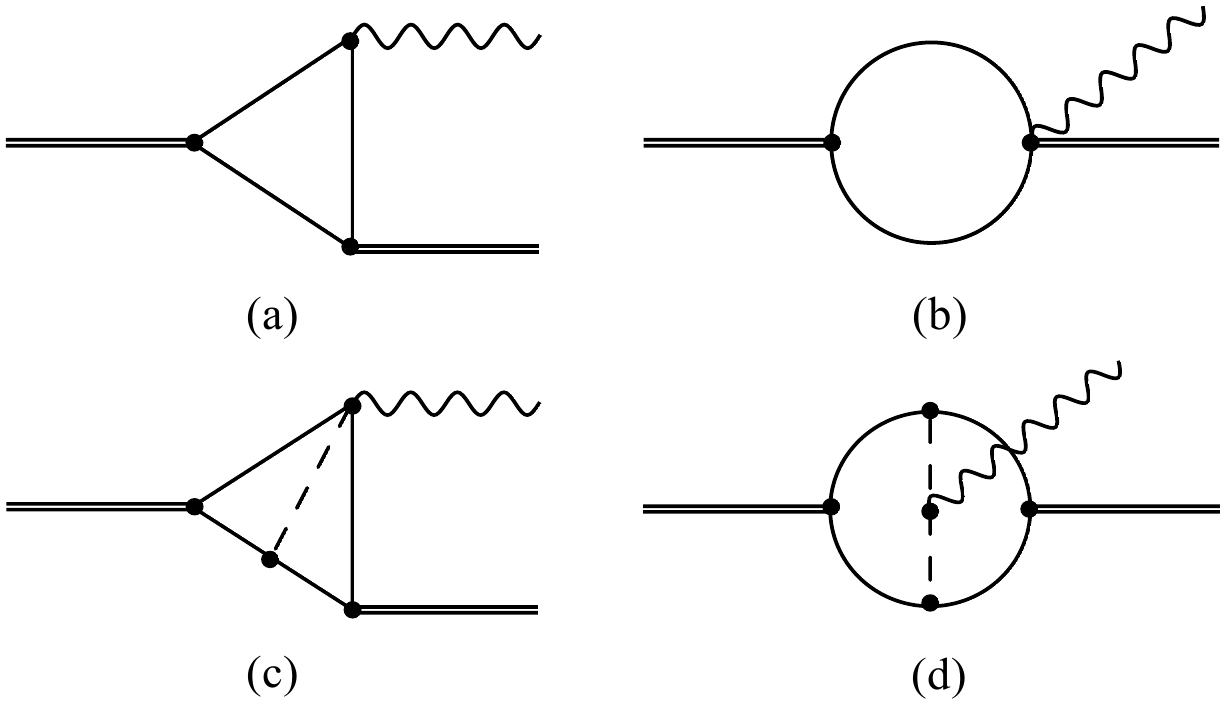}
    \caption{Feynman diagrams for the coupled-channel effects for the hindered
    M1 transitions between heavy quarkonia. The one-loop contributions are given
by (a) and (b). (c) and (d) are two typical two-loop diagrams.
The double, solid, wavy and dashed lines represent heavy quarkonia, heavy
mesons, photons, and pion, respectively. Adapted from Ref.~\cite{Guo:2011dv}.
\label{fig:M1loops}}
\end{figure}
In quark models the amplitude for such a transition is proportional to the
overlap of the wave functions of the initial and final heavy quarkonia, which is
tiny and quite model-dependent due to the different radial excitations --- this
is why they are called ``hindered''. This suppression is avoided in the
coupled-channel mechanism of heavy-meson loops.
In this mechanism, the initial and final $P$-wave heavy quarkonia couple to the
ground state pseudoscalar and vector heavy mesons in an $S$-wave. A few diagrams
contributing to this mechanism are shown in Fig.~\ref{fig:M1loops}. In (a), the
photon is emitted via its magnetic coupling to intermediate heavy mesons. In
(b), since the $S$-wave vertices do not have any derivative at LO, the photon
couples in a gauge invariant way to one of the vertices in the two-point loop
diagram. (c) and (d) are two typical two-loop diagrams. From the power counting
rules discussed in Sec.~\ref{sec:nreft1}, Fig.~\ref{fig:M1loops}~(a) provides
the leading contribution, while (b) is of higher order in the velocity counting
because there is one less nonrelativistic propagator. Their amplitudes scale as
\begin{equation}
  \Amp_{(a)}  \sim \frac{E_\gamma}{m_Q v}, \qquad \Amp_{(b)}  \sim 
\frac{E_\gamma v}{m_Q}\,,
\label{eq:M1Aab}
\end{equation}
respectively,
where $E_\gamma$ is the photon energy, and the dependence on the 
coupling constants is dropped. The $1/m_Q$ suppression comes from the fact that 
the polarization of a heavy (anti-)quark needs to be flipped in the M1 
transitions. For the two-loop diagram in (c), the amplitude scales as
\begin{equation}
   \Amp_{\rm (c)} \sim \frac{(v^5)^2}{(v^2)^5}
\frac{g^2}{(4\pi)^2F_\pi^2}  \frac{E_\gamma}{m_Q} M_H^2 = \frac{E_\gamma}{m_Q}
\left(\frac{g M_H}{\Lambda_\chi}\right)^2,
\label{eq:M1Ac}
\end{equation}
where the factor $1/(4\pi)^2$ appears because there is one more loop and the
hadronic scale $\Lambda_\chi=4\pi F_\pi\sim 1$ GeV was introduced as the hard
scale for the chiral expansion. The factor of $M_H^2$ was introduced to match
dimensions of the above equations to those of Eqs.~\eqref{eq:M1Aab}. Diagram (d)
has the same scaling as (c). Since the axial coupling constant $g\simeq0.6$ for
the charm case as determined from the width of $D^*\to D\pi$, and about 0.5 for
bottom~\cite{Flynn:2015xna}, one has $g M_D/\Lambda_\chi\lesssim1$ and $g
M_B/\Lambda_\chi\simeq2$. The value for $v$ defined as $(v_1+v_2)/2$ is about
0.4 for the transitions from the 2P to 1P charmonium states~\cite{Guo:2011dv},
and ranges from 0.3 to 0.2 for the transitions between $1P,2P$ and $3P$
bottomonia~\cite{Guo:2016yxl}. Hence, the two-loop diagrams are suppressed in
the charm sector, while they are of the same order as (a) for the bottom sector.
Therefore, one can make predictions for the charmonium transitions by
calculating the loops corresponding to (a). The results depend on a product of
two unknown coupling constants of the $1P$ and $2P$ charmonia to the charm
mesons. Taking model values for them, the decay width of the
$\chi_{c2}(2P)\to\gamma h_c(1P)$ is of $\mathcal{O}(100~\text{keV})$, two orders
of magnitude larger than the quark model prediction,
1.3~keV~\cite{Barnes:2005pb}.
Although quantitative predictions cannot be made for the bottomonium
transitions, it is expected that once such transitions would be observed they
must be due to coupled-channel effects as the partial widths were predicted to
be in the range from sub-eV to eV level in a quark model calculation that does
not include meson-loop effects~\cite{Godfrey:2015dia}. It is suggested in
\cite{Guo:2011dv,Guo:2016yxl} that the coupled-channel effects can be checked by
comparing results from both fully dynamical and quenched lattice QCD which has
and has no coupled-channel effects, respectively. Recent developments in lattice
QCD calculations of radiative decays~\cite{Dudek:2006ej,Dudek:2009kk,
Shultz:2015pfa,Briceno:2016kkp,Agadjanov:2014kha,Feng:2014gba,Leskovec:2016lrm,
Meyer:2011um,Owen:2015fra} should be helpful in illuminating this issue.

There are other heavy quarkonium transitions driven mainly by the
coupled-channel effects. A detailed study on the transitions between two
charmonia ($S$- and $P$-wave) with the emission of a pion or eta can be found in
Ref.~\cite{Guo:2010ak}. It is found that whether the coupled-channel effects
play a sizable role depends on the process. This is a result of the power
counting analysis; see the itemized discussion in Sec.~\ref{sec:nreft1}.
In particular, the  single-pion/eta transitions between two $S$-wave and
$P$-wave charmonia receive important contribution from charm-meson loops.
Therefore, the long-standing suggestion that the  $\psi'\to J/\psi\eta/\pi^0$
transitions can be used to extract the light quark mass
ratio~\cite{Ioffe:1979rv} needs to be reexamined. In fact, if we assume that
the triangle charm meson-loop diagrams saturate the transitions, the resulting
prediction of $\mathcal{B}(\psi'\to J/\psi\pi^0)/\mathcal{B}(\psi'\to
J/\psi\eta)$ is consistent with the experimental data. These transitions were
revisited considering both the loop and tree diagrams in
Ref.~\cite{Mehen:2011tp}.
Again based on the same power counting rules it was argued that the transitions
$\Upsilon(4S)\to h_b\pi^0/\eta$ have only a small pollution from the
bottom-meson loops, and are dominated by short-distance contribution
proportional to the light quark mass difference~\cite{Guo:2010ca}. They could be
used for the extraction of light quark mass ratio. Furthermore, the prediction,
made before the discovery of the $h_b(1P)$, on the branching fraction of the
order of $10^{-3}$ for the decay $\Upsilon(4S)\to h_b\eta$ was verified by the
Belle measurement, $(2.18\pm0.11\pm0.18)\times10^{-3}$~\cite{Tamponi:2015xzb}.

Parameter-free predictions can be made for ratios of partial widths of decays
dominated by the coupled-channel effects of heavy mesons in the same spin
multiplet, since all the coupling constants will get canceled in the ratios.
Such predictions on the hindered M1 transitions between $P$-wave heavy quarkonia
can be found in Refs.~\cite{Guo:2011dv,Guo:2016yxl}.

In Ref.~\cite{Guo:2012tg}, it is pointed out that coupled-channel effects can
even introduce sizable and nonanalytic pion mass dependence in heavy quarkonium
systems which couple to open-flavor heavy meson pairs in an $S$-wave.

To summarize this subsection, we stress that whether meson-loop effects are
important for the properties of quarkonia or not does not only depend on the
proximity to the relevant threshold, it is  also depends on the particular
transition studied.

\section{Hadronic molecules in lattice QCD}
\label{sec:lattice}

Lattice QCD is, in principle, the tool to calculate the spectrum of QCD from
first principles. There has been a remarkable progress in the last years in this
field, see
e.g.~\cite{Durr:2008zz,Baron:2010bv,Edwards:2011jj,Liu:2012ze,Liu:2016kbb}.
Still, the extraction of the properties of resonances and, in particular, of
hadronic molecules, from finite volume {calculations}  poses severe
challenges.
When QCD is put on an Euclidean space-time lattice {with a finite
space-time volume, asymptotic states cannot be defined and right-hand cuts are replaced by
poles, thus preventing a direct calculation of scattering and resonance
properties.}
This obstacle was overcome by L\"uscher a long time ago. He derived a relation
between the energy shift in the finite volume and the scattering phase shift in
the continuum \cite{Luscher:1990ux,Luscher:1986pf}, see also
Refs.~\cite{Wiese:1988qy,DeGrand:1990ip}.
This approach has become  known and used as L\"uscher's method.
More precisely, in order to determine the mass and width from the measured
spectrum, one first extracts the scattering phase shift by using the L\"uscher
equation. In the next step, using some parameterization for the $K$-matrix
(e.g., the effective range expansion), a continuation into the complex energy
plane is performed. As noted in Sec.~\ref{sec:Sproperties}, resonances
correspond to poles of the scattering $T$-matrix on the second Riemann sheet,
and the real and imaginary parts of the pole position define the mass and the
width of a resonance. A nice example is given by the $\rho$-meson, that has been
considered using L\"uscher's method, e.g., in
Refs.~\cite{Feng:2010es,Lang:2011mn,Aoki:2011yj,Dudek:2012xn}.
In these papers it has already been shown that even for such realistic
calculations of a well isolated resonance,
the inclusion of {hadron-hadron type interpolating operators} is mandatory,
it is simply not sufficient to represent the decaying resonance by properly
chosen quark bilinears {(for mesons)}.
For the discussion of hadronic molecules (or most other hadron resonances), this
method needs to be extended in various directions, such as considering higher
partial waves and spins
\cite{Bernard:2008ax,Luu:2011ep,Konig:2011nz,Konig:2011ti,Briceno:2012yi,
Gockeler:2012yj}, moving frames
\cite{Rummukainen:1995vs,Bour:2011ef,Davoudi:2011md,Fu:2011xz,Leskovec:2012gb,
Gockeler:2012yj}, multi-channel scattering
\cite{Liu:2005kr,Lage:2009zv,Bernard:2010fp,Doring:2011ip,Li:2012bi,Guo:2012hv},
including the use of unitarized chiral perturbation theory (and related methods)
\cite{Doring:2011vk,MartinezTorres:2011pr,Doring:2011nd,Albaladejo:2012jr,
Wu:2014vma,Hu:2016shf}, and three-particle final states
\cite{Kreuzer:2010ti,Polejaeva:2012ut,Briceno:2012rv,Hansen:2014eka,
Meissner:2014dea,Hansen:2015zga,Hansen:2015zta,Hansen:2016ync,Hansen:2016fzj,Guo:2017ism}.

Here, we will not attempt to review the lattice QCD approach to the hadron
spectrum in any detail but just focus on the bits and pieces that are relevant
for the investigation of possible hadronic molecules.
In Sec.~\ref{sec:reso} we summarize the L\"uscher method and its extension to
the multi-channel space, followed by a discussion of the compositeness criterion
in a finite volume, see Sec.~\ref{sec:compo}.
In Sec.~\ref{sec:qmdep}, we discuss how the quark mass dependence of certain
observables can be used to differentiate hadronic molecules from more compact
multi-quark states and in Sec.~\ref{sec:latres}, we briefly summarize pertinent
lattice QCD calculations for the possible molecular states containing charm
quarks. A short final subsection is devoted to certain states made of light
quarks only.

\subsection{Resonances in a finite volume}
\label{sec:reso}

\begin{figure}[t!]
\begin{center}
 \includegraphics[width=0.4\textwidth]{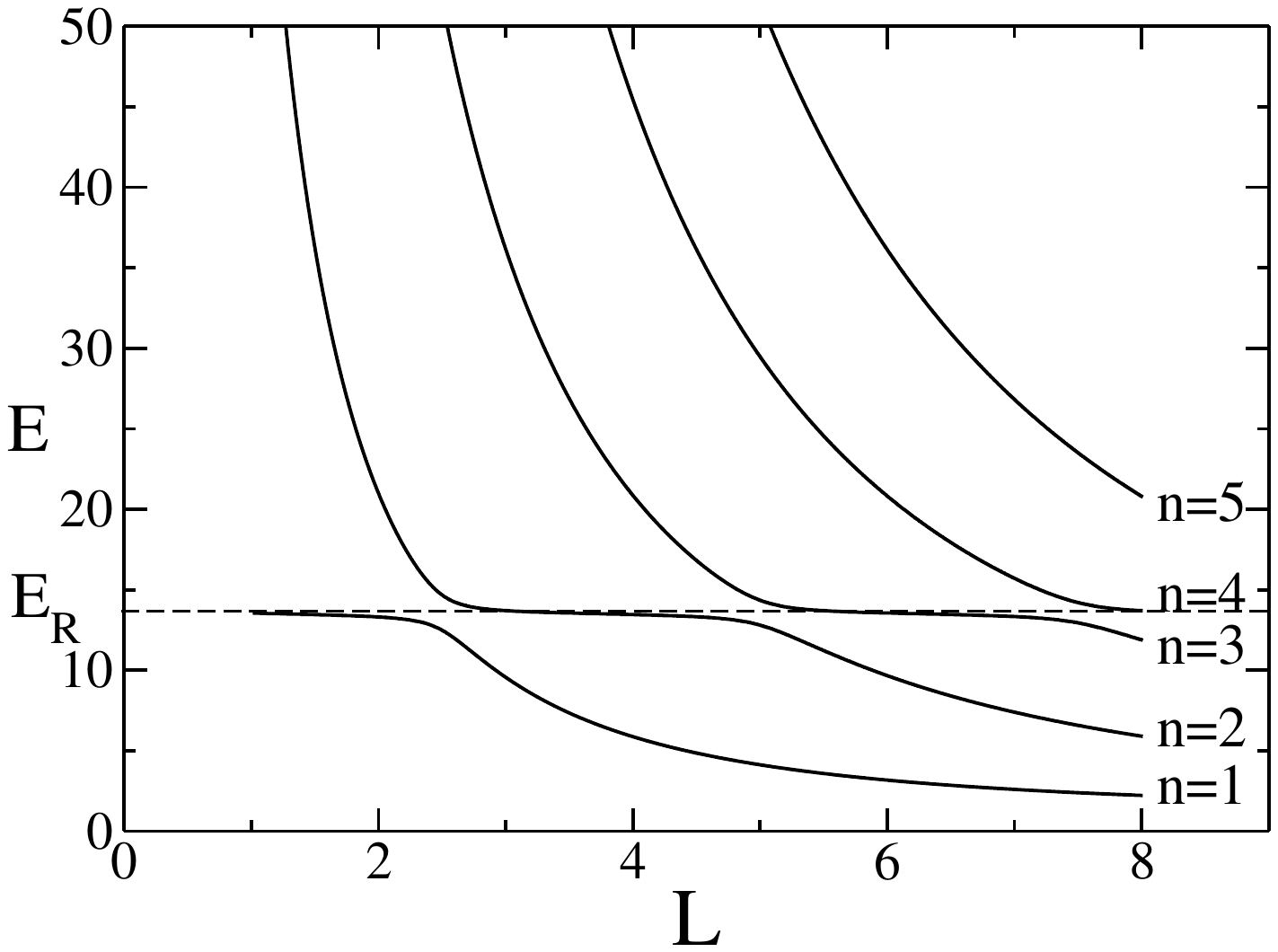}
\caption{Energy levels of an interacting two-particle system. In
case of a resonance in this system, the energy levels exhibit the
avoided level crossing (plateau) that allows to read off the 
resonance energy $E_R$ directly.} 
\label{fig:avoided}
\end{center}
\end{figure}

The essence of the L\"uscher approach can be understood in a simple
nonrelativistic model for the scattering of identical, spinless particles of
mass $m$ in 1+1 dimensions.
In the CM frame, the relative momentum is quantized according to $p =(2\pi/L)n$,
with $L$ the spatial lattice extension and $n$ an integer. In case of no
interactions between these particles, the energy of the two-particle system is
simply given by $E=2m+p^2/m$, which means that the free energy level-$n$ scales
as $n^2/L^2$ with the volume and thus levels with different $n$ do not
intersect. In the presence of interactions, this behaviour is modified. Let us
assume that this interaction leads to a narrow resonance at $\sqrt{s_R} = E_R -
i\Gamma_R/2$, that is $\Gamma_R \ll E_R$. In the infinite volume limit, this
interaction leads to a phase shift $\delta (p)$  in the asymptotic wave
function. Furthermore, in the presence of a resonance, the phase shift will
change by $\pi$ (known as Levinson's theorem~\cite{Levinson:1949zz}).
In a finite volume, this behavior translates into the boundary condition
\begin{equation}
p L + 2\delta(p) = 2\pi \, m~,~~~m\in {\mathbb Z}~ .
\end{equation}
This condition provides the link between the volume dependence of the
energy levels in the interacting system and the continuum phase shift.
If one follows an energy level inwards from the asymptotic region to
smaller lattice sizes, in the vicinity of a resonance, this boundary 
condition causes a visible distortion, the so-called {\em avoided level
crossing}, {\sl c.f.} Fig.~\ref{fig:avoided}. The plateau, where the energy of
the two-particle system is almost volume-independent, corresponds to the real
part of the pole $E_R$. The imaginary part of the pole is given by the slope
according to $\left.d \delta(p)/dE\right|_{E_R} = 2/\Gamma_R$. Clearly, this method can only
work when certain conditions are fulfilled. First, the method as described 
here is restricted to the elastic two-particle case. Second, one has to make
sure that the interaction range of the particles is much smaller than the
size of the box to make the notion of asymptotic states possible. Third,
to suppress polarization effects that arise from the interactions of the 
lightest particles in the theory with each other around the torus, one
has to choose $L$ such that $1/m \ll L$.

We now consider the extension of the L\"uscher method to the multi-channel
case, as most hadronic molecules are located close to a two-particle threshold
or between two close-by thresholds. To achieve this extension, an 
appropriate tool is a particular version of an NREFT,
because up to the energies where multi-particle inelastic states become
significant, such a framework is completely equivalent to the relativistic
field theory, provided the couplings in the nonrelativistic framework
are determined from matching to the relativistic $S$-matrix elements, for
details and further references, see \cite{Bernard:2008ax,Colangelo:2006va,Gasser:2011ju}. For the one-channel case, it was already shown in 
Ref.~\cite{Beane:2003yx}
that using such an NREFT, one obtains at a very simple and transparent 
proof of L\"uscher's formula.

To keep the presentation simple, we first consider a two-channel LSE in NREFT in
the infinite volume.
Let us consider antikaon-nucleon scattering in the region of the $\Lambda(1405)$
resonance, $\bar KN\to\bar KN, \Sigma\pi$.
The channel number 1 refers to $\bar KN$ and 2 to $\Sigma\pi$ with total isospin
$I = 0$.
The resonance $\Lambda (1405)$ is located between two thresholds, on the second
Riemann sheet, close to the real axis.
The two  thresholds are given by  $s_t = (m_N + M_K)^2$ and
$s'_t=(m_\Sigma+M_\pi)^2$. We work in the isospin limit and neglect the fact
that there are really two poles --- see Refs.~\cite{Oller:2000fj,Jido:2003cb}
and Sec.~\ref{sec:1405th}.\footnote{Note that in this  two-channel formulation
one only has one pole corresponding to one $\Lambda (1405)$. To deal with the
two-pole scenario requires the inclusion of more channels and explicit isospin
breaking.}
For energies above the $\bar KN$  threshold, $s > (m_N + M_K)^2$, the
coupled-channel LSE for the $T$-matrix elements $T_{ij}(s)$ in
dimensionally regularized NREFT reads (we only consider $S$-waves here)
\begin{eqnarray}
\label{eq:LSinfini} 
T_{11}  &=& H_{11} + H_{11} \, iq_1 T_{11} + H_{12} \,
iq_2 T_{21}~,\nonumber\\
T_{21}  &=& H_{21} + H_{21} \, iq_1 T_{11} + H_{22} \, iq_2 T_{21} \,, 
\end{eqnarray}
with
$q_1 = \lambda^{1/2} (s,m_N^2,M_K^2)/(2\sqrt{s})$, $q_2 = \lambda^{1/2}
(s,m_\Sigma^2,M_\pi^2)/(2\sqrt{s})$ and $\lambda(x,y,z)=x^2+y^2+z^2-2xy-2yz-2zx$
is the K\"all\'en function.
Furthermore, the $H_{ij}(s)$ denote the driving potential in the corresponding
channel, i.e. the matrix element of the interaction Hamiltonian between the free
two-particle states.
Continuation of the CM momentum $q_1$ below threshold $(m_\Sigma + M_\pi)^2 <s<
(m_N + M_K)^2$ is obtained via \beq iq_1 \to -\kappa_1 =
-\frac{(-\lambda(s,M_K^2,m_N^2))^{1/2}}{2\sqrt{s}} \ .
\eeq The resonance corresponds to a pole on the second Riemann sheet in the
complex $s$-plane. Its position is given by the solution of the secular 
equation \beq\label{eq:secular} \Delta (s) = 1 + \kappa_1^R \, H_{11} -
\kappa_2^R \, H_{22} - \kappa_1^R \kappa_2^R \, \left( H_{11} H_{22} -
H_{12}^2\right) \eeq with $\kappa_1^R = -(-\lambda
(s_R,m_N^2,M_K^2))^{1/2}/(2\sqrt{s_R})$ and $\kappa_2^R = (-\lambda
(s_R,m_\Sigma^2,M_\pi^2)^{1/2})/(2\sqrt{s_R})$.
The energy and width of the resonance are then  given by $\sqrt{s_R} = E_R - i
\Gamma_R/2$.

Consider next the same problem in a finite volume. Only discrete
values of the three-momentum are allowed, given by $\bm{k} = 
(2\pi/L)\bm{n}$, with $\bm{n}$ a triplet of integer numbers. 
Thus, we replace the three-momentum integration in the loops by a discrete
sum (see  Ref.~\cite{Bernard:2008ax} for more details).
The rotational symmetry is broken to a cubic symmetry, so mixing of 
different partial waves occurs. Here,  however, we only consider $S$-waves, 
neglecting the small mixing to higher partial waves. 
If necessary, the mixing can be easily included at a later 
stage, see e.g.~\cite{Bernard:2008ax,Doring:2012eu}.
The finite-volume version of the LSE
Eq.~(\ref{eq:LSinfini}) then takes the form
\beqa
T_{11}  &=& H_{11} - \frac{2  Z_{00}(1;k_1^2)}{\sqrt{\pi}L}\,  H_{11}
T_{11} - \frac{2 Z_{00}(1;k_2^2)}{\sqrt{\pi}L}\,  H_{12} T_{21}~,\nonumber\\
T_{21}  &=& H_{21} - \frac{2 Z_{00}(1;k_1^2)}{\sqrt{\pi}L}\,  H_{21}
T_{11} - \frac{2  Z_{00}(1;k_2^2)}{\sqrt{\pi}L}\, H_{22} T_{21}~,\nonumber\\
\eeqa
with 
\beqa
k_{1/2}^2 &=& \left(\frac{L}{2\pi}\right)^2 \, 
\frac{\lambda(s,M_{K/\pi}^2,m_{N/\Sigma}^2)}{4s}~,\nonumber\\
Z_{00} (1;k^2) &=& \frac{1}{\sqrt{4\pi}} \,\lim_{r\to 1}\sum_{\bm{n} \in 
{\mathbb R}^3}
\frac{1}{({\bm{n}\,}^2 - k^2)^r}~.
\eeqa
Here, we have neglected the terms that vanish exponentially at large 
$L$.
The secular equation that determines the spectrum can be brought into the
form
\beqa\label{eq:pseudophase}
&&\qquad  1 - \frac{2}{\sqrt{\pi} L} \,  Z_{00}(1;k_2^2)\, F(s,L) = 0~, 
\nonumber\\
&& F(s,L) = \left[ H_{22} -   \frac{2 Z_{00}(1;k_1^2)}{\sqrt{\pi}L}\,   
(H_{11}H_{22} - H_{12}^2)\right] \nonumber\\ 
&& \qquad \qquad \times \left[1 - \frac{2 Z_{00}(1;k_1^2)}{\sqrt{\pi}L}\,  
H_{11}\right]^{-1}.
\eeqa
This can be rewritten as 
\beqa\label{eq:1channel}
\delta(s,L) &=& -\phi(k_2) + n\,\pi~, \quad n = 0,1,2, \ldots  \nonumber\\
\phi(k_2) &=& -\arctan \frac{\pi^{3/2} \, k_2}{ Z_{00} (1;k_2^2)}~,
\eeqa
with
\beq
\tan\delta(s,L) = q_2(s) \,  F(s, L)~.  
\eeq
$\delta (s,L)$ is called the {\em pseudophase}.
The dependence of the pseudophase on $s$ and $L$
is very different from that of the usual scattering 
phase.
Namely, the elastic phase extracted from the lattice data by using
L\"uscher's formula is independent of the volume modulo terms that 
exponentially vanish at a large $L$. Further, the energies where the
phase passes through $\pi/2$ lie close to the real resonance locations.
In contrast with this, the pseudophase contains the function
$Z_{00}(1;k_1^2)$, which does not vanish exponentially at
a large $L$ and a positive $k_1^2$.
Moreover, 
the tangent of the pseudophase contains a tower of poles
at the energies given by the roots of the equation
$1-({2}/{\pi L})\,Z_{00}(1;k_1^2)\,H_{11}=0$. On the other hand,
in the infinite volume
this equation reduces to $1+\kappa_1^R\,H_{11}=0$, {\sl c.f.} with
Eq.~(\ref{eq:secular}), which has only one root below threshold very close to
the position of the $\Lambda(1405)$. Other roots in a finite volume 
stem from oscillations of $Z_{00}(1,k_1^2)$ 
between $-\infty$ and $+\infty$ when the variable $k_1^2$ varies 
along the positive semi-axis. {Their locations depend on $H_{11}$ and thus
contain information of the infinite-volume interaction}.
This is an effect of discrete energy levels in the ``shielded'' channel, {which is the channel with the lower
threshold in the coupled-channel system}.
The pseudophase depends on the three real functions $H_{11},H_{12},H_{22}$. Based on 
synthetic data it was shown in 
Ref.~\cite{Lage:2009zv} that a measurement of the lowest two eigenvalues
at energies between 1.4 and 1.5~GeV allows one to reconstruct the
pseudophase and to extract in principle the pole position. It was further
pointed out in that work that two-particle
thresholds also lead to an avoided level crossing, so the extraction
of the resonance properties from the corresponding plateaus in the
energy dependence of certain levels is no longer possible.
In the case of real data, taking into account the
uncertainties of each measurements, one has to measure more levels
on a finer energy grid. To obtain a sufficient amount of data in a
given volume, twisting and asymmetric boxes can also be helpful.
First such {calculations} have become available
recently and will be discussed below.

An alternative formulation, that allows the use of  fully relativistic
two-particle propagators and can easily be matched to the representation
of a given scattering amplitude based on unitarized chiral perturbation
theory (UCHPT) was worked out in~\cite{Doring:2011vk}. 
The method is based on the observation that
in coupled-channel UCHPT, certain
resonances are dynamically generated, e.g. the light scalar mesons in the
coupled $\pi\pi/\bar KK$ system. The basic idea is to
consider this approach  in a finite volume to
produce the volume-dependent discrete energy spectrum.
Reversing the argument, one is then able to fit the parameters of the
chiral potential to the measured energy spectrum on the
lattice and, in the next step, determine the resonance locations
by solving the scattering equations in the infinite
volume. By construction, this method fulfills the constraints from
chiral symmetry such as the appearance of Adler zeros at certain
unphysical points.
For recent developments using a relativistic framework, we refer
to~\cite{Briceno:2015csa,Briceno:2015dca,Briceno:2015tza}.

\subsection{Quark mass dependence}
\label{sec:qmdep}

To reduce numerical noise as well as to speed up algorithms, 
lattice {calculations}
are often performed at unphysical values of the light quark masses. 
While this at first sight may appear as a disadvantage, it is indeed
a virtue as it enables a new handle on investigating the structure
of certain states. However, in the case of multiple coupled channels,
one also has to be aware that thresholds and poles can move very
strongly as a function of the quark masses. This intricate interplay
between $S$-wave thresholds and resonances needs to be accounted for 
when one tries to extract the resonance properties.

To address the first issue, we specifically consider the charm-strange
mesons $D_{s0}^*(2317)$ and $D_{s1}(2460)$. As shown in \cite{Cleven:2010aw},
in the molecular picture describing these as $DK$ and $D^*K$ bound states,
a particular pion and kaon mass dependence arises. Consider first the
dependence on the light quark masses, that can be mapped onto the
pion mass dependence utilizing the Gell-Mann--Oakes--Renner 
relation~\cite{GellMann:1968rz},
$M_{\pi^\pm}^2 = B(m_u+m_d)$, that naturally arises in QCD as the leading 
term in the chiral expansion of the Goldstone boson mass. Here, $B$ is
related to the vacuum expectation value of the quark condensate. In fact, this
relation is fulfilled to better than 94\% in QCD~\cite{Colangelo:2001sp}.
As shown in Ref.~\cite{Cleven:2010aw}, the pion mass dependence of such
a molecular state is much more pronounced than for a simple $c\bar s$ state.
Even more telling and unique is, however, the kaon mass dependence.
For that, consider the  $M_K$ dependence of the mass of a bound state
of a kaon and some other hadron. The mass of such a kaonic bound state
is given by
\begin{equation}
M = M_K+M_h-E_B,
\end{equation}
where $M_h$ is the mass of the other hadron, and $E_B$ denotes the binding
energy. Although both $M_h$ and $E_B$ have some kaon mass dependence, it is 
expected to be a lot weaker
than that of the kaon itself. Thus, the important implication of this simple 
formula is that the leading
kaon mass dependence of a kaon-hadron bound state is {\em linear, and the 
slope is unity}. The only
exception to this argument is if the other hadron
is also a kaon or anti-kaon. In this case, the
leading kaon mass dependence is still linear but with the 
slope being changed to two.
Hence, as for the $DK$ and $D^*K$ bound states, one expects that
their masses are linear in the kaon mass, and the slope is 
approximately one. This expectation is borne out by
the explicit calculations performed in~\cite{Cleven:2010aw}.
Early lattice QCD attempts to investigate this peculiar kaon mass dependence
have led to inconclusive results~\cite{mcneile}. Other papers that discuss 
methods 
to analyze the structure of states based on their quark mass dependence
or the behavior at large number of colors are 
e.g.~\cite{Hanhart:2008mx,Pelaez:2010fj,Bernard:2010fp,Albaladejo:2012te,
Guo:2011pa,Nebreda:2011cp,Pelaez:2006nj,Guo:2015dha}.

The second issue we want to address briefly is the intricate interplay
of $S$-wave thresholds and resonance pole positions with varying quark masses,
as detailed in Ref.~\cite{Doring:2013glu}. In that paper, pion-nucleon
scattering in the $J^P = 1/2^-$ sector in the finite volume and at
varying quark masses based on UCHPT was studied. In the infinite volume,
both the $N(1535)$ and the  $N(1650)$ are dynamically generated
from the coupled channel dynamics of the isospin $I=1/2$ and strangeness
$S=0$ $\pi N, \eta N, K\Lambda$  and $K\Sigma$ system. Having fixed the
corresponding LECs in the infinite volume, one can straightforwardly
calculate the spectrum in the finite volume provided one knows the
octet Goldstone boson masses, the masses of the ground-state octet 
baryons and the meson decay constants. Such sets of data at different
quark masses are given by ETMC and QCDSF. ETMC provides masses
and decay constants for $M_\pi=269$~MeV and the kaon mass 
close to its physical value~\cite{Alexandrou:2009qu,Ottnad:2012fv}.
Quite differently, the QCDSF Collaboration~\cite{Bietenholz:2011qq}
obtains the baryon and meson masses  from an alternative approach to 
tune the quark masses. Most importantly, while the lattice size and 
spacing are comparable to those of the ETMC, the strange quark mass 
differs significantly from the physical value. The latter results in a
different ordering of the masses of the ground-state octet mesons and, 
consequently, in a different ordering of meson-baryon thresholds.
For the ETMC parameters,  all thresholds are moved to higher energies. 
The cusp at the $\eta N$ threshold has become more pronounced, but
no clear resonance shapes are visible. Indeed, going to the complex
plane, one finds that the poles are hidden {in the Riemann sheets which
are not directly connected to the physical one by crossing the cut at the
energies corresponding to the real parts of the poles}.
Using the QCDSF parameters, the situation is very different. In contrast to the ETMC case, a 
clear resonance signal is visible below the $K\Lambda$ threshold, 
that is the first inelastic channel in this parameter setup. Indeed, one
finds a pole  on the corresponding Riemann sheet. Unlike in the 
ETMC case, it is not hidden behind a threshold. Between
the $K\Lambda$ and the $K\Sigma$ threshold, there is only a hidden pole. 
The $K\Sigma$ and $\eta N$ thresholds are almost
degenerate, and on sheets corresponding to these higher-lying 
thresholds one only finds hidden poles. For more details, the reader
is referred to Ref.~\cite{Doring:2013glu}. In that paper, strategies
to overcome such type of difficulties are also discussed.

{It is worthwhile to mention that the composition of a hadron in general
may vary when changing the quark masses. However, as long as the quark masses are
not very different from the physical values, the quark mass dependence is rather
suggestive towards revealing the internal structure as different structures
should result in different quark mass dependence. }

\subsection{Measuring compositeness on lattice}
\label{sec:compo}

As discussed in Sec.~\ref{sec:weinberg}, the Weinberg compositeness criterion
offers a possibility to disentangle compact bound states from loosely bound
hadronic molecules. {By measuring the low-energy scattering observables in
lattice using the L\"uscher formalism discussed before, one can extract the
compositeness by using Eqs.~\eqref{eq:arwein}. For related work, see, e.g.,
Refs.~\cite{Suganuma:2007uv,MartinezTorres:2011pr,Ozaki:2012ce,
Albaladejo:2013aka}. It is pointed out in~\cite{Agadjanov:2014ana} that the use
of partially twisted boundary conditions is cheaper than studying the volume dependence in lattice
for measuring the compositeness.} 
The basic object in that work is the
scattering amplitude in the finite volume, which can be obtained from the
corresponding loop function $\tilde G_L^{\bm{\theta}}(s) = G(s)+\Delta
G_L^{\bm{\theta}}(s)$~\cite{Doring:2011vk}, where $\Delta G_L^{\bm{\theta}}$ can
be related to the modified L\"uscher function $Z_{00}^{\bm{\theta}}$ via
\begin{equation}
  \label{deltaG-luscher1}
  \Delta G_L^{\bm{\theta}}(s)=\frac1{8\pi\sqrt{s}}\left(
    ik - \frac{2}{\sqrt{\pi}L} Z_{00}^{\bm{\theta}}(1,\hat k^2)
  \right)+\cdots,
\end{equation}
where $\hat k = kL/(2\pi)$ and ellipsis denote terms that are exponentially
suppressed with the lattice size $L$~\cite{Doring:2011vk}. Here, in case 
of twisted boundary conditions, the momenta also depend on the twist angle
$\bm{\theta}$ according to 
$\bm{q}_n=(2\pi/L)\bm{n}+(\bm{\theta}/L),~0\leq\theta_i<2\pi$. 
In case of a bound state with mass $M$ in the infinite volume, the 
scattering amplitude should have a pole at $s=M^2$, with the corresponding 
binding momentum $k_B\equiv i\kappa$, $\kappa>0$, given by the equation
\begin{equation}
  \label{inf-vol-pole-eq}
  \psi(k_B^2)+\kappa = -8\pi M\Big[V^{-1}(M^2) - G(M^2)\Big] = 0\,,
\end{equation}
with $\psi(k^2)$ the analytic continuation of $k\cot\delta(k)$ for
arbitrary complex values of $k^2$. From this, it is straightforward
to evaluate the pole position shift,
\begin{eqnarray}
  \label{mass-shift}
\kappa_L - \kappa &=& \frac1{1-2\kappa\psi'(k_B^2)} \left[ -8\pi M_L \Delta
G_L^{\bm{\theta}} (M_L^2)\right. \nonumber\\
&& \qquad\qquad\qquad ~~~ + \left. \psi'(k_B^2)(\kappa_L-\kappa)^2 \right],
\end{eqnarray}
where the prime denotes differentiation with respect to $k^2$. 
This equation gives the bound state pole position $\kappa_L$ (and thus
the finite volume mass $M_L$) as a function of the infinite-volume 
parameters $g^2$ and $\kappa$. Having determined these parameters
from the bound state levels $\kappa_L$, one is then able to determine
the wave function renormalization constant $Z$ in the infinite volume.
In Ref.~\cite{Agadjanov:2014ana}, this procedure is scrutinized using
synthetic lattice data, for a simple toy model and a molecular model
for the charm scalar meson  $D_{s0}^*(2317)$. An important finding of this
work is that the extraction of $Z$ is facilitated by using twisted
boundary conditions, measuring the  dependence of the spectrum on the 
twist angle. Also, the limitations of this approach are discussed in 
detail. It remains to be seen how useful this method is for real 
lattice data. For related papers, also making use of twisted boundary
conditions to explore the nature of states, see e.g.
Refs.~\cite{Ozaki:2012ce,Briceno:2013hya,Korber:2015rce}.
A different approach to quantify compositeness in a finite volume has
recently been given in Ref.~\cite{Tsuchida:2017gpb}. Using this 
method, the $\bar KN$ component of the $\Lambda(1405)$
was found to be 58\%,  and the $\Sigma\pi$ and other components
also contribute to its structure. This is interpreted as a reflection
of the two-pole scenario of the  $\Lambda(1405)$.

\subsection{Lattice QCD results on the charm-strange mesons
and \texorpdfstring{$\bm{XYZ}$~}~states}
\label{sec:latres}

There have been quite a few studies of the charm-strange mesons and
some of the $XYZ$ states in lattice QCD. However, there are very few 
conclusive results at present, so we expect that this section will be 
outdated most quickly.

Let us consider first the charm-strange mesons. A pioneering lattice study of
the low-energy interaction between a light pseudoscalar meson and a charmed
pseudoscalar meson was presented in Ref.~\cite{Liu:2012zya}.
The scattering lengths of the five channels $D\bar K(I=0)$, $D\bar K(I=1)$, $D_s
K$, $D\pi(I=3/2)$ and $D_s\pi$ were calculated based on four ensembles with pion
masses of 301, 364, 511 and 617~MeV. These channels are free of contributions
from disconnected diagrams.
SU(3) UCHPT as developed in Ref.~\cite{Guo:2009ct} was used to perform the
chiral extrapolation. The LECs of the chiral Lagrangian were determined from a
fit to the lattice results. With the same set of LECs and the masses of the
involved mesons set to their physical values,  predictions for the other
channels including $DK(I=0)$, $DK(I=1)$, $D\pi(I=1/2)$ and $D_s\bar K$ were
made. In particular, it was found that the attractive interaction in the
$DK(I=0)$ channel is strong enough so that a pole is generated in the unitarized
scattering amplitude.  Within $1\sigma$ uncertainties of the parameters, the
pole is at $2315^{+18}_{-28}$~MeV, and it is always below the $DK$ threshold.
From calculating the wave function normalization constant, it was found that
this pole is mainly an $S$-wave $DK$ bound state (the pertinent scattering
length being close to $-1$~fm as predicted in~\cite{Guo:2009ct} for such a
molecular state using Eq.~\eqref{eq:arwein}).
Further, a much sharper prediction of the isospin breaking  decay width of the
$D_{s0}^*(2317)\to D_s\pi$ could  be given
\begin{equation}
\label{Eq:DecayWidth}
 \Gamma(D_{s0}^*(2317)\to D_s \pi) = (133\pm22)~{\rm keV}~,
\end{equation}
to be contrasted with the molecular prediction without lattice
data, $\Gamma(D_{s0}^*(2317)\to D_s \pi)=(180\pm110)$~keV~\cite{Guo:2008gp},
and typical quark model predictions for a $c\bar s$ charm scalar meson of the 
order of 10~keV, see 
e.g. Refs.~\cite{Godfrey:2003kg,Faessler:2007gv}. For a similar study 
using a covariant UCHPT instead
of the heavy-baryon formalism, see Ref.~\cite{Altenbuchinger:2013vwa}.

A systematic study of the charm scalar and axial mesons at lighter
pion masses ($M_\pi = 156$ and $266$~MeV) was performed in 
Refs.~\cite{Mohler:2012na,Mohler:2013rwa,Lang:2014yfa}. These data
were later reanalyzed with the help of finite-volume 
UCHPT~\cite{Torres:2014vna}. Most notably, the $DK$ scattering with 
$J^P=0^+$ was investigated in \cite{Mohler:2013rwa}, using $DK$ as well
as $c\bar s$ interpolating fields. Clear evidence of a bound state below
the $DK$ threshold was found and the corresponding scattering length
was $a_0 = -1.33(20)\,$fm, consistent with the molecular scenario.
The analysis of Ref.~\cite{Torres:2014vna} found a 70\% $DK$ ($D^*K$) 
component in the $D_{s0}^*(2317)\,(D_{s1}(2460))$ state.

The most systematic study in the coupled-channel $D\pi, D\eta$ and $D_s\bar K$
system with isospin $1/2$ and $3/2$ was reported in Ref.~\cite{Moir:2016srx}.
Using a large basis of quark-antiquark and meson-meson basis states, the
finite volume energy spectrum could be calculated to high precision, allowing
for the extraction of the scattering amplitudes in the $S$-, $P$- and $D$-waves.
With the help of the coupled-channel L\"uscher formalism and various
parameterizations of the $T$-matrix, three poles were found in the complex 
plane: a 
$J^P=0^+$ near-threshold bound state, $M_S=(2275.9\pm 0.9)\,$MeV,
with a large coupling to $D\pi$,
a deeply bound $J^P=1^-$ state,   $M_P=(2009\pm 2)\,$MeV,
and evidence for a $J^P=2^+$ narrow 
resonance coupled predominantly to$D\pi$,  $M_D=(2527\pm 3)\,$MeV.
An interesting observation was made in Ref.~\cite{Albaladejo:2016lbb}.
Using UCHPT, it was shown that there are in fact two 
($I=1/2, J^P=0^+$) poles  in the region of the $D_0^*(2400)$ in 
the coupled-channel $D\pi, D\eta, 
D_s\bar K$ scattering amplitudes. {They couple differently to the involved
channels and thus should be understood as two states.} Having all the parameters
fixed from earlier studies in Ref.~\cite{Liu:2012zya}, the energy levels for the coupled-channel system in a 
finite volume were predicted. These agree remarkably well with the
lattice QCD results in~\cite{Moir:2016srx}. The intricate interplay of 
close-by thresholds and resonance poles already pointed out 
in~\cite{Doring:2013glu} is also found, and it is stressed that more
high-statistics data are needed to better determine the higher mass pole.

We now turn to the $XYZ$ states. Consider first the $X(3872)$. There
have been a number of studies using diquark-diquark or tetraquark
interpolating fields over the years, but none of these has been 
conclusive, see e.g.~\cite{Chiu:2006hd,Yang:2012mya}. 
Evidence for a bound state with $J^{PC}=1^{++}$
$(11\pm7)$~MeV below the $D\bar D^*$ threshold was reported in 
Ref.~\cite{Prelovsek:2013cra}. 
This establishes a candidate for the  $X(3872)$ in addition to the
near-by scattering states  $D\bar D^*$ and $J/\psi \rho$. This computation
was performed at $M_\pi =266\,$MeV but in a small volume $L\simeq 2\,$fm. 
This finding was validated using the Highly Improved Staggered Quark 
action~\cite{Lee:2014uta}. Finally, a refined study allowing also
for the mixing of tetraquark interpolators with $\bar c c$ components
was presented in~\cite{Padmanath:2015era}.
A candidate for the $X(3872)$ with $I = 0$ is observed very 
close to the  experimental state only if both $\bar cc$ and $D\bar D^*$ 
interpolators are included. However, the candidate is 
not found if diquark-antidiquark and $D\bar D^*$ are used in the 
absence of $\bar c c$. Note that in 
Refs.~\cite{Jansen:2013cba,Garzon:2013uwa,Jansen:2015lha,Baru:2015tfa}
strategies for extracting the properties of the $X(3872)$ from 
finite-volume data (at unphysical quark masses) have been worked out.

Consider next the $Z_c(3900)$. Various lattice calculations have
been performed, which, however, did not lead to  conclusive results,
see e.g. 
Refs.~\cite{Prelovsek:2013xba,Prelovsek:2014swa,Chen:2014afa,
Ikeda:2016zwx}. 
For example, in the most recent work \cite{Ikeda:2016zwx} it
was argued that this state is most probably a threshold cusp. Also,
a systematic analysis of most of these data using a finite volume version of 
the framework in Ref.~\cite{Albaladejo:2015lob} did not allow
for a definite conclusion
on the nature of the $Z_c(3900)$~\cite{Albaladejo:2016jsg}.

The Chinese Lattice QCD Collaboration has also studied $D^* \bar D_1$ 
\cite{Meng:2009qt,Chen:2016lkl} and $D^*\bar D$ scattering 
\cite{Chen:2015jwa}
with the aim of investigating the structure
of the $Z_c(4430)$ and $Z_c(4025)$, respectively.  These studies were
mostly exploratory and no definite statements could be drawn.

\subsection{Lattice QCD results on hadrons built from light quarks}
\label{sec:latres2}

Here we summarize briefly some very recent results on hadrons made 
entirely of light $u,d,s$ quarks, more precisely, the scalar mesons 
$f_0(500)$  and $a_0(980)$ as well as $\Lambda(1405)$.

The  first determination of the energy dependence of the isoscalar $\pi\pi$
elastic scattering phase shift and the extraction of the $f_0(500)$ based on
dynamical QCD using the methods described above was given by the Hadron Spectrum
Collaboration in Ref.~ \cite{Briceno:2016mjc}.
From the volume dependence of the spectrum the $S$-wave phase shift up to the
$K\bar K$ threshold could be extracted. The calculations were performed at pion
masses of 236 and 391~MeV. The resulting amplitudes are described in terms of a
scalar meson which evolves from a bound state below the $\pi\pi$ threshold at
the heavier quark mass to a broad resonance at the lighter quark mass.
{This is } in line with the prediction of Ref.~\cite{Hanhart:2008mx} based
on UCHPT to one loop.
 Earlier, the same collaboration had analyzed the coupled channel
$\pi\eta, K\bar K, \pi\eta'$ system with isospin $I=1$ and extracted properties
of the $a_0(980)$ meson~\cite{Dudek:2016cru}. {The model-independent
lattice data on energy levels were reanalyzed using UCHPT in
Refs.~\cite{Guo:2016zep,Doring:2016bdr}.} In particular, Ref.~\cite{Guo:2016zep}
pointed out some ambiguities in the $I=1$ solution.

There have been quite a few studies of the $\Lambda(1405)$ as a simple
three-quark baryon state by various lattice collaborations. In view of the
intricacies of the coupled channel $K^- p$ scattering discussed earlier,
we will not further consider these as coupled-channel effects must be 
considered.
An exception is the analysis of Ref.~\cite{Hall:2014uca}  based on 
the PCAS-CS ensembles~\cite{Aoki:2008sm} 
with  three-quark sources allowing for scalar and vector diquark configurations
that leads to the vanishing of the strange magnetic form factor of the 
$\Lambda(1405)$
at the physical pion mass. It is argued that this can only happen if the
$\Lambda(1405)$ is mostly an antikaon-nucleon molecule. This is further 
validated
by applying a finite-volume Hamiltonian approach to the measured energy 
levels~\cite{Wu:2014vma}. This lattice QCD result appears to be at odds with the
accepted two-pole scenario. However, as pointed out in the UCHPT analysis of 
Ref.~\cite{Molina:2015uqp}, these results exhibit some shortcomings. It is
argued in that work, that what is really discussed in \cite{Hall:2014uca} is 
the heavier of the two poles. In particular the complete absence of the 
$\pi \Sigma$ threshold in these data is discussed, as this threshold  would couple
stronger to the lighter pole. This effect is presumably due to the neglect of 
the baryon-meson interpolating fields in Ref.~\cite{Hall:2014uca}. The required
operators are also specified in~\cite{Molina:2015uqp}. It will be interesting to see
lattice QCD {calculations} including all the relevant channels and required
interpolating fields. {We also point out that
better methods to calculate the matrix elements of unstable states has been
given in~\cite{Bernard:2012bi,Briceno:2015tza}.}

\section{Phenomenological manifestations of hadronic molecules}
\label{sec:6}

A large number of  theoretical studies on the recently discovered exotic
candidates focus on the computation of masses (for recent reviews see
Refs.~\cite{Brambilla:2014jmp,Chen:2016qju,Lebed:2016hpi,Esposito:2016noz}).
However, from the discussions in Secs.~\ref{sec:3} and \ref{sec:4}, it is
clear that the internal structure and especially the molecular nature of a
physical state manifests itself predominantly in some properly identified
dynamical production and decay processes. For near-threshold hadronic molecules,
the pertinent observables are provided by a set of low-energy quantities: the
scattering length and effective range for the constituent-hadron system or,
equivalently, the effective coupling of the hadronic molecular candidate to its
constituents, since these quantities are heavily intertwined as demonstrated in
Sec.~\ref{sec:3}. As discussed in detail there, the probability to find the
two-hadron component in the physical state, $(1-\lambda^2)$, can be extracted
directly from these quantities.
However, due to the presence of various energy scales driven by different
physics aspects, not all production and decay processes are sensitive to the
effective coupling as will be discussed with various examples mainly on the
$XYZ$ states in this section.
In addition, the implications of heavy quark spin and flavor symmetries on the
spectrum of hadronic molecules as well as the interplay between hadronic
molecules and nearby triangle singularities are presented.

\subsection{Long-distance production and decay mechanisms}
\label{sec:6-long}

As in Sec.~\ref{sec:3}, we denote the wave function of a hadronic molecule
candidate as $\Psi$. In order to allow for a quantitatively controlled analysis,
the state  must be located close to the relevant two-body threshold of $h_1$ and
$h_2$.
Then the long-distance momentum scale is given by $\gamma=\sqrt{2\mu E_B}$, {\sl
c.f.} Eq.~(\ref{eq:gamdef}), with $\gamma\ll\beta$, where $\beta$ is the inverse
of the range of forces. We define two classes of production and decay processes,
namely
\begin{itemize}
  \item long-distance processes, in which the momenta of all particles in the
CM frame of $h_1h_2$ are of $\order{\gamma}$;
  \item short-distance processes, which involve particles with a momentum
$\gtrsim \beta$ in the CM frame of $h_1h_2$.
\end{itemize}
It is shown in this section that only the former are sensitive to the molecular
component of the state investigated. The complications in the latter will be
discussed in the next section.

\subsubsection{Decays into the constituents and transitions between molecular states}

The long-distance processes involving hadronic molecules can be computed using
the EFT machinery introduced in Sec.~\ref{sec:4}. When only
 the degrees of freedom with momenta of $\order{\gamma}$ are kept, the EFT is
XEFT or, more generally, \nreftii.
When all particle energies are much smaller than $\beta^2/(2\mu)$ in the
$h_1h_2$ CM frame, the amplitudes involving pure molecular states for the
pertinent processes  are at LO determined by the scattering length
universality~\cite{Braaten:2004rn}. Decay amplitudes are then proportional to
the effective coupling $g_{\rm eff}$ defined in Eq.~\eqref{eq:ga} which can also
be expressed in terms of the scattering length.
Clearly such an approach cannot be simply applied to predominantly compact
states since then  $g_{\rm eff}$  becomes small and short-distance mechanisms
become more important than hadronic loops.

For instance, as soon as  the $\X$  is treated as a $D\bar D^*$ molecule, the
most important long-distance processes are its decays into $D^0\bar D^{*0}\to
D^0[\bar D^0\pi^0/\bar D^0\gamma]$, discussed in Sec.~\ref{sec:4-nreft2xeft}.
The decay rates and the momentum distributions of the final states serve as a
good probe for the structure of the $\X$~\cite{Voloshin:2003nt}.\footnote{These
decays were also discussed in
Refs.~\cite{Swanson:2003tb,Voloshin:2005rt,Meng:2007cx, Fleming:2007rp,
Liang:2009sp, Stapleton:2009ey, Baru:2011rs,Guo:2014hqa, Polosa:2015tra}.}
Higher order corrections can be calculated using \nreft~detailed in Sec.~\ref{sec:nreft1}.
For the $\Y$, the most important process for the detection of its $D_1\bar D$
component would be the decay into $D_1\bar D\to [D^*\pi/D^*\gamma]\bar
D$~\cite{Cleven:2013mka,Qin:2016spb}.

It may happen that two of the particles in the final states in the above
mentioned three-body decays form another hadronic molecule in the final state.
The transition of a shallow bound state into a light
particle and another shallow bound state receives two enhancements: large
coupling constants for the vertices involving the molecular states, and the
$1/v\simeq 2/(v_1+v_2)$ factor as shown in Sec.~\ref{sec:nreft1}, where $v_1$ and $v_2$ denote the relative velocities of the heavy mesons before
and after the emission of the light particle (see Fig.~\ref{fig:triangle} and
Eq.~\eqref{eq:Ipc}).

The possibility of a near-threshold pole in the $D\bar D$ final state
interaction and its possible influence on the $\X\to D^0\bar D^0\pi^0$
transition was studied in Ref.~\cite{Guo:2014hqa}, {\sl c.f.}
Fig.~\ref{fig:XDDpi_FSI}. Note that experimental information on this
distribution does not exist yet.\footnote{A few calculations based on
phenomenological models suggested the possible existence of a $D\bar D$ bound
state~\cite{Wong:2003xk,Zhang:2006ix,FernandezCarames:2009zz,Zhang:2009vs,
Liu:2010xh,Li:2012mqa}.} However, the interplay of hadronic molecules in the
final and the initial state might well have been observed already as detailed in
the remaining parts of this section.

\begin{figure}[tb]
    \begin{center}
     \includegraphics[width=\linewidth]{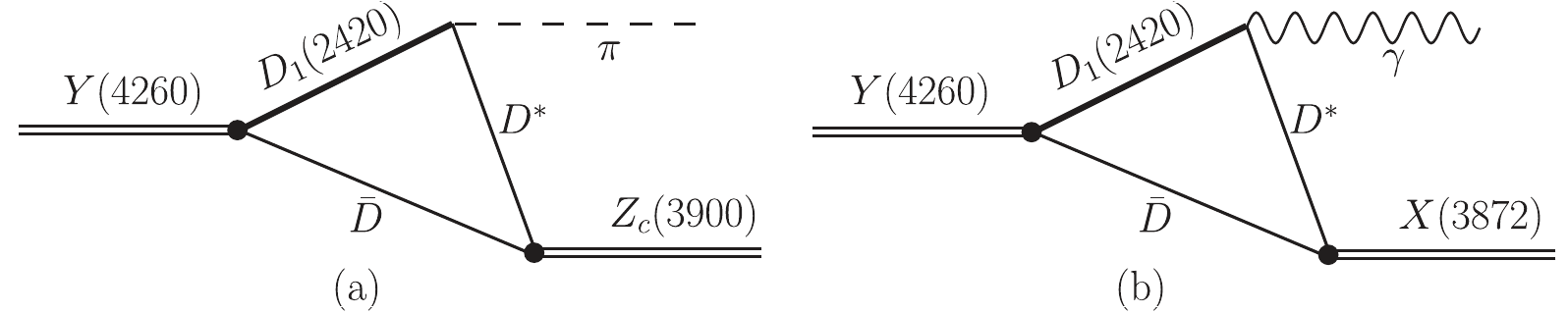}\\
     \caption{Schematic diagrams for the decays of the $\Y$ to $\Z\pi$ and
to $\X\gamma$ assuming that $D_1 \bar D-c.c.$ dominates the dynamics. The
diagrams from the charge conjugated channel are not shown.
     \label{fig:ytoxz}}
    \end{center}
  \end{figure}

The $D_1(2420)\bar D$ threshold in the $J^{PC}=1^{--}$ channel is the closest
$S$-wave open-charm threshold that the $\Y$ can couple to. It is at the same
time the lowest $S$-wave open-charm threshold with vector quantum numbers, which
provides a natural explanation why the first (established) exotic vector state
is significantly heavier than the $X(3872)$.
Assuming that $\Y$ is a  $D_1\bar D$ molecule and $\X$ and $Z_c(3900)$ are
$D^*\bar D$ molecules with $J^{PC}=1^{++}$ and $1^{+-}$, respectively, the
decays of $\Y$ into  $Z_c\pi$ and $\X\gamma$ will occur through the mechanisms
shown in Fig.~\ref{fig:ytoxz} (a) and (b), where the type of the light particle
accompanying the hadronic molecular state is controlled by the positive and
negative $C$ parity of the $\pi^0$ and the photon, respectively.
Since $v\lesssim 0.1$ for both transitions, $1/v$ indeed provides a large
factor, shown as the solid lines in Fig.~\ref{fig:pc}. Therefore,  a copious
production of $\Z$ from  $\Y\to \Z \pi$ transitions as observed at both BESIII
and Belle~\cite{Ablikim:2013mio,Liu:2013dau} appears naturally within this
dynamical picture~\cite{Wang:2013cya}.
In addition, $if$ the above explanation is indeed correct, also the $\X$ must be necessarily
produced with a large rate for the production
 in $\Y$ radiative decays~\cite{Guo:2013nza}.
 Indeed, assuming that the $\Y$ and
$\X$ are pure bound states of $D_1\bar D$ and $D^0\bar D^{*0}$,\footnote{It has
been emphasized in
Sec.~\ref{sec:4-nreft2xeft} that the charged charm mesons need to be taken
into account for the $\X$ in the framework of \nreft~since they are below the
hard scale and should be treated explicitly. The reason for neglecting them here
is that the rate for the $D_1^0\to D^{*0}\gamma$ is at least one
order of magnitude larger than that for the $D_1^+\to D^{*+}\gamma$ from
nonrelativistic quark model
calculations~\cite{Fayyazuddin:1994qu,Godfrey:2005ww,Close:2005se}.}
respectively,
one can get the coupling strengths using the relation in
Eq.~\eqref{eq:residue_bs} with $\lambda^2=Z=0$
\begin{eqnarray}
  |g_{\text{NR},X}^{}| &=& (0.20\pm0.20\pm0.03)~\text{GeV}^{-1/2} \,,
  \nonumber\\ |g_{\text{NR},Y}^{}| &=&
(1.26\pm0.09\pm0.66)~\text{GeV}^{-1/2} \,,
\end{eqnarray}
where we have taken the $\X$ binding energy to be $(90\pm90)$~keV, the first
errors are from the uncertainties of the binding energies, and
the second ones are from the approximation of Eq.~\eqref{eq:residue_bs}
due to neglecting terms suppressed by $\gamma/\beta$. Here, the inverse of the
range of forces is conservatively estimated by $\beta\sim \sqrt{2\, \mu\,
\Delta_\text{th}}$, with $ \Delta_\text{th}$ the difference
between the threshold of the components and the next close one, which is
$M_{D^{*+}} + M_{ D^+ } - M_{D^{*0}} - M_{ D^0 }$ for the $\X$ and $M_{D_1} +
M_{ D^* } - M_{D_1} - M_{ D }$ for the $\Y$, respectively.
The partial width for the $\Y\to\X\gamma$ in \nreft~can
reach a few tens of keV depending on the exact value of the $\X$ binding
energy~\cite{Guo:2013nza}.
The predicted copious production of the $\X$ in the $\Y$ radiative decays
was later confirmed by BESIII~\cite{Ablikim:2013dyn}.

It is worthwhile to mention that if the CM energy of the $e^+e^-$ collisions is
very close to the $D_1(2420)\bar D$ threshold at around 4.29~GeV,\footnote{The
production of an $S$-wave pair of $D_1(2420)$ and $\bar D$, which are
$j_\ell^P=3/2^+$ and $j_\ell=1/2^-$ states, respectively, breaks
HQSS~\cite{Li:2013yka}. This point was in fact already noticed in the classical
papers of the Cornell model, see Table~VI in Ref.~\cite{Eichten:1978tg} and
Table~VIII in Ref.~\cite{Eichten:1979ms}. However, in the energy region about
4.2~GeV HQSS breaking can be sizable~\cite{Wang:2013kra}.} the production of the
$\Z$ and $\X$ via the mechanism under consideration gets even more enhanced
since the kinematical condition for the TS discussed in Sec.~\ref{sec:4-3ploop}
is then (nearly) satisfied.
However, the resulting enhancement is balanced by the fact that the energy is
away from the peak of the $\Y$ spectral function.

It should be mentioned that some experimental observations consistent with
hadronic molecules are also claimed to be consistent with other models. For
instance, treating both the $\X$ and $\Y$ as tetraquarks also leads to a sizable
width for the radiative decay $\Y\to\X\gamma$ ~\cite{Chen:2015dig}.
However, large transitions to the constituents of the hadronic molecules appear
to be unique signatures for the molecular states.
In addition, their importance leaves visible imprints in the line shapes of
states like  the$Y(4260)$~\cite{Cleven:2013mka,Qin:2016spb}.

\subsubsection{More on line shapes}
\label{sec:morelineshapes}

The line shape of a hadronic molecule near its constituent threshold
reflects a long-distance phenomenon and can be used as a criterion for
establishing their nature.
The energy dependence of a hadronic molecule production line shape
generally does not appear to be trivial.

The data available at present for the line shapes of $X(3872)$ appear to be
insufficient for a unique conclusion about the pole locations of the state.
For example, a simultaneous fit of the line shape of the $X(3872)$ in the
$J/\psi\pi\pi$ and the $D^{*0}\bar{D}^0$ invariant mass distributions employing
a generalized  Flatt\'e parametrization~\cite{Hanhart:2007yq} revealed that
$X(3872)$ is a virtual state.
However, as soon as an explicit quarkonium pole is included in the analysis, the
authors of Refs.~\cite{Zhang:2009bv,Kalashnikova:2009gt,Meng:2014ota} find that
$X(3872)$ is the $2^3P_1$ charmonium with a large coupling to the
$D^0\bar{D}^{*0}$ channel --- in the light of the discussion of
Sec.~\ref{sec:wein} one needs to conclude that also in this case the $X(3872)$
has a sizeable molecular admixture. It should also be stressed that in
Ref.~\cite{Hanhart:2007yq} the width of the $D^*$ was omitted, which might
distort the line shapes ({\sl c.f.} Sec.~\ref{sec:lineshapes}) as was pointed
out in Refs.~\cite{Stapleton:2009ey,Braaten:2007dw}. According to these
analyses, once this effect is included, the fit favors a bound state solution.
Another study based on an improved Flatt\'e formula~\cite{Artoisenet:2010va}
notices that the current data can accept $X(3872)$ as both a  $D^0\bar{D}^{*0}$
hadronic molecule or the fine-tuned $2^3P_1$ charmonium coupled with  the
$D^0\bar{D}^{*0}$ channel. Also the more recent analysis of
Ref.~\cite{Kang:2016jxw} finds solutions with either a bound state pole or
virtual states.
Thus, to further investigate the nature of $X(3872)$ a high resolution scan of
its line shapes, especially within a few $\mev$ of the $D^0\bar{D}^{*0}$
threshold, is necessary.

A study of the line shapes of the two $Z_b$ states in $h_b(1P,2P)\pi$ channels
is presented in Ref.~\cite{Cleven:2011gp} based on the \nreft~framework
discussed in Sec.~\ref{sec:4}. By fitting to the $h_b(1P,2P)\pi$ invariant mass
distribution, it is found that the current data are consistent with the two
$Z_b$ states being $B\bar{B}^*$ and $B^*\bar{B}^*$ bound states, respectively.
Their line shapes in the elastic channels are also studied in the
XEFT/\nreftii~(see Sec.~\ref{sec:4-XEFT}) framework with the HQSS breaking
operator included explicitly in Ref.~\cite{Mehen:2013mva} where, however, no
pole locations were extracted. The line shapes of the two $Z_b$ states were
investigated in both elastic and inelastic channels in
Refs.~\cite{Hanhart:2015cua,Guo:2016bjq} based on separable
interactions, see Fig.~\ref{fig:Zb5Sfull}. An
explicit calculation revealed that the line shapes get distorted very little
when a non-separable interaction is included, such as the one-pion-exchange
potential. Similar studies of the $Z_c(3900)$ line shape in both $J/\psi\pi$ and
$D\bar D^*$ can be found in
Refs.~\cite{Albaladejo:2015lob,Zhou:2015jta,Pilloni:2016obd}. Based on these
results one needs to state that  also the data currently available for the
$Z_c(3900)$ are insufficient to pin down the pole locations.

The situation that it is not easy to extract the pole locations is partly
because that the observed peaks for the $\X$, $\Zc$ and $Z_b$ states are too
close to the  thresholds. Being tens of MeV below the $D_1\bar D$ threshold,
which, however, also means a larger uncertainty, the $\Y$ situation is
different. One sees clearly a nontrivial structure around the $D_1\bar D$
threshold in the $\Y$ line shape in the $J/\psi \pi\pi$ invariant mass
distribution, {\sl c.f.} Fig.~\ref{fig:Y4260-BESIII}. It indicates that the
coupling to the $D_1\bar D$ plays an essential role in understanding the $\Y$ in
line with the analysis in Sec.~\ref{sec:lineshapes}.
Correspondingly the molecular picture predicts a highly nontrivial energy
dependence for $e^+e^-\to D_1\bar D\to [D^*\pi]D$~\cite{Cleven:2013mka},
{\sl c.f.} the middle panel in Fig.~\ref{fig:Y4260LineShape}, that awaits
experimental confirmation at, e.g., BESIII.

\subsubsection{Enhanced isospin violations in molecular transitions}
\label{sec:isospinviol}

As argued above in Sec.~\ref{sec:wein} for molecular states the coupling of the
pole to the continuum
channel that forms the
state is large. As a consequence of this loop effects get very important that
can lead to very much
enhanced isospin violations if the molecular state is close to the relevant
threshold.
To be concrete we start with a detailed discussion of the implications of this
observation
for the $f_0$--$a_0$ mixing, first discussed in Ref.~\cite{Achasov:1979xc}.
\begin{figure}[t]
\begin{center}
\epsfig{file=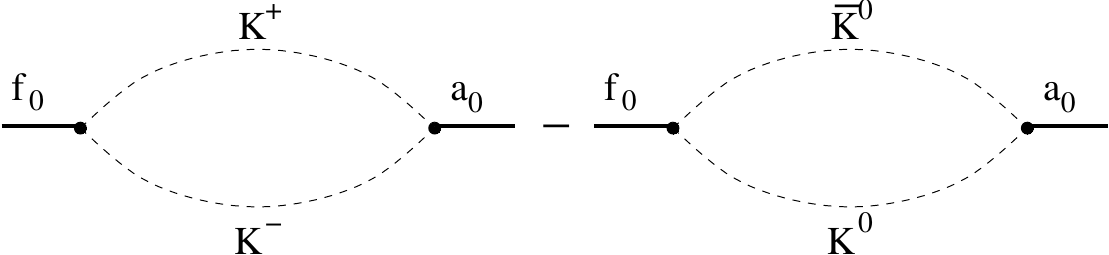, height=2.cm}
\caption{\it Graphical illustration of the leading contribution to the $f_0-a_0$
mixing
    matrix element.}
\label{lcsb}
\end{center}
\end{figure}
If both the isoscalar $f_0(980)$ as well as the isovector $a_0(980)$ were $\bar
KK$ molecular
states, the leading mixing effect of the two scalar mesons would the difference
of the loop functions
of charged and neutral kaons as
depicted in Fig.~\ref{lcsb} (in isospin conserving transitions the sum enters).
In the near-threshold regime we may approximate the loops by their leading
energy dependence provided by the respective unitarity cuts:
\begin{eqnarray}\nonumber
\langle f_0 |T| a_0\rangle &=& ig_{f_0K\bar K}g_{a_0K\bar
    K}\sqrt{s}\left( p_{K^0}-p_{K^+} \right) \\ & & \qquad  \qquad  \qquad +
{\cal
    O}\left({p_{K^0}^2-p_{K^+}^2}\right) \, ,
\label{llam}
\end{eqnarray}
where $p_K$ denotes the modulus of the relative momentum of the kaon pair.
Obviously, this leading contribution model-independently provides a measure of
the product of the effective couplings of $a_0$ and $f_0$ to the kaons and
therefore, as discussed in Sec.~\ref{sec:weinberg},   to the molecular admixture
of both states.
In the isospin limit  the loops cancel exactly. However, as soon as the masses
of the two-hadron states are different due to isospin violations
($M_{K^0}-M_{K^+}=4$ MeV), there appears an offset in the thresholds and the
mentioned cancellation is no longer complete.
This results in a transition between different isospins with all its strength
located  in between the two thresholds.
In Refs.~\cite{Wu:2007jh,Hanhart:2007bd} it was predicted that this very
peculiar effect should show up prominently in the transition $J/\psi \to \phi
\pi^0\eta$, if both $f_0(980)$ and $a_0(980)$ are $K\bar K$ molecular states,
since only then the coupling of the states to the $K\bar K$ is sufficiently
strong for the effect to be observable. A few years later the predicted very
narrow signal was measured at BESIII~\cite{Ablikim:2010aa} providing strong
evidence for a prominent molecular admixture in these light scalar mesons.

Another prominent example where a molecular nature of a near-threshold state
leads to a natural explanation of a huge isospin violation is the equal decay
rate of the $X(3872)$ to the isoscalar $\pi\pi\pi J/\psi$ and the isovector
$\pi\pi J/\psi$ channels, where in both cases the few-pion systems carry vector
quantum numbers and thus may be viewed as coming from the decay of an $\omega$
and a $\rho^0$ meson, respectively. The argument goes as follows:
The mass of the $X(3872)$ is located 7~MeV below the nominal $\omega J/\psi$
threshold, but 5~MeV above the nominal $\rho^0 J/\psi$ threshold. In addition,
the width of the $\omega$ is only 8~MeV such that the decay of the $X(3872)$
into the $\rho^0 J/\psi$ channel is strongly favored kinematically. Therefore,
the $X(3872)$ cannot have a significant isovector component in its wave function
since otherwise it would significantly more often decay into the $\rho^0 J/\psi$
than into the $\omega J/\psi$ channel.
However, a calculation for an isoscalar $X(3872)$ that predominantly decays via
$D^*\bar D$ loops naturally gives the experimental branching ratios for the
pions-$J/\psi$ channels simply because the close proximity of the mass of the
$X$ to the neutral $D^*\bar D$-threshold automatically produces an enhanced
isospin violation in the necessary strength, if the $X(3872)$ is a $D^*\bar D$
molecule, since only then the coupling to continuum is strong
enough~\cite{Gamermann:2009fv}.
An isoscalar nature of the $X(3872)$ is also required from a study of other
possible decay channels~\cite{Mehen:2015efa}, and its effective couplings to the
charged and neutral channels are basically the
same~\cite{HidalgoDuque:2012pq,Guo:2014hqa}.

As the last example in this context we would like to mention the hadronic width
of the isoscalar  $D_{s0}^*(2317)$.
This state is located below the $DK$ threshold and as such can decay strongly
only via isospin violation  into the isovector $D_s\pi $ channel.
The two most prominent decay mechanisms of $D_{s0}^*(2317)$ are an isospin
conserving transition into $D_s\eta $, followed by the isospin violating $\pi
\eta$ mixing amplitude, and the isospin violating difference between a $D^0K^+$
and a $D^+K^0$ loop (subleading contributions to this transition were studied in
Ref.~\cite{Guo:2008gp}).
The former mechanism should be present regardless the nature of the state, which
typically leads to widths of the order of 10~keV~\cite{Colangelo:2003vg}.
The latter mechanism, however, is large, if indeed the $D_{s0}^*(2317)$ were a
$DK$ molecule.
In fact typical calculations for molecular states give a width of the order of
100~keV~\cite{Faessler:2007gv,Lutz:2007sk,Liu:2012zya} --- for more details we
refer to Sec.~\ref{sec:latres}.
Thus, if this admittedly small width could be measured, e.g., at
$\overline{\text{P}}$ANDA, its value would provide direct experimental access to
the nature of $D_{s0}^*(2317)$.

\subsubsection{Enhanced production of hadronic molecules and conventional
hadrons due to triangle singularities}
\label{sec:6-ts}

From the analysis in Sec.~\ref{sec:4-3ploop}, we see that the TS on the
physical boundary is always located close to the threshold
of the intermediate particles. Furthermore, its effect is most pronounced
if  the two intermediate particles are in an $S$-wave, since otherwise the
centrifugal barrier suppresses small momenta. Hadronic
molecules are located naturally near-thresholds as well, and in all cases
considered in this review couple in
$S$-wave to its constituents. Therefore, in the course of this review there are
two important aspects of TS
that need to be discussed:
On the one hand, a TS may lead to a pronounced structure in experimental
observables that could be mistaken as a state; on the other hand, a TS can
enhance the production of a hadronic molecule in a given reaction. Note that
also the production of the conventional
hadrons can by strongly enhanced by a TS within small  energy regions.
 This is accompanied by a significant distortion of the line shapes since
 the location of the TS depends on the invariant masses of the external states.
 For example, the signal in the $\eta\pi\pi$ channel interpreted as $\eta(1405)$
 and the signal in the $K\bar{K}\pi$ interpreted as $\eta(1475)$ could
 find their origin in a single pole accompanied by a
TS~\cite{Wu:2011yx,Wu:2012pg}.
 We note that although
$P$-wave couplings are present in the triangle loop for the $\eta(1405/1475)$
decays,
the perfect satisfaction of the TS condition, Eq.~\eqref{eq:trianglesing},
causes detectable effects in these decays.

Since the  $K\bar K^*$ system can contribute to both $I=0$ and $I=1$ channels
with $J^{PC}=1^{++}$ and $1^{+-}$,
 one expects that
the TS may cause enhancements also in these channels.
In the following we list those possible enhancements and their quantum
numbers which can be searched for in experiment:
\begin{eqnarray}\label{KKstar-TS}
f_1(1420), &\ 0^+, \ 1^{++}:& \ \frac{1}{\sqrt{2}}(K^*\bar{K}-K\bar{K}^*)
\nonumber\\
&& \to K\bar{K}\pi, \ \eta\pi\pi, [3\pi];\nonumber\\
a_1(1420), &\ 1^-, \ 1^{++}:& \ \frac{1}{\sqrt{2}}(K^*\bar{K}-K\bar{K}^*)
\nonumber\\
&&  \to K\bar{K}\pi, \ 3\pi, \ [\eta\pi\pi];\nonumber\\
\tilde{h}_1(1420), &\ 0^-, \ 1^{+-}:& \
\frac{1}{\sqrt{2}}(K^*\bar{K}+K\bar{K}^*) \nonumber\\
&&  \to \rho\pi, \ \omega\eta, \ (\phi\eta), [\omega\pi], \ [\rho\eta], \
[\phi\pi];\nonumber\\
\tilde{b}_1(1420), &\ 1^+, \ 1^{+-}:& \
\frac{1}{\sqrt{2}}(K^*\bar{K}+K\bar{K}^*) \nonumber\\
&&  \to \phi\pi, \ \omega\pi, \rho\eta, \ [\rho\pi], \ [\omega\eta],
\end{eqnarray}
where $\tilde{h}_1(1420)$ and $\tilde{b}_1(1420)$ refer to the TSs whether or
not there exists resonances around.
Note that the $f_1(1420)$ needs to be taken into
account for the angular distribution in the $J/\psi\to\gamma 3\pi$
process~\cite{BESIII:2012aa}, and a detailed
partial wave analysis suggests the presence of the $f_1(1420)$ resonance
together with a TS mechanism~\cite{Wu:2012pg},
while it is argued in Ref.~\cite{Debastiani:2016xgg} that the $f_1(1420)$ is
the manifestation of the $f_1(1285)$ at higher energies due to the TS.
The $a_1(1420)$ has been reported by the
COMPASS Collaboration in $\pi^-p\to \pi^-\pi^-\pi^+ p$ and $\pi^-\pi^0\pi^0
p$~\cite{Adolph:2015pws} and can be well explained by the TS
mechanism~\cite{Liu:2015taa,Ketzer:2015tqa}. The decay channels in the square
brackets are $G$-parity violating and those in the round brackets are limited by
the phase space. One notices that there are states observed in the relevant mass
regions, namely, $a_1(1260)$, $f_1(1285)$, $h_1(1170)$, and
$b_1(1235)$~\cite{Olive:2016xmw}. Although most of these states have masses
outside
of the TS favored mass region, {\sl i.e.} $1.385\sim 1.442$ GeV~\cite{Liu:2015taa},
the $h_1(1380)$ is located at the edge of the TS kinematics and some detectable
effects could be expected~\cite{Guo:2013nza,Ablikim:2013dyn}. Notice that when
there is a TS in action,  the peak position for a resonance could be shifted
towards its location.
Some structures around thresholds of a pair of other light hadrons were also suggested to be due to a TS, for instance, the $f_2(1810)$ around the $K^*\bar
K^*$ threshold~\cite{Xie:2016lvs} and the $\phi(2170)$ around the $N\bar \Delta$
threshold~\cite{Lorenz:2015pba}.

In the heavy quarkonium sector,  the most famous example for an enhanced
production rate via the TS is the observation of the $Z_c(3900)$ at the mass
region of $Y(4260)$~\cite{Wang:2013cya,Wang:2013hga,Liu:2013vfa,
Szczepaniak:2015eza,Gong:2016jzb, Pilloni:2016obd}.
In addition, the sensitivity of the TS to the kinematics of the reaction might
well be the reason why the $Z_c(4020)$ is not seen in the same
decay~\cite{Wang:2013cya,Wang:2013hga}.
As discussed in Sec.~\ref{sec:6-long}, the same mechanism also enhances the
transition $Y(4260)\to \gamma X(3872)$~\cite{Guo:2013nza} and suggests that the
rate for $e^+e^-\to\gamma X_2$, which can be used to search for the spin-2
partner of the $\X$, $X_2$ (see Sec.~\ref{sec:4-interactions}), gets most
enhanced if the $e^+e^-$ collision energy is between 4.4 and
4.5~GeV~\cite{Guo:2014ura}.\footnote{This suggestion is based on the assumption
that the $X_2$ mass is very close to the $D^*\bar D^*$ threshold. If its mass is
tens of MeV below the threshold as suggested in~\cite{Baru:2016iwj}, then the TS
would be further away from the physical boundary and the production would get
less enhanced. } A candidate for the analogue of $Y(4260)$ in the bottomonium
sector is $\Upsilon(11020)$, since it is located close to the $B_1\bar B$
threshold~\cite{Wang:2013hga,Bondar:2016pox}. Here, the TS could
affect both the $Z_b(10610)$ and $Z_b(10650)$~\cite{Wang:2013hga}, although at
the mass of $\Upsilon(11020)$ the production of the lower $Z_b$ state is more
favored since the corresponding TS is closer~\cite{Bondar:2016pox}.
Based on the current statistics in Belle~\cite{Bondar:2016pox}, it is difficult
to judge whether there is one peak or two peaks present in Fig.~\ref{fig:Zb6S}.
In Refs.~\cite{Pakhlov:2014qva,Uglov:2016nql} the structure identified as the
charged exotic $Z_c(4430)$ observed in the $\pi \psi'$ final states by
Belle~\cite{Mizuk:2009da,Chilikin:2013tch} and LHCb~\cite{Aaij:2015zxa} was
claimed to be not connected to a pole but to owe its existence to the presence
of a TS.

Suggestions to search for resonance-like structures due to the TS in the heavy
meson and heavy quarkonium mass regions can be found
in~\cite{Liu:2014spa,Liu:2015cah,Liu:2015taa,Liu:2017vsf}. In particular, the
very recent BESIII results on the $\pi^\pm\psi'$ invariant mass distributions
for the $e^+e^-\to \pi^+\pi^-\psi'$ process at different CM energies seem in
line with the predictions made in~\cite{Liu:2014spa}.
Some of the strongly favored triangle loops are listed in
Table~\ref{tab:trianglesingularity}.

\begin{table*}
\caption{The triangle loops $[M_1M_2M_3]$ corresponding to
Fig.~\ref{fig:triangle}
which have shown large impact on the production of hadronic molecules
and conventional hadrons in experiment are listed in the first column.
The second column is the measured process with the final states in the bracket.
The checkmarks in the last column indicate that the triangle singularity of the
corresponding process locates in the physical region, {\sl i.e.} satisfying
Eq.\eqref{eq:trianglesing}. Although  the singularity of the process without
check mark is not located in the physical region, since it is not far away, it
can still enhance the corresponding production rate significantly.
 }
\begin{ruledtabular}
\begin{tabular}{ l c c }
$[M_{1}M_{2}M_{3}]$ & $A\to BC(\to\text{final states})$ &
Eq.\eqref{eq:trianglesing}\tabularnewline
\hline
$[K^{*}KK]$ & $\eta(1405/1475)\to
a_{0}(980)\pi(\to3\pi)$~\cite{Wu:2011yx,Wu:2012pg} & $\checkmark$\tabularnewline
 & $a_{1}(1420)\to
f_{0}(980)\pi(\to\eta\pi\pi)$~\cite{Ketzer:2015tqa,Liu:2015taa} &
$\checkmark$\tabularnewline
\hline
$[D_{1}DD^{*}]$ & $Y(4260)\to X(3872)\gamma$ ~\cite{Guo:2013nza} &
\tabularnewline
 & $Y(4260)\to
Z_{c}(3900)\pi$~\cite{Wang:2013cya,Wang:2013hga,Szczepaniak:2015eza,
Gong:2016jzb} & \tabularnewline
\hline
$[\Lambda(1890)\chi_{c1}p]$ & $\Lambda_{b}\to
P_{c}(4450)K$~\cite{Guo:2015umn,Liu:2015fea,Bayar:2016ftu} &
$\checkmark$\tabularnewline
\hline
$[D_{s3}(2860)\Lambda_{c}(2595)D]$ & $\Lambda_{b}\to
P_{c}(4450)K$~\cite{Liu:2015fea} & \tabularnewline
\end{tabular}
\end{ruledtabular}
\label{tab:trianglesingularity}
\end{table*}

The $P_c(4450)$  structure observed by LHCb~\cite{Aaij:2015tga} in
$\Lambda_b$ decays, no matter what its nature is, also contains a TS
contribution, as long as it strongly couples in an
$S$-wave~\cite{Bayar:2016ftu} to either
$\chi_{c1}p$~\cite{Guo:2015umn,Liu:2015fea,Guo:2016bkl,
Meissner:2015mza}  or $\Lambda_c(2595)\bar D$~\cite{Liu:2015fea}.
Recent discussions on the role of TSs in the
light baryonic sector can be found in Refs.~\cite{Wang:2016dtb,Roca:2017bvy,
Debastiani:2017dlz,Samart:2017scf}.

It is worthwhile to emphasize that once the kinematics for a process (nearly) satisfies
the TS condition given in Eq.~\eqref{eq:trianglesing},  the
enhancement in reaction rates due to TS contribution is always there, and can
produce a narrow peak once the relevant coupling is in an
$S$-wave~\cite{Bayar:2016ftu}.
The only question is whether it is strong enough to produce the observed or an
observable structure. Complications due to the interference between
the TS and a tree-level contribution is discussed
in~\cite{Schmid:1967ojm,Goebel:1982yb} in the single-channel case, and
in~\cite{Anisovich:1995ab,Szczepaniak:2015hya} for coupled channels.
The key to distinguishing whether a structure is
solely due to a TS or originates from a genuine resonance is the sensitivity of
the TS on kinematics: If in reactions with different kinematics the same structure is
observed, it most likely reflects the existence of a pole (resonance).

\subsection{Short-distance production and decay mechanisms}
\label{sec:6-short}

In Sec.~\ref{sec:6-long} we discussed both decay and production mechanisms
sensitive to the long-range parts of the wave function of a state and therefore
sensitive to its molecular nature. Here we demonstrate that there are
also decays and production reactions that do not allow one to extract the
molecular component of a given state. We start with an example of the former
to then switch to the latter.

It was claimed long ago~\cite{Swanson:2004pp} that the ratio \[
\frac{\mathcal{B}(\X\to \gamma\psi') }{\mathcal{B}(\X\to \gamma J/\psi) }, \]
with the measured value given in Eq.~\eqref{eq:sec2X3872R}, is very sensitive to
the molecular component of the $\X$ wave function. In particular using vector
meson dominance and  a quark model, in Ref.~\cite{Swanson:2004pp} it was
predicted to be about $4\times10^{-3}$ if
the $\X$ is a hadronic molecule with a dominant $D^0\bar D^{*0}$ component plus
a small admixture of the $\rho J/\psi$ and $\omega J/\psi$. However, as
demonstrated in Ref.~\cite{Guo:2014taa}, when radiative decays of $\X$ are
calculated using \nreft~(see Sec.~\ref{sec:nreft1}) field theoretic consistency
calls for the inclusion of a counter term at LO. In other words, the transitions
are controlled by short-distance instead of long-distance dynamics and therefore
do not allow one to extract any information on the molecular component of the
$\X$ wave function.

There also have been many claims that production rates of multiquark states
in high-energy collisions are sensitive to a molecular admixture of those
states.
The production of the $\X$ at hadron colliders was debated
in~\cite{Bignamini:2009sk,Artoisenet:2009wk,Bignamini:2009fn,Artoisenet:2010uu,
Butenschoen:2013pxa,Esposito:2013ada,Meng:2013gga}, and that
of the spin and flavor partners of the $\X$ was discussed
in~\cite{Bignamini:2009fn,Guo:2013ufa,Guo:2014sca}. For the production of the
$\X$ in $B$ decays, we refer
to~\cite{Braaten:2004ai,Braaten:2004fk,Meng:2005er,Fan:2011aa,Meng:2013gga}.
The production of the $\X$ in heavy ion collisions was discussed in
\cite{Cho:2013rpa,Torres:2014fxa} and by the ExHIC Collaboration including other
hadronic molecular candidates~\cite{Cho:2010db,Cho:2011ew,Cho:2015exb}.
In Ref.~\cite{Larionov:2015nea} it was proposed that the hadronic
molecular component of the $\X$ could be extracted from its production in
antiproton-nucleus collisions. Prompt productions of
diquark-antidiquark tetraquarks at the LHC were studied
in~\cite{Guerrieri:2014gfa}.  Here, we discuss to what extent
 high-energy reactions can be used to
disentangle the structure of a near-threshold state.

The underlying physics for the short-distance production and decay processes of
a shallow hadronic molecule  is characterized by vastly different scales.
This allows for a derivation of
factorization formulae for the corresponding
amplitudes~\cite{Braaten:2004ai,Braaten:2004fk,Braaten:2005jj,Braaten:2006sy}.

\begin{figure}[t]
    \centering
    \includegraphics[width=\linewidth]{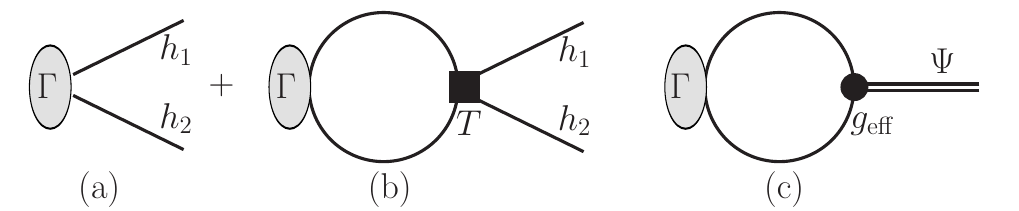}
    \caption{Production of a pair of hadrons (a+b) and the hadronic molecule
formed by them (c) from a source $\Gamma$. Here the shaded area, the solid
lines and the double line denote the source, the constituent hadrons and the
hadronic molecule, respectively.
\label{fig:production}}
\end{figure}

We illustrate the production of a $h_1h_2$ pair and a near-threshold state
with wave function
$\Psi$ in Fig.~\ref{fig:production}, where $\Gamma$ denotes a short-distance
source. The pair of $h_1$ and $h_2$ can be produced directly at short
distances shown as diagram (a), and through rescattering shown as diagram (b).
If the rescattering strength were weak, the production would be well approximated by only the
$\Gamma$ term. There would also be no drastic energy dependence in the
near-threshold region, and the strength of the $S$-wave cusp exactly at the
threshold would not be strong enough to produce a narrow
peak~\cite{Guo:2014iya}.
However, if the rescattering is strongly attractive, the amplitude $T$
possesses a pole, which we assume to be located close to the threshold with a small binding
energy
$E_B$. Then one gets for  the production amplitude of a given decay channel $j$ of
the state of interest:
\begin{eqnarray}
    \mathcal{M}_j(\bm k; E) &=&  \Gamma_j^\Lambda(\bm k;E)  \\
    &+&    \sum_i \int_\Lambda\! \frac{d^3q}{(2\pi)^3}
\Gamma_i^\Lambda(\bm{q};E)\, G_i(\bm q;E)\,   T_{ij}(\bm q, \bm k; E)
    \nonumber
    \label{eq:Aproduction_pair}
\end{eqnarray}
where $G_i$ ($\Gamma_i^\Lambda$) is the propagator (short-distance production
amplitude) for the $i$-th intermediate channel and $\Lambda$ denotes the cutoff of
the integral. In principle all dynamical degrees of freedom below the energy
scale $\Lambda$ should be accounted for. The short-distance contribution
$\Gamma_i^\Lambda$ also serves to absorb the $\Lambda$ dependence, and as a
result $\mathcal{M}$ is $\Lambda$ independent.

Let us consider the kinematic situation that the CM momentum of
$h_1$ and $h_2$ is very small, $\sim \gamma$, and $\Lambda$ is much larger than
$\gamma^2/(2\mu)$ but still small enough to prevent other channels from being
dynamical. In this case, the intermediate state for diagram (b) is $h_1 h_2$.
The LO term of the momentum expansion of $\Gamma^\Lambda$ is simply a constant.
The nonrelativistic two-body propagator is $G(\bm q;E) = \left[ E - \bm
q^2/(2\mu) + i\epsilon \right]^{-1}$, and the $T$-matrix is given by
Eq.~\eqref{eq:T1c}. Thus, one obtains
\begin{equation}
  \mathcal{M}(\bm k;E)
  = \Gamma^\Lambda \left[ 1+ \frac{{\Lambda}/{\sqrt{2\pi}}
  -\sqrt{-2\mu E} + \order{\Lambda^{-1}}}{ \gamma -\sqrt{-2\mu E} } \right].
  \label{eq:Mfactorization}
\end{equation}
If $\Gamma^\Lambda\propto \Lambda^{-1}$, the LO $\Lambda$ dependence will be
absorbed~\cite{Braaten:2004ai}, and the factorization
formula~\cite{Braaten:2004ai,Braaten:2004fk,Braaten:2005jj} for the production
of the low-momentum $h_1h_2$ pair follows:
\begin{equation}
  \mathcal{M}(\bm k;E)
  = \frac{\Gamma\,\mu}{(2\pi)^{3/2}} T_\text{NR}^{}(E) + \order{\Lambda^{-1}}\,,
  \label{eq:factorization}
\end{equation}
where $\Gamma\equiv\Gamma^\Lambda\Lambda$ is the short-distance part, and the
long-distance part
$T_\text{NR}^{}(E)=(2\pi/\mu)\left(\gamma -\sqrt{-2\mu E} \right)^{-1} $ is
provided by the scattering $T$-matrix. From the derivation
in~\cite{Braaten:2006sy}, it becomes
clear that the
short-distance part is the Wilson coefficient of the operator production
expansion in the EFT.
A similar factorization formula was derived in~\cite{Guo:2014ppa} with the help
of chiral symmetry for high-energy productions of kaonic bound states predicted
in~\cite{Guo:2011dd}.

The amplitude for the production of the
near-threshold state  is obtained from Eq.~\eqref{eq:factorization} by
replacing $T_\text{NR}^{}(E)$ by the square root of its residue
$g_\text{NR}^{2}$
given in Eq.~\eqref{eq:gNR} and multiplying the factor $1/\sqrt{2\mu}$ to
account for the difference in normalization factors
\begin{equation}
  \mathcal{M}_\Psi
  = \frac{\Gamma\,\sqrt{\mu}}{4(\pi)^{3/2}} g_\text{NR} +
\order{\Lambda^{-1}}\,.
  \label{eq:factorizationX}
\end{equation}
Hence, the production rate, $\propto
g_\text{NR}^2\propto \sqrt{E_B}$, {\sl c.f.}  Eq.~\eqref{eq:residue}, seems
suppressed for very loosely bound states,
which is consistent with the common
intuition~\cite{Braaten:2004fk,Artoisenet:2009wk}.
In particular, one expects a suppression of
a loosely bound state in high-energy reactions.
The factorization explained above is the foundation for the proposal to extract
the short-distance production mechanism of $\X$ in $B_c$ semi-leptonic and
hadronic decays~\cite{Wang:2015rcz}.

Note that it is a straightforward consequence of Eq.~(\ref{eq:factorizationX})
that in ratios of short-distance production rates for two
hadronic molecules related to each other through some symmetry the long-distance
part containing the information of the structure of the states cancels, and the remaining part solely reflects the difference in the
short-distance dynamics and phase space. It could be misleading if such ratios
are taken as evidence in support of or against dominantly composite nature of
given states.

All the derivations of the factorization formula are based on the LO \nreftii~so
far, which allows self-consistently only the possibility that $\Psi$ is a
composite system of $h_1$ and $ h_2$, see the discussion below
Eq.~\eqref{eq:gNR}. If we go to higher orders, momentum-dependent terms need to
be kept in the potential as well as the short-distance production amplitude.
The probability of $\Psi$ to be a $h_1h_2$ composite system would be
$1-\lambda^2<1$, and thus one would also need to introduce a contact
production term for $\Psi$.  These new terms parameterize the
short-distance physics though a detailed dynamics which depends on the more
fundamental theory that cannot be specified within the EFT. Such contributions
could
be interpreted as a short-distance core of the physical wave function $\Psi$.
 The factorization and the related renormalization at higher orders for
the near-threshold production of $h_1 h_2$ and $\Psi$ in short-distance
processes remain to be worked out.

The long-distance contribution in Eqs.~\eqref{eq:factorization} and
\eqref{eq:factorizationX} is calculable in
NREFT, and the short-distance contribution is subject to the more fundamental
theories which are QCD and/or electroweak theory.  For inclusive high-energy
hadron collisions, one is not able to calculate the short-distance contribution
model-independently (for estimates using Monte Carlo event generators,
see~\cite{Bignamini:2009sk,Artoisenet:2009wk,Bignamini:2009fn,Artoisenet:2010uu,
Esposito:2013ada,Guo:2013ufa,Guo:2014ppa,Guo:2014sca}), and thus
only the order of magnitude of the production
cross sections can be estimated. For
the production of hadronic molecules in heavy meson decays such as $B$ decays,
again one normally is only able to get an order-of-magnitude estimate at the
best due to the nonperturbative nature of QCD which dominates the hadronic
effects in the short-distance part.
The production rate of the $\X$ in $B$ decays was estimated in
Ref.~\cite{Braaten:2004ai,Braaten:2004fk}. In particular,
Ref.~\cite{Braaten:2004ai} predicted
that the branching fraction of  $B^0\to \X K^0$ should be much smaller than
that of  $B^+\to\X K^+$ assuming that the $\X$ is a $D^0\bar D^{*0}$
hadronic molecule. The prediction seems to be in contradiction with the later
measurements~\cite{Olive:2016xmw} summarized in Sec.~\ref{sec:2} giving a
ratio of around 0.5, {\sl c.f.} Eq.~(\ref{eq:sec2X3872RB}).
However, as already mentioned before the $\X$ is to a very good approximation an isoscalar state.
Its couplings to the neutral $D^0\bar D^{*0}$ and the charged $D^+D^{*-}$
channels is almost the same even if the isospin breaking is taken into
account~\cite{Gamermann:2009fv,Guo:2014hqa}.
Therefore Eq.~(\ref{eq:factorizationX}) needs to be generalized to coupled channels.
In particular  the production of $\Psi$ gets modified to
\begin{equation}
  \mathcal{M}_\Psi
  = \frac{1}{4(\pi)^{3/2}} \sum_i \Gamma_i\,\sqrt{\mu_i} g_{\text{NR},i} +
\order{\Lambda^{-1}}\,,
  \label{eq:factorizationX_multichannel}
\end{equation}
where the summation runs over all possible intermediate channels below the
cutoff $\Lambda$. Notably,
$g_{\text{NR},i}$ denotes the coupling of the $\Psi$ state to the $i$-th
channel.  As a result, the $\X$
production rates in neutral and charged $B$ decays should be similar. This may
also be understood as that the short-distance parts in~\cite{Braaten:2004ai}
should include the charged channel and were not properly estimated.

We now turn to the discussion of production rates of shallow bound states
like $\X$ at hadron colliders. It was claimed  that
the cross section for the inclusive $X(3872)$ production at high $p_T$ at the
Tevatron in $\bar pp$ collisions is too large to be consistent with the
interpretation of $X(3872)$ as a $D^0\bar D^{*0}$
molecule~\cite{Bignamini:2009sk}. The reasoning was based on the following
estimate for an upper
bound of the cross section:
\begin{eqnarray}
\sigma(\bar pp\to X) &\sim& \left| \int d^3{\bm k}\,\langle X|D^0\bar D^{*0}
({\bm k})\rangle\langle D^0\bar D^{*0}({\bm k})|\bar pp\rangle\right|^2
\nonumber \\
&\simeq& \left| \int_{\cal R} d^3{\bm k}\,\langle X|D^0\bar D^{*0}({\bm
k}) \rangle\langle D^0\bar D^{*0}({\bm k})|\bar pp\rangle\right|^2 \nonumber
\\
&\leq& \int_{\cal R} d^3 {\bm k} \left|\Psi({\bm k})\right|^2
\int_{\cal R} d^3 {\bm k}\left|\langle D^0\bar D^{*0}({\bm k})|\bar
pp\rangle\right|^2 \nonumber \\
&\leq& \int_{\cal R} d^3 {\bm k}\left|\langle D^0\bar D^{*0}({\bm
k})|\bar pp\rangle\right|^2 \nonumber \\ &\sim&
\sigma(\bar pp\to X)^{\rm max} \ ,
\label{eq:keyargument}
\end{eqnarray}
where $\cal R$ means that the momentum integration only receives support up to
some characteristic scale
$\cal R$. The upper bound quoted above depends drastically on the value of
$\cal R$. A value of ${\cal R}=35$~MeV$\simeq\gamma$, the binding momentum of
the $\X$, was chosen in~\cite{Bignamini:2009sk}. The so-estimated upper
bound of 0.071~nb is orders of magnitude smaller than the Tevatron result
of 37 to 115~nb.

\begin{figure}[t]
    \centering
    \includegraphics[width=\linewidth]{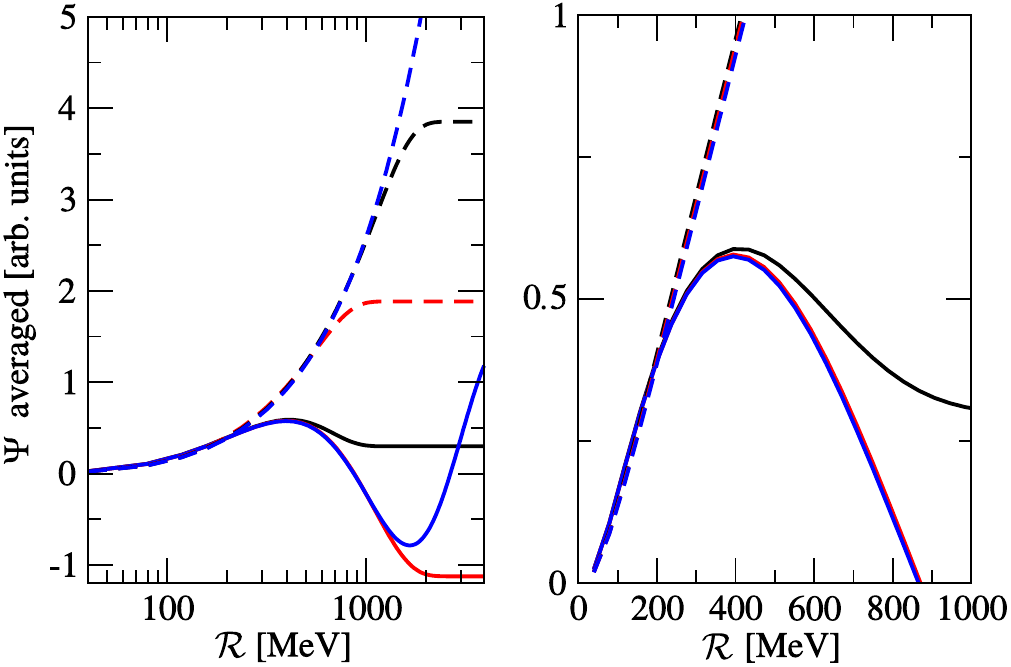}
    \caption{Averaged deuteron wave functions for various cutoff values:
$\Lambda = 0.8, 1.5, 4$ GeV shown as black, red, blue curves, respectively. The
solid (dashed) lines show the result for the wave functions with (without)
one-pion exchange. Taken from~\cite{Albaladejo:2017}.
\label{fig:wfaveraged}}
\end{figure}

However, for the derivation in Eq.~\eqref{eq:keyargument} to be valid  $\cal
R$ must be large enough that the wave function of the bound state
gets largely probed for otherwise the symbol between the first and the second
integral needs to be changed
from $\simeq$ to $\gg$ and the whole line of reasoning gets spoiled. But this requirement calls for
 values of $\cal R$ much larger than the binding momentum.
 To demonstrate this claim we switch to the deuteron wave function.
 Fig.~\ref{fig:wfaveraged} shows
the averaged deuteron wave function calculated from
$\bar \Psi_\Lambda({\cal R}) = \int_{\cal R} d^3 \bm{k} \,
\Psi_\Lambda(\bm{k})$,
where the subindex $\Lambda$ indicates that a regulator needs to be specified to
get a well defined wave function (for more details,
see~\cite{Nogga:2005fv}).
The right panel in the figure is a zoom in linear scale to the relevant $\cal R$
range. From the figure, it is clear that $\bar \Psi_\Lambda({\cal R})$ is far
from being saturated for ${\cal R}\simeq$45~MeV which is the deuteron binding
momentum. One needs to take ${\cal R}\sim 2M_\pi\sim$300~MeV, the order of the
inverse range of forces as pointed out
in~\cite{Artoisenet:2009wk,Artoisenet:2010uu} based on rescattering arguments,
to get a reasonable estimate so that the second line in
Eq.~\eqref{eq:keyargument} can be a good approximation of the first line. With
such a large support $\cal R$, the upper bound becomes
consistent with the Tevatron
measurement~\cite{Bignamini:2009sk,Artoisenet:2009wk,Artoisenet:2010uu}.
In addition, as discussed in case of the $B$ decays, the charged channels need
to be considered as well.

\begin{table}[t]
  \caption{Integrated cross sections (in units of nb) reported
in~\cite{Albaladejo:2017} for the inclusive $pp/\bar
p\to X(3872)$ processes in comparison with the
CDF~\cite{Bauer:2004bc} and CMS~\cite{Chatrchyan:2013cld} data converted into
cross sections~\cite{Guo:2014sca}. The ranges of the results cover those
obtained using both Pythia and Herwig. Here, we have
converted the experimental data into cross sections~\cite{Guo:2014sca}. }
\begin{ruledtabular}
\begin{tabular}{ l  c  c  c   }
  $\sigma(pp/\bar p\to X)$   & $\Lambda=0.1$~GeV   & $\Lambda=[0.5,1]$~GeV  &
Experiment \\
 \hline
 Tevatron   & 0.05-0.07         & 5-29   &  37-115   \\%
 LHC 7      & 0.04-0.12         & 4-55   &  13-39    \\%
\end{tabular}
\label{tab:crossec}
\end{ruledtabular}
\end{table}
In fact, with the factorization derived above, it was suggested that one can
estimate the cross section by combining the use of Monte Carlo event
generators, such as Pythia~\cite{Sjostrand:2007gs} and
Herwig~\cite{Bahr:2008pv},
and EFT to get the short-distance and long-distance contributions,
respectively~\cite{Artoisenet:2009wk,Artoisenet:2010uu,Guo:2014sca}. In
Table~\ref{tab:crossec}, we show the estimates obtained
in~\cite{Albaladejo:2017}. Indeed, if a small cutoff is used,
$\Lambda=0.1$~GeV,  and only the neutral charmed
mesons are considered by using~\cite{Guo:2014ppa,Guo:2014sca}
\begin{eqnarray}
  \sigma(pp/\bar p\to X)
  &\approx& C\, 2\pi^2 \left|\int \frac{d^3{\bm
  k}}{(2\pi)^3} \Psi_X({\bm k})\right|^2 \, \nonumber\\
  &=& C\, 2\pi^2 \left| g_{\text{NR},X}^{} \Sigma_\text{NR}^{}(-E_B)\right|^2,
\end{eqnarray}
the obtained cross sections are orders of magnitude smaller than the data, in
line with the observation in~\cite{Bignamini:2009sk}.
Here $C=d\sigma[D^0\bar D^{*0}]/dk/k^2$, playing the role of
$|\Gamma^\Lambda|^2$ in Eq.~\eqref{eq:Mfactorization}, is a constant determined
from fitting to the
differential cross section of the direct production of the charmed-meson pair
from Pythia and Herwig, $\Psi_X(\bm{k})$ is the momentum-space wave function of
the $\X$, $g_{\text{NR},X}^{}$ is the coupling of the $\X$ to $D^0\bar D^{*0}$,
$\Sigma_\text{NR}(-E_B)$ is the loop function in Eq.~\eqref{eq:SigmaGaussian}
but keeping the full $\Lambda$ dependence from the Gaussian regulator. However,
when a larger cutoff in the range of, e.g., $[0.5,1.0]$~GeV is used, the cross
sections become consistent with both the CDF and CMS measurements. One
important point is that the charged $D^+D^{*-}+c.c.$ channel needs to be taken
into account for this case as discussed above. This is because the binding
momentum for the charged
channel $\gamma_\pm\simeq126$~MeV is well below the cutoff so that the charged
charmed mesons should also play a dynamical role.

Finally, in Ref.~\cite{Esposito:2015fsa}, the cross sections for the production
of light (hyper-)nuclei at small $p_T$ at ALICE~\cite{Adam:2015vda} were
extrapolated to large $p_T$, and it was found that they are much smaller than
the $\X$ production at large $p_T$ at CMS~\cite{Chatrchyan:2013cld}.
Since light (hyper-)nuclei are loosely bound states of baryons, the authors
concluded that loosely bound states are hardly produced at high $p_T$, and
therefore disfavored the hadronic molecular interpretation of the $\X$.
However, although the long-distance contributions for these productions can be
managed in the EFT or universality framework, the short-distance contribution
for the $\X$ is completely different from that for light nuclei. This leads to a
different energy dependence of the cross sections for the light nuclei and the
$\X$, and makes such a direct comparison questionable.
One essential difference is: At short distances, the $\X$ can be produced
through $c\bar c$ or $u\bar u (d\bar d)$, which hadronizes into a pair of
charmed mesons at larger distances, while the minimal quark number in the light
nuclei is always $3N$, with $N$ the number of baryons, giving rise to a
suppression.
Therefore, it is natural that the $\X$ production cross section at very high
$p_T$ is orders of magnitude larger than that of light nuclei. This point also
leads to the critique~\cite{Guo:2016fqg},\footnote{For a response, see
\cite{Brodsky:2016uln}.} which was appreciated in~\cite{Voloshin:2016phx},
against the use of constituent counting rules in hard exclusive processes as a way to identifying mulqituark states with a
hidden-flavor $q\bar q$
pair~\cite{Kawamura:2013iia,Kawamura:2013wfa,Brodsky:2015wza,Chang:2015ioc}.

To summarize, the production and decay processes with a large energy release
involve both long and short distance scales. Only the long-distance part is
sensitive to the low-energy quantities, thus to the hadronic molecular
structure, and can be dealt with in the EFT framework. On the contrary, high
energy production rates depend crucially on what happens at short distances,
which is often unknown though in principle could be extracted from other
reactions depending on the same short-distance physics.
However, despite that it is hard to calculate the integrated production rates,
the differential invariant mass distributions around the near-threshold states
provide a direct access to their line shapes and precise, high-resolution data
on those are urgently called for. This has been discussed in general in
Sec.~\ref{sec:lineshapes} and also in Sec.~\ref{sec:morelineshapes}.

\subsection{Implications of heavy quark spin and flavor symmetries}

It turns out that HQSS, and especially its breaking, is also an important
diagnostic tool when it comes to understanding the structure of certain
states~\cite{Cleven:2015era}. The most straightforward example to illustrate
this point is the spin doublet made of $D_{s0}^*(2317)$ and $D_{s1}(2460)$. On
the one hand, both are not only significantly lighter than the prediction of the
quark model as reported in, e.g., Ref.~\cite{DiPierro:2001dwf}, also their spin
splitting differs. On the other hand, the mass difference of the two states
agrees exactly to the mass difference between the $D$ and the $D^*$, which would
be a natural result if $D_{s0}(2317)$ and $D_{s1}^*(2460)$ were $DK$ and $D^*K$
molecular states, respectively, since at LO in the heavy quark expansion the
$DK$ interaction agrees to the $D^*K$ interaction~\cite{Kolomeitsev:2003ac}.
In complete analogy it was argued in Ref.~\cite{Guo:2009id} that, if indeed the
$J^{PC}=1^{--}$ state $Y(4660)$ were a bound system of $\psi'$ and $f_0(980)$ as
conjectured in Ref.~\cite{Guo:2008zg}, there should exist a $J^{PC}=0^{-+}$
state which is $\eta_c' f_0(980)$ bound system. The mass difference of the
latter state to the $\eta_c' f_0(980)$ should agree to that of $Y(4660)$ to the
$\psi' f_0(980)$ threshold,\footnote{This prediction receives support from a
calculation using QCD sum rules~\cite{Wang:2009cw}.} and it was even possible to
estimate the decay width of the state as well as a most suitable discovery channel.
For more discussion on the implications of HQSS for the spectrum of exotic
states we refer to Ref.~\cite{Cleven:2015era}, where also predictions from other
approaches are contrasted to those of the molecular picture.

\begin{table*}[tb]
\caption{\label{tab:partners} Possible spin and flavor partners of heavy-flavor
hadronic molecules. For the experimentally established states, the masses and
decay modes are taken from~\cite{Olive:2016xmw}. The predicted partners are
denoted by question marks.
The predictions from~\cite{Guo:2013sya} are those computed with a 0.5~GeV
cutoff, and the result from~\cite{Baru:2016iwj} is taken from the results with
the cutoff limited between 0.8 and 1.0~GeV.
}
\begin{ruledtabular}
\begin{tabular}{l c c c c }
        $J^{P(C)}$ & State & Main component & Mass (MeV) & (Expected) main decay
        mode(s) \\\hline
       $0^+$ & $D_{s0}^{*}(2317)$ & $DK$ & $2317.7\pm0.6$  & $D_s^+\pi^0$ \\
       $1^+$ & $D_{s1}(2460)$ & $D^*K$ & $2459.5\pm0.6$  &
$D_s^{*+}\pi^0,D_s^{(*)+}\gamma$ \\
       $0^+$ & $B_{s0}^{*}(?)$ & $B\bar K$ & $5730\pm16$  &
       $B_s^{*0}\gamma, B_s^0\pi^0$
       \\
       $1^+$ & $B_{s1}(?)$ & $B^*\bar K$ & $5776\pm16$  &
       $B_s^{(*)0}\gamma, B_s^{*0}\pi^0$
       \\\hline
       $1^-$ & $D_{s1}^*(2860)$ & $D_1(2420)K$ & $2859\pm27$ & $DK$, $D^*K$ \\
       $2^-$ & $D_{s2}^*(?)$ & $D_2(2460)K$ &
$2910\pm9$~\cite{Guo:2011dd} &
       $D^*K,D_s^*\eta$ \\
       $1^-$ & $B_{s1}^* (?)$ & $B_1(5720)\bar K$ &
$6151\pm33$~\cite{Guo:2011dd} & $B^{(*)} \bar K$, $B^{(*)}_s\eta $ \\
       $2^-$ & $B_{s2}^*(?)$ & $B_2(5747)\bar K$ &
$6169\pm33$~\cite{Guo:2011dd} &
       $B^{*} \bar K$, $B^{*}_s\eta$ \\
       \hline
       $1^{++}$ & $X(3872)$ & $D\bar D^*$ & $3871.69\pm0.17$ & $D^0\bar D^0\pi^0$,
       $J/\psi\pi\pi$, $J/\psi\pi\pi\pi$
       \\
       $2^{++}$ & $X_2 (?)$ & $D^*\bar D^*$ & $4012^{+4}_{-5}$~\cite{Guo:2013sya} & $D\bar
       D^{(*)}, J/\psi\omega$
       \\
        &      & &  $3980\pm20$~\cite{Baru:2016iwj}
         &
         \\
       $1^{++}$ & $X_b(?)$ & $B\bar B^*$ &  $10580^{+9}_{-8}$~\cite{Guo:2013sya} &
       $\Upsilon(nS)\omega,\chi_{bJ}\pi\pi$
       \\
       $2^{++}$ & $X_{b2}(?)$ & $B^*\bar B^*$ & $10626^{+8}_{-9}$~\cite{Guo:2013sya} &
       $B\bar B^{(*)},\Upsilon(nS)\omega,\chi_{bJ}\pi\pi$ \\
        $2^{+}$ & $X_{bc}(?)$ & $D^*B^*$& $7322^{+6}_{-7}$~\cite{Guo:2013sya} &
$DB,DB^*,D^*B$ \\
   \end{tabular}
\end{ruledtabular}
\end{table*}

More predictions can be made for heavy-flavor hadronic molecules by using heavy
quark flavor symmetry. The LO predictions are rather straightforward if there is
only a single heavy quark in the system. For instance, one would expect the
$D_{s0}^*(2317)$ as a $DK$ bound state to have a bottom partner, a $\bar B K$
bound state, with almost the same binding energy. This prediction together with
the one for the bottom partner of the $D_{s1}(2460)$ are given in the fourth and
fifth rows in Table~\ref{tab:partners}, where the error of 16~MeV accounts for
the use of heavy quark flavor symmetry and is estimated as $2\left( M_D+M_K-
M_{D_{s0}^*(2317) } \right)\times \Lambda_\text{QCD}(m_c^{-1}-m_b^{-1}) $. Such
simple predictions are in a remarkable agreement with the lattice results of the
lowest-lying $0^+$ and $1^+$ bottom-strange mesons:
$(5711\pm13\pm19)$~MeV for the $B_{s0}^*$ and $(5750\pm17\pm19)$~MeV for the
$B_{s1}$~\cite{Lang:2015hza}. This agreement may be regarded as a further
support of the hadronic molecular nature of the $D_{s0}^*(2317)$ and
$D_{s1}(2460)$ states.
For more complicated predictions of these two states using various versions of
UCHPT, we refer to
Refs.~\cite{Kolomeitsev:2003ac,Guo:2006fu,Guo:2006rp,Cleven:2010aw,
Altenbuchinger:2013vwa,Torres-Rincon:2014ffa,Cleven:2014oka}.

A more detailed discussion of spin symmetry partners of hadronic molecules
formed by a pair of $S$-wave heavy mesons can be found in
Sec.~\ref{sec:4-interactions}. Some of the spectroscopic consequences of HQSS
and heavy quark flavor symmetry for hadronic molecules are listed in
Table~\ref{tab:partners}.

\subsection{Baryon candidates for hadronic molecules}
\label{sec:1405th}

For a discussion of and references for charmed baryons and the $P_c$
pentaquark structures as
possible hadronic molecules we refer to Sec.~\ref{sec:2}.
The closing part of this section will be used to describe the most
recent developments about the $\Lambda(1405)$ which basically
settled the debate on the nature of this famous state. What is described
below is yet another example how the interplay of high quality data
and systematic theoretical investigations allows one to identify
the nature of certain states.

As already stressed in Sec.~\ref{sec:lam1405exp}, there are good reasons to
classify the $\Lambda(1405)$ as  an exotic particle, since it does not at all
fit into the pattern of the otherwise in this mass range quite successful quark
models.

\begin{figure}[t!]
\begin{center}
\includegraphics[width=0.45\textwidth]{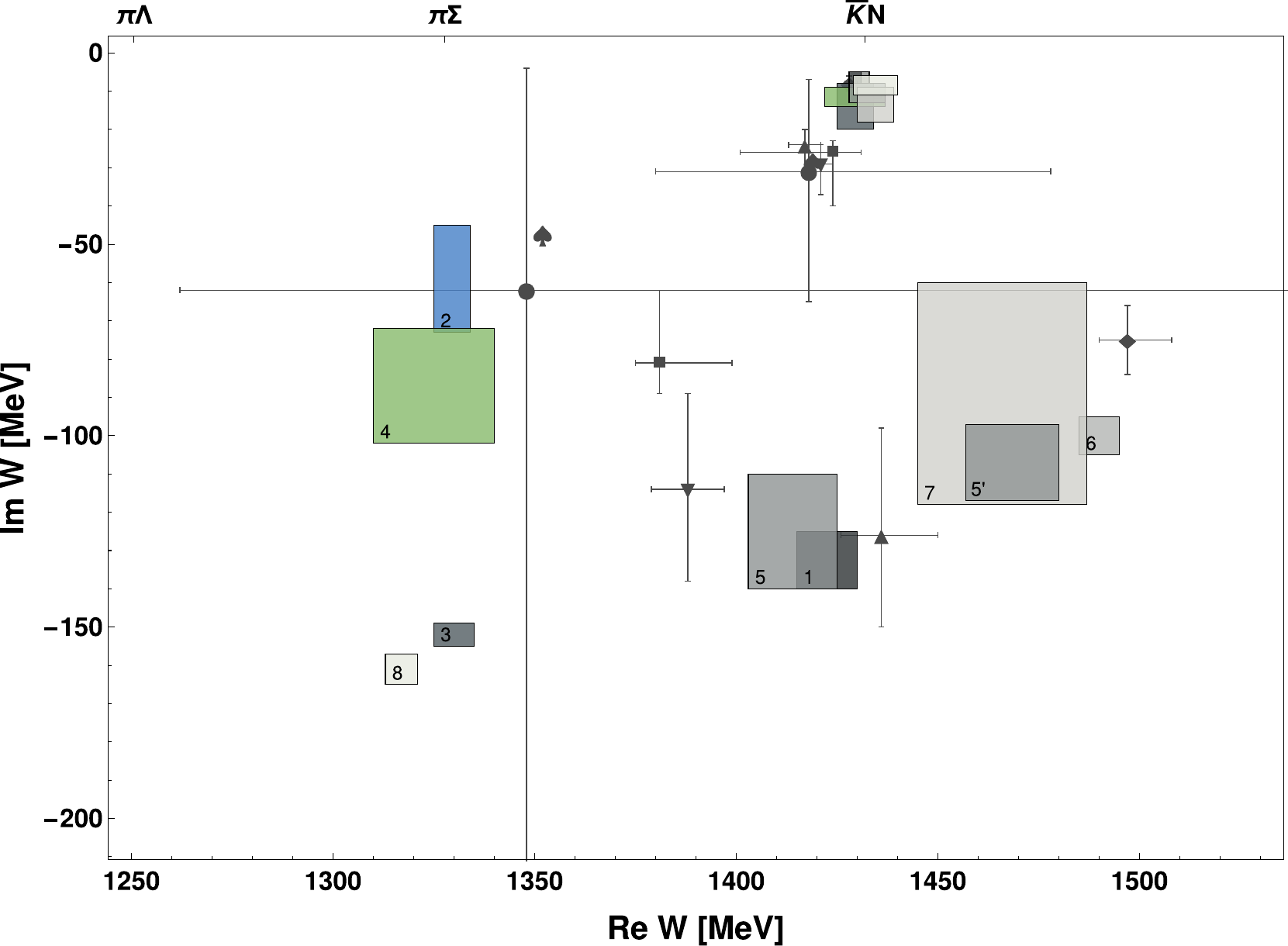}
\caption{Positions of the two poles of the $\Lambda(1405)$ in the
complex plane. Filled cirlces:~\cite{Borasoy:2006sr},
filled squares:~\cite{Ikeda:2012au}, triangles:~~\cite{Guo:2012vv},
boxes:~\cite{Mai:2014xna}, spades:~\cite{Roca:2013av}. The labels and
colors are described in the text. Figure courtesy of Maxim Mai.
}
\label{fig:twopole}
\end{center}
\end{figure}

Using information on $\bar KN$ scattering the existence of the $\Lambda(1405)$
was predicted by Dalitz and Tuan~\cite{Dalitz:1959dn,Dalitz:1960du}  before its
observation.
Already this study highlights the importance of the $\bar K N$ dynamics for the
$\Lambda(1405)$. In most modern investigations it appears as a dynamically
generated state through coupled-channel effects among all the ten isospin
channels ($K^-p$, $\bar{K}^0n$, $\pi^0\Lambda$, $\pi^0\Sigma^0$,
$\pi^+\Sigma^-$, $\pi^-\Sigma^+$, $\eta\Lambda$, $\eta\Sigma^0$, $K^+\Xi^-$,
$K^0\Xi^0$) or some of them, in other words, a hadronic molecule. This
coupled-channel problem has been studied by solving the LSE or Bethe--Salpeter
equation with interaction kernels derived from chiral perturbation theory within
a given accuracy. This procedure was first proposed in Ref.~\cite{Kaiser:1995eg}
and further refined in Ref.~\cite{Oller:2000fj} and various follow-ups.
The major finding of Ref.~\cite{Oller:2000fj} was the fact that there are indeed
two poles, one stronger coupled to the $\bar KN$ channel and the other to
$\Sigma\pi$, which should thus be understood as two distinct states.
Both poles are located at the second Riemann sheet, and have shadow poles in
the third one~\cite{Oller:2000fj}.
This two-pole scenario can be understood by considering the SU(3) limit and its
subsequent breaking, see~\cite{Jido:2003cb}. Presently, various groups have
performed calculations to the NLO
accuracy~\cite{Ikeda:2012au,Guo:2012vv,Mai:2012dt}, see  also the recent
comparison of all these works in Ref.~\cite{Cieply:2016jby}
 and the mini-review by the PDG~\cite{Olive:2016xmw}. Including
the photoproduction data on $\gamma p\to K^+ \Sigma\pi$ from
CLAS~\cite{Moriya:2013eb}, one finds that the heavier of the poles is fairly
well pinned down, while the lighter one still shows some sizable spread in its
mass and width~\cite{Mai:2014xna}.
All this is captured nicely in Fig.~\ref{fig:twopole}. Fitting only the
scattering and threshold ratio data with the NLO kernel, one finds two poles,
but with very limited precision~\cite{Borasoy:2006sr}.
Adding the precise kaonic hydrogen data, the situation changes markedly, as
shown by the different solutions found by various groups,
see~\cite{Ikeda:2012au,Mai:2012dt,Guo:2012vv}.
Still, as first pointed out in Ref.~\cite{Mai:2014xna}, even with these data
there is a multitude of solutions with almost equal $\chi^2$, as in depicted by
the boxes labeled $1,\ldots,8$ in Fig.~\ref{fig:twopole}.
However, the photoproduction data severely constrain this space of solutions.
From the eight solutions only two survive, the blue (solution~2) and the green
(solution~4) boxes in the figure from Ref.~\cite{Mai:2014xna} as well as the
modified LO solution depicted by the spades from Ref.~\cite{Roca:2013av}. This
again is a nice example that only through an interplay of various reactions one
is able to pin down the precise structure of hadronic molecules (or other
hadronic resonances). Clearly, more data on $\pi\Sigma$ mass distributions would
be needed to further sharpen these conclusions, see e.g.~\cite{Ohnishi:2015iaq}.

\section{Summary and outlook}
\label{sec:sum}

In this review, we have discussed the experimental indications for and
the theoretical approaches to hadronic molecules, which are a particular
manifestation of non-conventional states in the spectrum of QCD. The 
observation that these multi-hadron bound states appear close to or in 
between two-particle thresholds allows one to write down nonrelativistic
effective field theories. This gives a systematic access to the production,
decay processes and other reactions involving hadronic molecules. In the last
decade or so, through precise measurements of the spectrum of QCD invloving 
charm and bottom quarks, more and more potential hadronic molecules have
been observed. We have shown how explicit calculations of various decay modes
can be used to test this scenario. This is the only way to eventually
disentangle hadronic molecules from other multi-quark states like, e.g., 
tetraquarks. More detailed and accurate measurements are therefore called
for, complemented by first-principle lattice QCD calculations with
parameters close to the physical point and accounting for the 
involved coupled-channel dynamics related. More than 60 years after
Weinberg's groundbreaking work on the question whether the deuteron is an
elementary particle, we are now in the position to identify many more of such
loosely bound states in the spectrum of QCD and to obtain a deeper 
understanding of the mechanism underlying the appearance and binding 
of hadronic molecules.

\section*{Acknowledgments}

We are grateful to all our collaborators for sharing their insights into the
topics discussed here.
This work is supported in part by DFG and NSFC through funds provided to the
Sino-German CRC 110 ``Symmetries and the Emergence of Structure in QCD" (NSFC
Grant No.~11621131001, DFG Grant No.~TRR110), by NSFC (Grant Nos.11425525,
11521505 and 11647601), by the Thousand Talents Plan for Young Professionals, by
the CAS Key Research Program of Frontier Sciences (Grant No.~QYZDB-SSW-SYS013),
by the CAS President's International Fellowship Initiative (PIFI) (Grant
No.~2017VMA0025), and by the National Key Basic Research Program of China under
Contract No. 2015CB856700.

\bibliography{hmreview-fk}

\end{document}